\documentclass[%
 amsmath,amssymb,
 reprint,%
]{revtex4-2}

\usepackage[colorlinks=true, allcolors=blue]{hyperref}
\usepackage{graphicx}
\usepackage{siunitx}
\usepackage{xcolor}
\usepackage{ulem}

\usepackage{dcolumn}
\usepackage{bm}

\begin{document}

\preprint{AIP/123-QED}

\title[]{Multiscale studies of delayed afterdepolarizations I: A comparison of two biophysically realistic mathematical models for human ventricular myocytes}

\author{Navneet Roshan}
 \affiliation{Centre for Condensed Matter Theory, Department of Physics, Indian Institute of Science, Bangalore, 560012, India \\}
\author{Rahul Pandit}
\affiliation{Centre for Condensed Matter Theory, Department of Physics, Indian Institute of Science, Bangalore, 560012, India \\}
 \email{rahul@iisc.ac.in}

\date{\today}

\begin{abstract}
Focal arrhythmias, which arise from delayed afterdepolarizations (DADs), are observed in various pathophysiological heart conditions; these can lead to arrhythmias and sudden cardiac death. A clear understanding of the interplay of electrophysiological factors of cardiac myocytes, which lead to DADs, can suggest pharmacological targets that can eliminate DAD-induced arrhythmias. Therefore, we carry out multiscale investigations of two mathematical models for human-ventricular myocytes, namely, the ten Tusscher-Panfilov TP06~\cite{ten2006alternans} model and the HuVEC15 model~\cite{himeno2015human} of Himeno, \textit{et al.}, at the levels of single myocytes, one- and two-dimensional (1D and 2D) tissue, and anatomically realistic bi-ventricular domains. By using continuation analysis, we uncover steady- to oscillatory-state transitions in the Ca\textsuperscript{2+} concentrations and show that they lead to DADs. We demonstrate that the Sarco/endoplasmic reticulum Ca\textsuperscript{2+}-ATPase (SERCA) pump uptake rate and the Ca\textsuperscript{2+} leak through the ryanodine-receptor (RyR) channel impact this transition significantly. We show that the frequencies and amplitudes of the DADs are key features that can be used to classify them into three types, at the single-myocyte level. By carrying out detailed parameter-sensitivity analyses, we identify the electrophysiological parameters, in the myocyte models, that most affect these key features. We then obtain stability (or phase) diagrams that show the regions of parameter space in which different types of DADs occur. We demonstrate that the Na\textsuperscript{+}/Ca\textsuperscript{2+} exchanger plays a protective role by suppressing DADs in the TP06 model. We present representative tissue simulations of the spatiotemporal evolution of waves of electrical activation, in these models, to illustrate how arrhythmogenic premature ventricular complexes (PVCs) emerge from patches of DAD cells, when we pace the tissue.  We discuss the implications of our results for some DAD-induced ventricular arrhythmias, which we examine in detail in the companion Paper II.

\end{abstract}
\maketitle

\section{\label{sec:Intro}Introduction}
Cardiac diseases are the leading cause of mortality~\cite{nowbar2019mortality} in the industrialized world; and focal arrhythmias, engendered by afterdepolarizations, are one of the factors responsible for these deaths. These afterdepolarizations can occur during the recovery and diastolic phases of an action potential (AP); the former are known as early afterdepolarizations (EADs)~\cite{zimik2015comparative,volders1997similarities,vandersickel2014study} and the latter as delayed afterdepolarization (DADs); both EADs and DADs lead to an abnormal increase in the 
transmembrane potential $V_{\rm{m}}$ of a cardiac myocyte. We focus principally on DADs. 
Not only have DADs been found in human-cardiac-myocyte models~\cite{verkerk2001ionic}, but they have also been discovered in various excitable cell types across different species~\cite{kass1982fluctuations,marban1986mechanisms,rizzi2008unexpected,song2008increase,stambler2003characterization,verkerk2001ionic,wongcharoen2007effects,xie2006dioxin}. Given that DADs can lead to life-threatening cardiac arrhythmias, it is essential to understand their dependence on electrophysiological and cardiac-tissue properties. By studying two biophysically realistic human-ventricular-tissue mathematical models, with comprehensive descriptions of ionic channels and intracellular calcium-ion dynamics, and carrying out \textit{in silico} multiscale simulations - from single-cell to tissue levels in idealised and anatomically realistic domain - we uncover several properties of DAD-induced arrhythmias, which cannot be studied easily in \textit{in vitro, ex vivo, and in vivo} experiments. 
Therefore, our work complements such experimental studies. In particular,
we elucidate the dependence of DADs on cardiomyocyte parameters and uncover the spatiotemporal evolution of premature ventricular complexes (PVCs) that arise from regions of cardiac tissue with clumps of DAD cells.

We focus on DADs that are associated with human ventricular myocytes. These DADs have been observed in nonischemic heart failures~\cite{pogwizd2004cellular}, exercise-induced catecholaminergic polymorphic ventricular tachycardias (CPVTs)~\cite{leenhardt2012catecholaminergic}, Purkinje fibers that survive after myocardial infarctions~\cite{lazzara1973electrophysiological}, acidosis~\cite{orchard1994acidosis,lascano2013role}, hypertrophied failing hearts~\cite{vermeulen1994triggered}, digitalis toxicity~\cite{ferrier1973cellular,rosen1973mechanisms}, and increase in  catecholamines~\cite{priori1990mechanisms}. The underlying cause of the DADs are spontaneous Ca\textsuperscript{2+} releases (SCRs) at the sub-cellular level. These SCRs usually occur during intra-cellular Ca\textsuperscript{2+} overload~\cite{vassalle2004calcium}. To generate Ca\textsuperscript{2+} overload, various methods are in use in experiments~\cite{killeen2007arrhythmogenic,wu1995palmitoylcarnitine,verkerk2001ionic,verkerk2000injury}. We find that, in the models we study, enhancing the L-type calcium-channel current $I_{\rm{CaL}}$ suffices to get Ca\textsuperscript{2+} overload.
Once a Ca\textsuperscript{2+}-overload condition has been attained, the leak of the Ca\textsuperscript{2+} in the sarcoplasmic reticulum (SR), through the ryanodine-receptors (RyRs) channels, plays an important role in the development of SCRs~\cite{wleklinski2020molecular}. This leak~\cite{fabiato1983calcium} then leads to SCRs by modulating the opening probability of ryanodine receptors (RyRs) through Ca\textsuperscript{2+} in certain subspaces [e.g., the subspace SS in the TP06 model shown schematically in Fig.~\ref{fig:myocyte_schema}(a) and parts of the junctional space (jnc) in the HuVEC15 model shown schematically in Fig.~\ref{fig:myocyte_schema}(b)]. The SCRs increase $V_{\rm{m}}$ by activating the Na\textsuperscript{+}/Ca\textsuperscript{2+} exchangers (NCX or NaCa)~\cite{matsuda1997na+} in the forward mode, in which the electrogenic NCX extrudes one Ca\textsuperscript{2+} out of a myocyte and exchanges it with three Na\textsuperscript{+} ions.

Based on their effects on $V_{\rm{m}}$, DADs can be classified into different types, including subthreshold, suprathreshold, and oscillatory types. However, their distinguishing features need to be defined clearly; and investigations are required to establish the parameters that influence these features. We initiate a detailed study of these parameters in two biophysically realistic mathematical models for human ventricular myocytes, namely, the ten Tusscher-Panfilov TP06~\cite{ten2006alternans} model and the HuVEC15~\cite{himeno2015human} model of Himeno, \textit{et al.}.

Mechanisms of SCRs are the subject of experimental~\cite{knollmann2006casq2,palade1983spontaneous} and modeling studies~\cite{shiferaw2003model,colman2019arrhythmia}. 
Cardiomyocyte models ~\cite{walker2017estimating,colman2019arrhythmia}, with a detailed network of calcium release units (CaRUs) and sarcoplasmic reticulum (SR) combined with the stochastic description of RyRs, obtain \textit{inter alia} probability distributions of timing and amplitude of SCR events; averaged quantities, like SCR waveforms, which follow from such models, can then be employed in cardiac-tissue simulations. A detailed study of such models is computationally expensive as it tries to resolve sub-cellular scales. In particular, it is not possible to account simply, in such an approach, for evolving parameters that are found, e.g., in studies of acidosis ~\cite{lascano2013role}. Therefore, it is computationally advantageous to use common-pool myocyte models, in which each myocyte is treated as a grid point, and each physiological parameter can be varied suitably. Some other studies (e.g., Ref.~\cite{xie2010so}) utilize a threshold-driven (or \textit{commanded}) RyR opening in common-pool models to generate SCRs and DADs; but such approaches do not capture the effect of parameters [such as the Sarco/endoplasmic reticulum Ca\textsuperscript{2+}-ATPase (SERCA) pump uptake rate], on the SCR properties (e.g., the timing and strength of the SCRs).

Only a handful of common-pool myocyte models can give rise to SCRs and DADs. The following human-ventricular myocyte models are potential candidates for the triggering of DADs: the Iyer07~\cite{iyer2007mechanisms}, Fink08~\cite{fink2008contributions}, TP06~\cite{ten2006alternans}, and HuVEC15~\cite{himeno2015human}. We have been able to chart out the parameter regimes in which DADs occur in the latter two models, which we compare in detail. We use the Ca{\textsuperscript{2+}}-oscillation hypotheses, proposed in Ref.~\cite{fink2011ca2+}, to test the ability of these models to generate DADs. 
Furthermore, we identify the key features of SCRs and DADs that help us to classify DADs into three types.

 Before we present the details of our study, we give a qualitative overview of our principal results. We carry out an equilibrium-continuation analysis of the Ca\textsuperscript{2+}-subsystem in the TP06 model to uncover transitions from steady to oscillatory behaviors. We then characterize different types of DADs in the TP06 and HuVEC15 myocyte models. We identify the parameters that affect, sensitively, key features of the DADs in these models by using parameter-sensitivity analyses. This allows us (a) to obtain representative stability (or phase)
diagrams that show the regions of parameter space in which different types of DADs occur and (b) to explore how the interplay of these parameters leads to such DADs. We investigate a mechanism in which the NCX plays a protective role by suppressing the emergence of DADs in the TP06 model. Finally, we show, via detailed tissue simulations, in 1D, 2D, 3D, and anatomically realistic domains, into which we introduce patches with DAD myocytes, how PVCs emerge from such patches when we pace the tissue.
   
   We have organized the rest of this paper as follows. In Sec.~\ref{sec:Mat-and-Methods} we describe the models we use and the numerical and theoretical methods that we employ. Section~\ref{sec:results} is devoted to a detailed discussion of our results. Section~\ref{sec:disc_and_concl} discusses our results in the context of earlier numerical and experimental studies.

\section{\label{sec:Mat-and-Methods}Models and Methods}
\subsection{\label{subsec:Models}Models}

\begin{figure*}
     \centering
      \includegraphics[width=\textwidth]{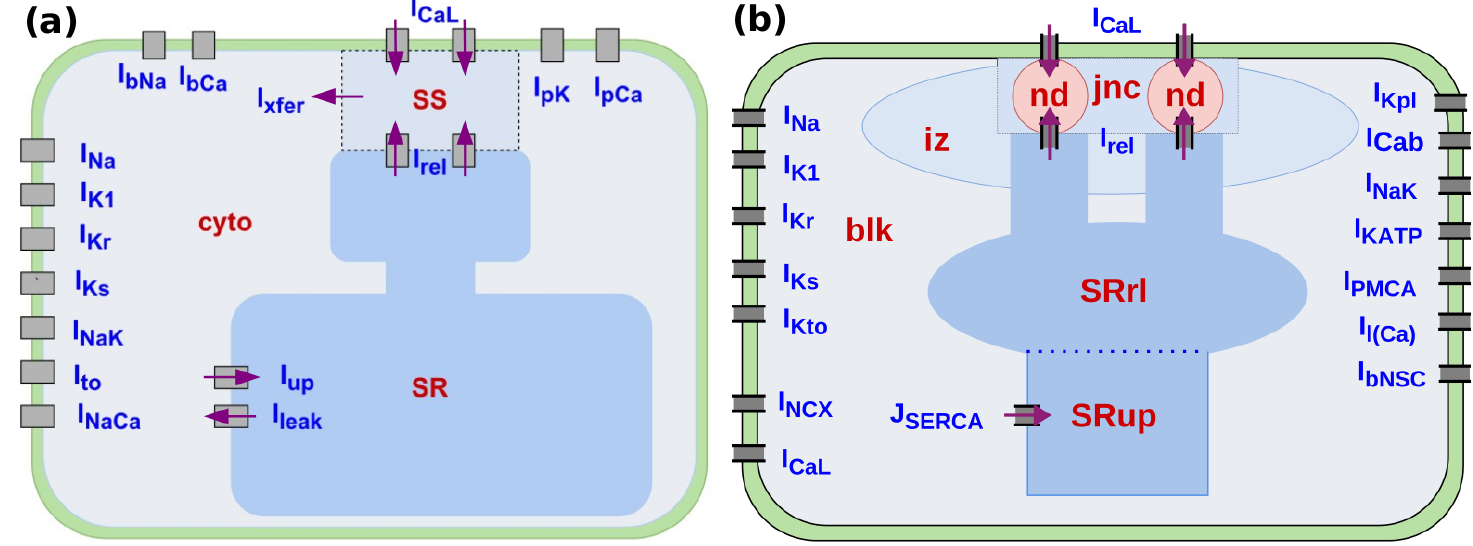}
      \caption{ (Color online) Schematic diagrams comparing the Ca\textsuperscript{2+} subsystems in the TP06 and HuVEC15 human-ventricular-myocyte models. (a) The TP06 myocyte volume is divided into three Ca\textsuperscript{2+} compartments: the sub-space (SS), the sarcoplasmic reticulum (SR), and the cytosol (CYTO); $I_{\rm{rel}}$ is the Ca\textsuperscript{2+} release rate from the SR to the SS; the RyRs and $I_{\rm{CaL}}$ open into the SS. $I_{\rm{up}}$ is the SERCA pump uptake rate of Ca\textsuperscript{2+} from the CYTO to the SR; $I_{\rm{leak}}$ is the Ca\textsuperscript{2+} leak from the SR to the CYTO; and $I_{\rm{xfer}}$ is the diffusion flux of Ca\textsuperscript{2+}  ions from the SS to the CYTO; $I_{\rm{NCX}}$ and $I_{\rm{CaL}}$ are transmembrane currents. (b) The HuVEC15 is a tightly coupled $I_{\rm{CaL}}$ and RyR model and it has more Ca\textsuperscript{2+} compartments than there are in the TP06 model; these are the junctional space (jnc), the intermediate zone (iz), the nano domain (nd), and the bulk space (blk). The sarcoplasmic reticulum (SR) has two sub-compartments, namely, SR\textsubscript{up} and SR\textsubscript{rl},  with Ca\textsuperscript{2+} concentrations $[Ca^{2+}]_{\rm{SR\textsubscript{up}}}$ and $[Ca^{2+}]_{\rm{SR\textsubscript{rl}}}$, respectively; the jnc is the region where the RyR and $I_{\rm{CaL}}$ channel openings meet; $J_{\rm{SERCA}}$ is the Ca\textsuperscript{2+} uptake rate from the CYTO to the SR; $I_{\rm{NCX}}$ and $I_{\rm{CaL}}$ are transmembrane currents, that are part of Ca\textsuperscript{2+}- subsystem.}
      \label{fig:myocyte_schema}
\end{figure*}

 To describe the human-ventricular-myocyte action potential (AP) and its Ca\textsuperscript{2+} subsystem, we use the TP06~\cite{ten2006alternans} and HuVEC15~\cite{himeno2015human} mathematical models, which adopt different approaches for 
 the modeling of the Ca\textsuperscript{2+} subsystem. The the schematic diagrams in Fig.~\ref{fig:myocyte_schema} 
 illustrate the differences between these models.

The sarcoplasmic reticulum (SR) is a single compartment in the TP06 model; in contrast, in the HuVEC15 model, this is divided into two compartments, namely, the SR uptake (SR\textsubscript{up}) and the SR release (SR\textsubscript{rl}) compartments. In the TP06 model, the ionic-diffusion space, i.e., the space inside the myocyte and outside the SR, is divided into the subspace (SS) and the cytosol (CYTO); whereas, in the HuVEC15 model, the ionic diffusion space is divided into three compartments, a junctional space (jnc), an intermediate zone (iz), and a bulk space (blk). The volume of the subspace in the TP06 model is $0.33\%$ of the myocyte volume (V\textsubscript{c}), whereas the volume of the junctional space in the HuVEC15 model is slightly larger ($\simeq 0.8\%$ of the myocyte volume (V\textsubscript{c})). To model the RyR calcium release, the TP06 model uses a reduced version of the four-state Markov model of CICR [developed in Refs.~\cite{shannon2004mathematical} and \cite{stern1999local}]; the RyR opening and closing dynamics incorporate the influences of calcium in the SS and the SR, namely, $[Ca^{2+}]_{\rm{SS}}$ (the trigger) and $[Ca^{2+}]_{\rm{SR}}$ (the load). The CaRU in the HuVEC15 model relies on the tightly coupled $I_{\rm{CaL}}$ and RyR model [developed by Hinch, \textit{et. al}~\cite{fink2008contributions}]; the calcium release from the SR is modeled as a regenerative activation of RyRs; this relies on two coupled states, closed (c) and open (o), to represent the opening and closing dynamics; these states depend on the calcium concentration in a small subdomain called the nano-domain (nd), which exists in the junctional space (jnc), between the $I_{\rm{CaL}}$ and RyR clusters.

The TP06 and HuVEC15 models account for $12$ and $14$ transmembrane ionic currents, respectively.
We use the following ordinary differential equation (ODE) for the single-myocyte $V_{\rm{m}}$:
\begin{align}
\label{eq:ODE}
\begin{split}
\dfrac{\partial V_{\rm{m}}}{\partial t} = & -\dfrac{I_{\rm{stim}} + I_{\rm{model}}}{C_{\rm{m}}}\,; 
\end{split}
\end{align}
and we use the following partial differential equation (PDE) for the spatio-temporal evolution of $V_{\rm{m}}$ in cardiac tissue:
\begin{align}
\label{eq:PDE}
\begin{split}
 \dfrac{\partial V_{\rm{m}}}{\partial t} = & -\dfrac{I_{\rm{stim}} + I_{\rm{model}}}{C_{\rm{m}}} + D\nabla^2 V_{\rm{m}}\,;
\end{split}
\end{align}
here, $t$ is the time, $C_{\rm{m}}$ is the capacitance per unit area of the myocyte membrane, $I_{\rm{stim}}$ is the externally applied current stimulus to the myocyte, $I_{\rm{model}}$ is the sum of all transmembrane ionic currents in the model, and $D$ is the diffusion constant, which is taken to be a scalar for simplicity (except when we employ an anatomically realistic bi-ventricular domain). For the TP06 and HuVEC15 models, we use, respectively,
\begin{eqnarray}
\label{eq:TP06currents}
I_{\rm{TP06}}&=&I_{\rm{Na}}+I_{\rm{CaL}}+I_{\rm{K1}}+I_{\rm{Kr}}+I_{\rm{Ks}}+I_{\rm{to}} \nonumber \\
&+& I_{\rm{pK}} + I_{\rm{bCa}}+I_{\rm{NaCa}}+I_{\rm{NaK}}+I_{\rm{bNa}}+I_{\rm{pCa}}
\end{eqnarray}
and
\begin{eqnarray}
\label{eq:HUVEC15currents}
I_{\rm{HuVEC15}}&=&I_{\rm{Na}}+I_{\rm{CaL}}+I_{\rm{K1}}+I_{\rm{Kr}}+I_{\rm{Ks}}+I_{\rm{Kto}} +I_{\rm{Kpl}} \nonumber \\ 
&+&I_{\rm{Cab}} + I_{\rm{NCX}}+ I_{\rm{NaK}}+I_{\rm{KATP}} +I_{\rm{PMCA}} \nonumber \\ 
&+&I_{\rm{l(Ca)}}+I_{\rm{bNSC}};
\end{eqnarray}
the currents for the TP06 (Eq.~\ref{eq:TP06currents}) and HuVEC15 (Eq.~\ref{eq:HUVEC15currents}) models are defined in Table~\ref{tab:ask_tp_currents}; and Refs.~\cite{ten2006alternans,himeno2015human} describe in detail the ODEs for ion-channels and gating variables in the TP06 and HuVEC15 models, respectively.
 
 \subsection{\label{subsec:deint}Numerical Integration of ODEs and PDEs}

 We integrate the ODEs for $V_{\rm{m}}$ by using the forward-Euler method, the gating-variable ODEs in the TP06 model by employing the Rush-Larsen~\cite{rush1978practical} scheme, and those for the HuVEC15 model via the generalized Rush-Larsen (GRL1) scheme~\cite{marsh2012secrets}. 

The time step we use to integrate the above ODEs and PDEs is $0.02$ ms; and the spatial-grid size is $0.025$ cm. Here, the diffusion constant is $D=0.00154 ~{\rm cm}^2/{\rm s}$, for the TP06 model, and $D=0.0012 ~~{\rm cm}^2/{\rm s}$, for the HUVEC15 model; these values lead to conduction velocities of $67~{\rm cm}/{\rm s}$ and $62~{\rm cm}/{\rm s}$ for the TP06 and HuVEC15 models, respectively. We employ three-, five-, and seven-point stencils for the Laplacians in our one-dimensional (1D), two-dimensional (2D), and three-dimensional (3D) simulations. In our 1D, 2D, and 3D simulations, we use the following domain sizes, respectively: a cable with 256 grid points; a rectangle with $512\times220$ grid points that is $12.8\times5.5 ~{\rm cm}^2$; and a cuboid with $512\times220\times20$ grid points that is $12.8\times5.5 \times 2 ~{\rm cm}^3$; in 3D we also carry out representative simulations for an anatomically realistic bi-ventricular domain. To trigger DADs, in the middle of the 1D cable, we introduce contiguous 30 (TP06 model) or 60 (HuVEC15 model) grid points with DAD myocytes; in 2D, we use a circular DAD-myocyte-clump  (henceforth, a DAD clump) of radius 40 (TP06 model) and 80 (HuVEC15 model) grid points; in a 3D cuboid domain, we use a cylindrical DAD clump of radius 40 (TP06 model) or 80  (HuVEC15 model) and a height of 20 grid points for both the models. For the simulations in the anatomically realistic human-bi-ventricular geometry, we obtain the DTMRI data from Ref.~\cite{winslow2011cardiovascular} for the human bi-ventricular geometry enclosed in a cubical box;
for this geometry, we use the phase-field approach [see, e.g., Refs.~\cite{fenton2005modeling,rajany2021effects,majumder2016scroll}]. We model the DAD clump in human bi-ventricular geometry as an overlapping region of a DAD sphere, with a radius of 80 grid points, located in the cubical box, and the human-bi-ventricular geometry [see Fig.~\ref{fig:whole_heart}(a)].

\begin{table*}[ht]
	\centering
	\begin{tabular}{|c|c|c|c|}
	\hline
	 \multicolumn{2}{|c|}{\textbf{TP06 currents}} & \multicolumn{2}{c|}{\textbf{HuVEC15 currents}} \\

\hline
$I_{\rm{Na}}$ & Fast Na\textsuperscript{+} & $I_{\rm{Na}}$ & Na\textsuperscript{+} (Fast and Late)   \\
\hline
 $I_{\rm{CaL}}$ & L-type Ca\textsuperscript{2+} & $I_{\rm{CaL}}$ & L-type Ca\textsuperscript{2+}   \\
\hline
$I_{\rm{K1}}$ & Inward rectifier K\textsuperscript{+} & $I_{\rm{K1}}$ & Inward rectifier K\textsuperscript{+}   \\
\hline
$I_{\rm{Kr}}$ & Rapid delayed rectifier K\textsuperscript{+} & $I_{\rm{Kr}}$ & Fast delayed rectifier K\textsuperscript{+}  \\
\hline
$I_{\rm{Ks}}$ & Slow delayed rectifier K\textsuperscript{+} & $I_{\rm{Ks}}$ & Slow  delayed rectifier K\textsuperscript{+}   \\
\hline
$I_{\rm{to}}$ & Transient outward K\textsuperscript{+} & $I_{\rm{Kto}}$ & Transient outward K\textsuperscript{+}    \\
\hline
$I_{\rm{pK}}$ & Plateau  K\textsuperscript{+} & $I_{\rm{Kpl}}$ & Plateau K\textsuperscript{+}  \\
\hline
 $I_{\rm{bCa}}$ & Background Ca\textsuperscript{2+} & $I_{\rm{Cab}}$ & Background Ca\textsuperscript{2+}    \\
\hline
$I_{\rm{NaCa}}$ & Na\textsuperscript{+}-Ca\textsuperscript{2+} exchanger & $I_{\rm{NCX}}$ & Na\textsuperscript{+}-Ca\textsuperscript{2+} exchanger  \\
\hline
$I_{\rm{NaK}}$ & Na\textsuperscript{+}- K\textsuperscript{+} ATPase & $I_{\rm{NaK}}$ & Na\textsuperscript{+}-K\textsuperscript{+} pump   \\
\hline
 $I_{\rm{bNa}}$ & Na\textsuperscript{+} background & $I_{\rm{KATP}}$ & ATP-sensitive potassium   \\
\hline
 $I_{\rm{pCa}}$ & Plateau Ca\textsuperscript{2+}  & $I_{\rm{PMCA}}$ & Plasma membrane Ca\textsuperscript{2+} -ATPase   \\
\hline
& & $I_{\rm{l(Ca)}}$ & Ca\textsuperscript{2+}-activated background cation   \\
\hline 
   & & $I_{\rm{bNSC}}$ & Background non-selective cation \\
 \hline
	\end{tabular}

	\caption{Lists of the ionic currents used in the TP06 and the HuVEC15 models for human-ventricular myocytes; Refs.~\cite{ten2006alternans} and \cite{himeno2015human} describe the details of the ODEs for the TP06 and HuVEC15 models, respectively.} 
	\label{tab:ask_tp_currents}
\end{table*}

In our studies, we scale the conductances and fluxes to model the up-regulation and down-regulation of various ion channels as in Ref.~\cite{mulimani2022silico}; e.g., to scale $G_{\rm{CaL}}$, we use $G_{\rm{CaL}} = S_{\rm{GCaL}} \times G_{\rm{CaL0}}$, where $G_{\rm{CaL0}}$ is the control value for $G_{\rm{CaL}}$ and $S_{\rm{GCaL}}$ is the scale factor for $G_{\rm{CaL}}$. For the tissue studies with the TP06 model, we use $I_{\rm{leak}} = 0$.

\subsection{\label{subsec:Caprotocol}Ca\textsuperscript{2+} overload protocol}

DADs are transient phenomena; they occur during Ca\textsuperscript{2+} overload in cardiac myocytes. Therefore, we scale $G_{\rm{CaL}}$ in mathematical models for cardiac myocytes to model this overload. An increase in $G_{\rm{CaL}}$ also increases the APD; to compensate for this increase in the APD, we increase $G_{\rm{Kr}}$ by a proportional factor. The scale factors for the representative case of runs R1-R4, for both TP06 and HuVEC15 models, are given in Table~\ref{tab:ICaLs_and_Ikr}; for the other values of $G_{\rm{CaL}}$, we use straight-line fits to get a suitable value of $G_{\rm{Kr}}$ (see the Appendix). In Fig.~\ref{fig:compare_APs_L_and_kr} we show plots of the AP (top row) for the (a) TP06 and (b) HuVEC15 models for various values of the scale factors $S_{\rm{GCaL}}$ and $S_{\rm{GKr}}$; these values are chosen so that the SR Ca\textsuperscript{2+} can be increased [Figs.~\ref{fig:compare_APs_L_and_kr}  (c) and (d)] without introducing any significant changes in the APDs.

\begin{figure}
     \centering
         \includegraphics[width=0.45\textwidth]{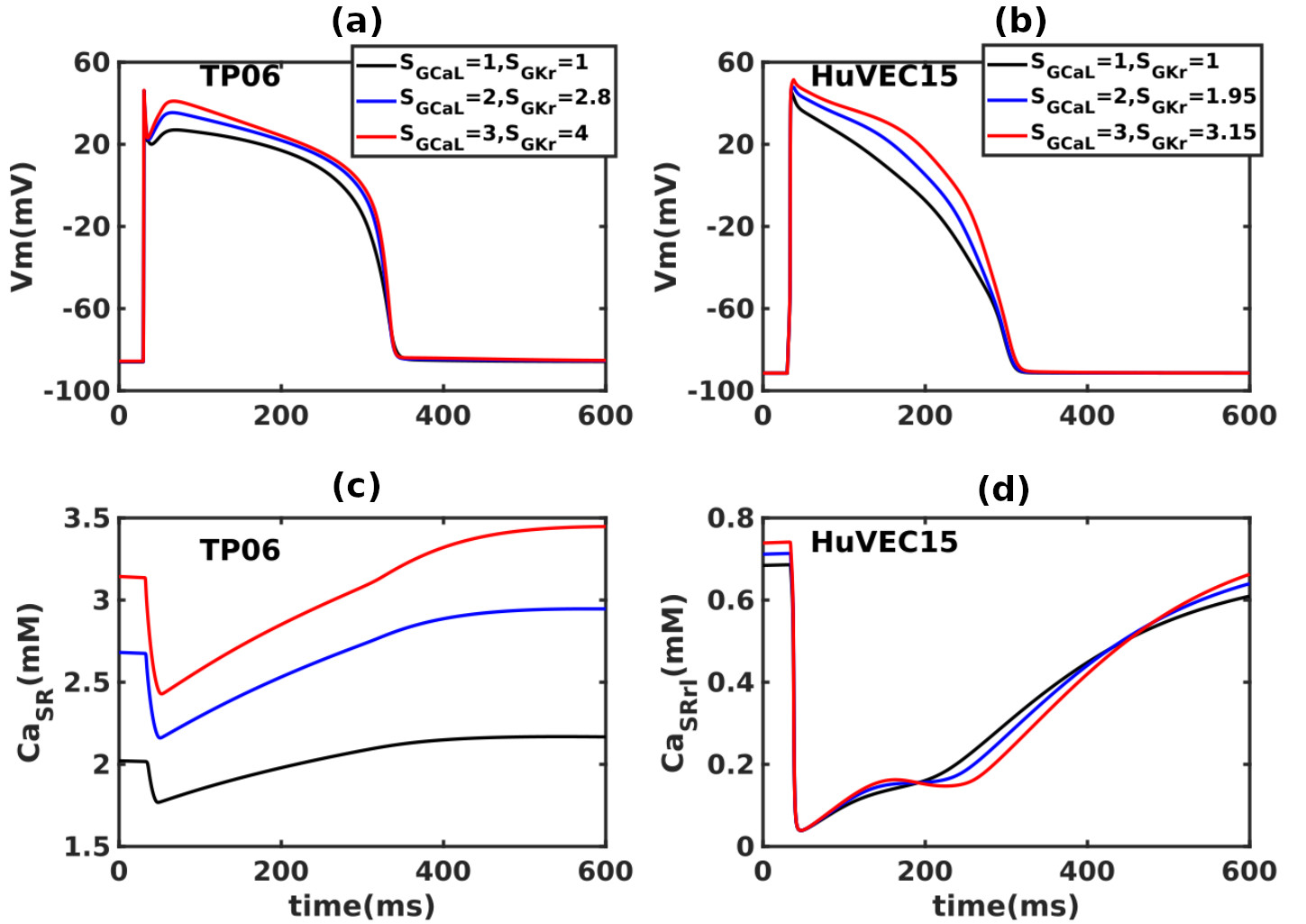}
         \caption{ (Color online)\textbf{The method for Ca\textsuperscript{2+} overload:} We increase $G_{\rm{CaL}}$, by changing $S_{\rm{GCaL}}$, to enhance the Ca\textsuperscript{2+} load; and we increase $G_{\rm{Kr}}$, by changing $S_{\rm{GKr}}$,  to counterbalance the rise in the APD because of the enhancement of $G_{\rm{CaL}}$. The resulting Ca\textsuperscript{2+}-overload protocol increases the Ca\textsuperscript{ 2+} content in the SR; in particular, this protocol increases $[Ca^{2+}]_{\rm{SR}}$ without changing the APD significantly. (a) and (b): are plots of APs for different values of $S_{\rm{GCaL}}$ and $S_{\rm{GKr}}$ from the TP06 and HuVEC15 models, respectively. (c) and (d): SR Calcium concentrations for the same factors of $S_{\rm{GCaL}}$ and $S_{\rm{GKr}}$ for both these models.}
         \label{fig:compare_APs_L_and_kr}
\end{figure}

\begin{table}[ht]

	\centering
	\begin{tabular}{|c|c|c|c|c|} 
	\hline
	Sr. & \multicolumn{2}{c|}{TP06} & \multicolumn{2}{c|}{HuVEC15} \\
	\cline{2-3}  \cline{4-5}
	no. & $S_{\rm{GCaL}}$ & $S_{\rm{GKr}}$ & $S_{\rm{GCaL}}$ & $S_{\rm{GKr}}$ \\
	\hline
	R1 & 1 & 1   & 1 & 1 \\
	\hline
	R2 & 2 & 2.8 & 2 & 1.95 \\
	\hline
	R3 & 3 & 4   & 3 & 3.15 \\
	\hline 
	R4 & 4 & 5.1 & 4 & 4.7 \\
	\hline
	
	\end{tabular}
	\caption{ The scale factors $S_{\rm{GCaL}}$ and $S_{\rm{GKr}}$ that we use for runs R1-R4 for the TP06 and HuVEC15 models
	(see text and Fig.~\ref{fig:compare_APs_L_and_kr}).}

	\label{tab:ICaLs_and_Ikr}
\end{table}

\subsection{The Ca\textsuperscript{2+} subsystem and numerical continuation}
\label{Ca-subsys}

The authors of Ref.~\cite{fink2011ca2+} have proposed that oscillations in the calcium-ion concentration, in the  Ca\textsuperscript{2+} subsystem of a cardiomyocyte model, might be associated with DADs. Therefore, we consider the Ca\textsuperscript{2+} subsystems, for the TP06 and HuVEC15 models, as follows: the Ca\textsuperscript{2+} subsystem contains equations for the  Na\textsuperscript{+}-Ca\textsuperscript{2+} exchanger, the SERCA  pump, the RyR release channels, and for the Ca\textsuperscript{2+} concentrations in various compartments. Note that the ODEs for the $V_{\rm{m}}$ and associated ion-channel dynamics are not part of the Ca\textsuperscript{2+} subsystem; their inclusion complicates the analysis, because we must then consider the full set of ODEs for these models.

The Ca\textsuperscript{2+} subsystem for the TP06 model has $4$ ODEs and that for the HuVEC15 Ca\textsuperscript{2+} model has $21$ ODEs. We provide the detailed ODEs for the Ca\textsuperscript{2+} subsystem for the TP06 model in the Appendix.

In our studies of numerical continuation, we follow~\cite{kuznetsov2019codim}. The ODEs we consider, e.g., the four ODEs for the TP06 model (see the Appendix), are of the form $\dot{u} = F(u,\alpha)$, with $u \in \mathbb{R}^n$ and $\alpha \in \mathbb{R}^m$. An equilibrium
$u_0$ satisfies $F(u_0,\alpha_0) = 0$; and the Jacobian matrix $\mathcal{A} = F_u(u_0,\alpha_0)$ has eigenvalues $\lambda_1, \lambda_2, \ldots , \lambda_N$. A Hopf bifurcation is characterised by the appearance of two, purely imaginary eigenvalues~\cite{kuznetsov2019codim}. The Matlab package Matcont~\cite{dhooge2003matcont}, which is a numerical-continuation toolbox for ODEs, allows us to obtain the manifold of equilibrium points (we have checked our results by also using the 
package XPPAUT~\cite{ermentrout2001xppaut}). We illustrate this, for the TP06 model, in  Subsection~\ref{subsec:eqcont}.

\subsection{\label{subsec:param-sens}Parameter-Sensitivity Analysis}

    The key features in $V_{\rm{m}}$ during DADs are: (a) the frequency of the DADs, because of multiple spontaneous calcium-ion releases; and (b) the amplitude of the DADs. We perform parameter-sensitivity analyses, as in Ref.~\cite{sobie2009parameter}, to obtain the principal model parameters that influence features (a) and (b) significantly 
    in the TP06 and  HuVEC15 models. In particular, we choose random scale factors, for the maximal conductances and the calcium fluxes, from a log-normal distribution that has a median value of $1$; and we use the standard-deviation parameter $\sigma=0.1$ to control the ranges of variation for these parameters. In this manner we generate $1000$ randomly chosen factors for each one of the conductances and fluxes; we use these for the inputs into our parameter-sensitivity analysis. For the TP06 model we use $14$ inputs; and for the HuVEC15 model we have $17$ inputs. Next, we compute the APs for these models, for a given set of input values, by stimulating the model myocyte with a train of $500$ stimuli (square current pulses of height $-52 ~\text{pA/pF}$  and duration $1 ~\text{ms}$ for the TP06 model and height $-12 ~\text{pA/pF}$ and duration $2.5 ~\text{ms}$ for the HuVEC15 model) with a pacing frequency of $1$ Hz. For each set of randomly chosen parameter inputs, we save the last ten APs and calculate the average amplitude and frequency of the DADs; these are our two outputs. With these outputs, we perform parameter-sensitivity analysis to obtain the parameters that influence these outputs sensitively,
	by constructing input and output matrices from these input and output data. With $n$, the number of samples, and $p$, the number of model parameters, we build the $n \times p$ input matrix $\mathbf{X}$. We also construct the $n \times m$ output matrix $\mathbf{Y}$, with $m$ the number of outputs ($m=2$ here). By using the matrices $\mathbf{X}$ and $\mathbf{Y}$, we perform a partial-least-squares (PLS) regression to calculate the regression coefficients $\mathbf{B_{PLS}}$.

\section{\label{sec:results}Results}
We present our results in the following subsections: Subsection~\ref{subsec:eqcont} is devoted to the equilibrium-continuation analysis of the Ca\textsuperscript{2+}-subsystem. In Subsection~\ref{subsec:types} we present and characterize different types of DADs in the TP06 and HuVEC15 myocyte models. Subsection~\ref{subsec:crucial-params} deals with the results of 
our parameter-sensitivity analyses, which allow us to identify the parameters that affect, sensitively, the frequencies
and amplitudes of the DADs in these models. We present in Subsection~\ref{subsec:phases} representative stability (or phase)
diagrams, that show the regions of parameter space in which different types of DADs occur. We explore, in Subsection~\ref{subsec:interplay}, how the interplay of these parameters leads to such DADs. In Subsection ~\ref{subsec:protective}, we elaborate on a mechanism in which the NCX plays a protective role by suppressing the emergence of 
DADs in the TP06 model. In the final Subsection ~\ref{subsec:tissue} we discuss our tissue simulations, in 1D, 2D, 3D, and anatomically realistic domains into which we introduce patches with DAD myocytes.

\subsection{\label{subsec:eqcont}Nonlinear analysis of the Ca\textsuperscript{2+}-subsystem}

\begin{figure}[h]
     \centering
         \includegraphics[width=0.45\textwidth]{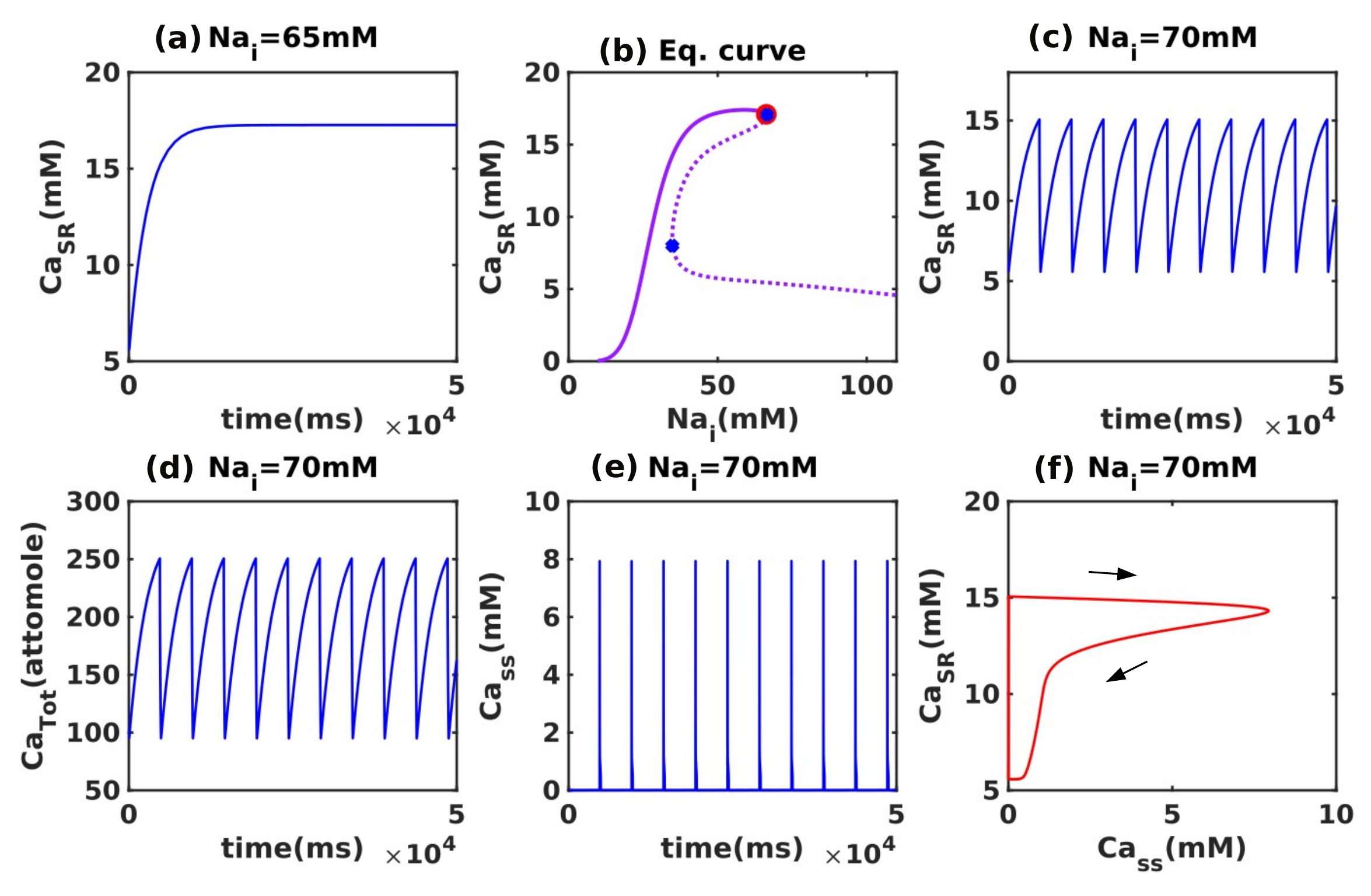}
         \caption{(Color online) \textbf{Equilibrium continuation and Ca\textsuperscript{2+} oscillations:} (a) Plot of 
         $Ca_{\rm{SR}}$ versus time $t$, at $Na_{\rm{i}}=65$ ~\text{mM}, showing saturation at long time to an equilibrium concentration; by using equilibrium continuation (see text) we detect an Andropov-Hopf bifurcation in the TP06 Ca\textsuperscript{2+}-subsystem.  (b): equilibrium curve exhibiting the dependence of fixed points on the parameter $Na_{\rm{i}}$. The solid line represents the stable equilibrium values. The red dot at $Na_{\rm{i}}=66.48$ ~\text{mM} shows the transition of the system to an oscillatory state, and the blue dots at the turning points show the emergence and vanishing of two extra equilibrium points.  (c): $Ca_{\rm{SR}}$ oscillations at $Na_{\rm{i}}=70$ ~\text{mM}; (d): total calcium content (free and buffered) $Ca_{\rm{Tot}}$ oscillations at $Na_{\rm{i}}=70$ ~\text{mM}; (e):  $Ca_{\rm{SS}}$ oscillations at $Na_{\rm{i}}=70$ ~\text{mM}; (f): $Ca_{\rm{SR}}$ and $Ca_{\rm{SS}}$ oscillations.}
         \label{fig:6_plots}
\end{figure}

In Fig.~\ref{fig:6_plots}(a) we demonstrate, by numerical integration of the ODEs for the Ca\textsuperscript{2+}-subsystem
in the TP06 model [see the Appendix], that, as time increases, $Ca_{\rm{SR}}$ achieves a steady-state equilibrium value if the control parameter $Na_{\rm{i}}=65$~\text{mM}. We start with this steady-state value and perform equilibrium continuation, by using the Matlab package Matcont~\cite{dhooge2003matcont}, to obtain the dependence of this equilibrium value on $Na_{\rm{i}}$, which we show via the purple full curve in Fig.~\ref{fig:6_plots}(b). We also find that a pair of neutral-saddle equilibrium points appear beyond a threshold value of $Na_{\rm{i}}$ that we indicate by a blue point in Fig.~\ref{fig:6_plots}(b), in which the dashed purple curve shows the dependence of the neutral-saddles on $Na_{\rm{i}}$; this curve meets the purple equilibrium at the red Hopf critical point. Beyond this critical value of $Na_{\rm{i}}$, there are no critical points and the long-time behavior of
$Ca_{\rm{SR}}$ is oscillatory, because of a limit cycle that results from an Andropov-Hopf bifurcation. In Fig.~\ref{fig:6_plots}(c) we present an illustrative plot of the temporal oscillations in $Ca_{\rm{SR}}$ at $Na_{\rm{i}}=70$~\text{mM}. Oscillations also occur in
the total calcium content (free and buffered) $Ca_{\rm{Tot}}$ and $Ca_{\rm{SS}}$ as we show in Figs.~\ref{fig:6_plots}(d) and (e), respectively, for the
representative value $Na_{\rm{i}}=70$ ~\text{mM}; the underlying limit cycle's projection is shown in 
Fig.~\ref{fig:6_plots}(f) via the red, closed curve in the  $Ca_{\rm{SR}}$ - $Ca_{\rm{SS}}$ plane. Thus, we have 
shown that the Ca\textsuperscript{2+}-subsystem in the TP06 model can show calcium oscillations, which lead, in turn,
to DADs in the myocyte AP. We expect that the Ca\textsuperscript{2+}-subsystem for the HuVEC15 model 
shows similar oscillations, but, given the large number of
ODEs in this subsystem, it is not easy to get convergence to equilibrium points by using the Matlab Matcont package.

\begin{figure}
     \centering
         \includegraphics[width=0.48\textwidth]{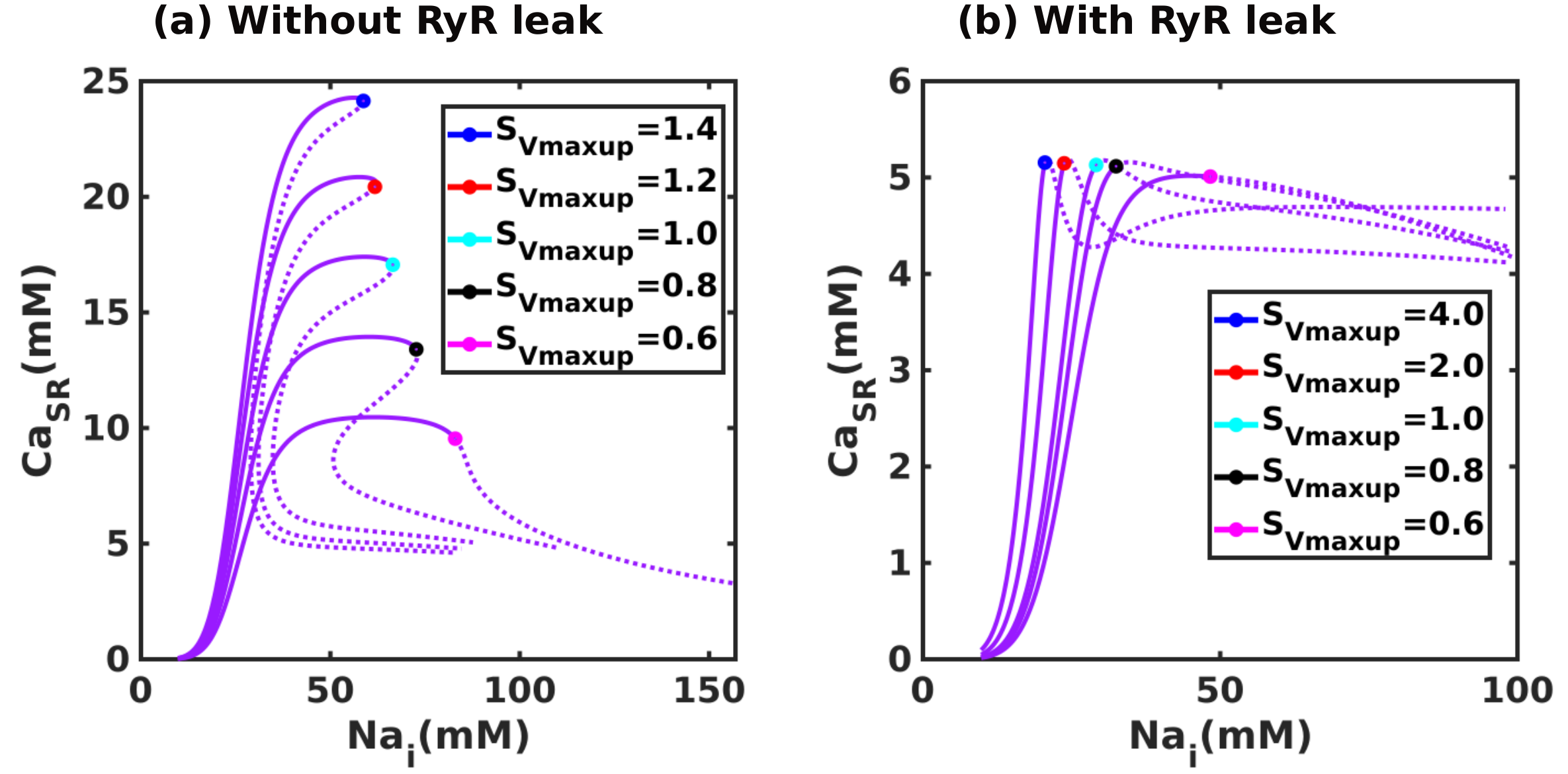}
         \caption{(Color online) \textbf{Andronov-Hopf bifurcations and their SERCA dependence:} To detect the change in the Hopf-bifurcation point with other parameters, such as the SERCA-pump uptake rate $S_{\rm{Vmaxup}}$, we perform equilibrium continuation for different values of $S_{\rm{Vmaxup}}$. (a): $Ca_{\rm{SR}}$ versus $Na_{\rm{i}}$ plots of equilibrium-continuation curves in purple and Hopf points on them; (b): $Ca_{\rm{SR}}$ versus $Na_{\rm{i}}$ plots of equilibrium-continuation curves and Hopf points on them for various values of $S_{\rm{Vmaxup}}$, in the presence of the RyR leak. In both the plots, an increase in $S_{\rm{Vmaxup}}$ shifts the Hopf point towards large values of $Ca_{\rm{SR}}$ and low values of $Na_{\rm{i}}$; by comparing (a) and (b), we can infer how the RyR leak affects the $Ca_{\rm{SR}}$ and $Na_{\rm{i}}$ requirements for this Andronov-Hopf bifurcation.}
         \label{fig:Hopf}
\end{figure}

In Figs.~\ref{fig:Hopf} (a) and (b) we show, for the TP06 Ca\textsuperscript{2+}-subsystem, how the 
equilibrium-continuation curves and Hopf point, shown in Fig.~\ref{fig:6_plots}(b), change as we tune the scale factor $S_{\rm{Vmaxup}}$ (for $V_{\rm{maxup}}$), both without and with the RyR leak, respectively. In particular, the Hopf point moves 
as we change the scale factor $S_{\rm{Vmaxup}}$; the Hopf points are shown in blue, red, cyan, black, and purple for $S_{\rm{Vmaxup}} = 1.4, 1.2, 1.0, 0.8,$ and $0.6,$ respectively.
By comparing Figs.~\ref{fig:Hopf} (a) and (b), we see that the RyR leak lowers the values of 
$Ca_{\rm{SR}}$ and $Na_{\rm{i}}$ at which the Hopf-point occurs.
Furthermore, we find that the NCX does not have a sizeable effect on the equilibrium-continuation curves and Hopf point. 
In summary, from our equilibrium-continuation analysis we infer that the TP06 model should be capable of triggering DADs and that the addition of the RyR significantly reduces the Ca\textsuperscript{2+}-overload requirement for oscillations. These key insights, about the ranges of various physiological parameters and state variables, inform our study of DADs in the next Subsection. \\

\subsection{\label{subsec:types}Types of DADs in the TP06 and HuVEC15 models}

\begin{figure}
     \centering
      \includegraphics[width=0.45\textwidth]{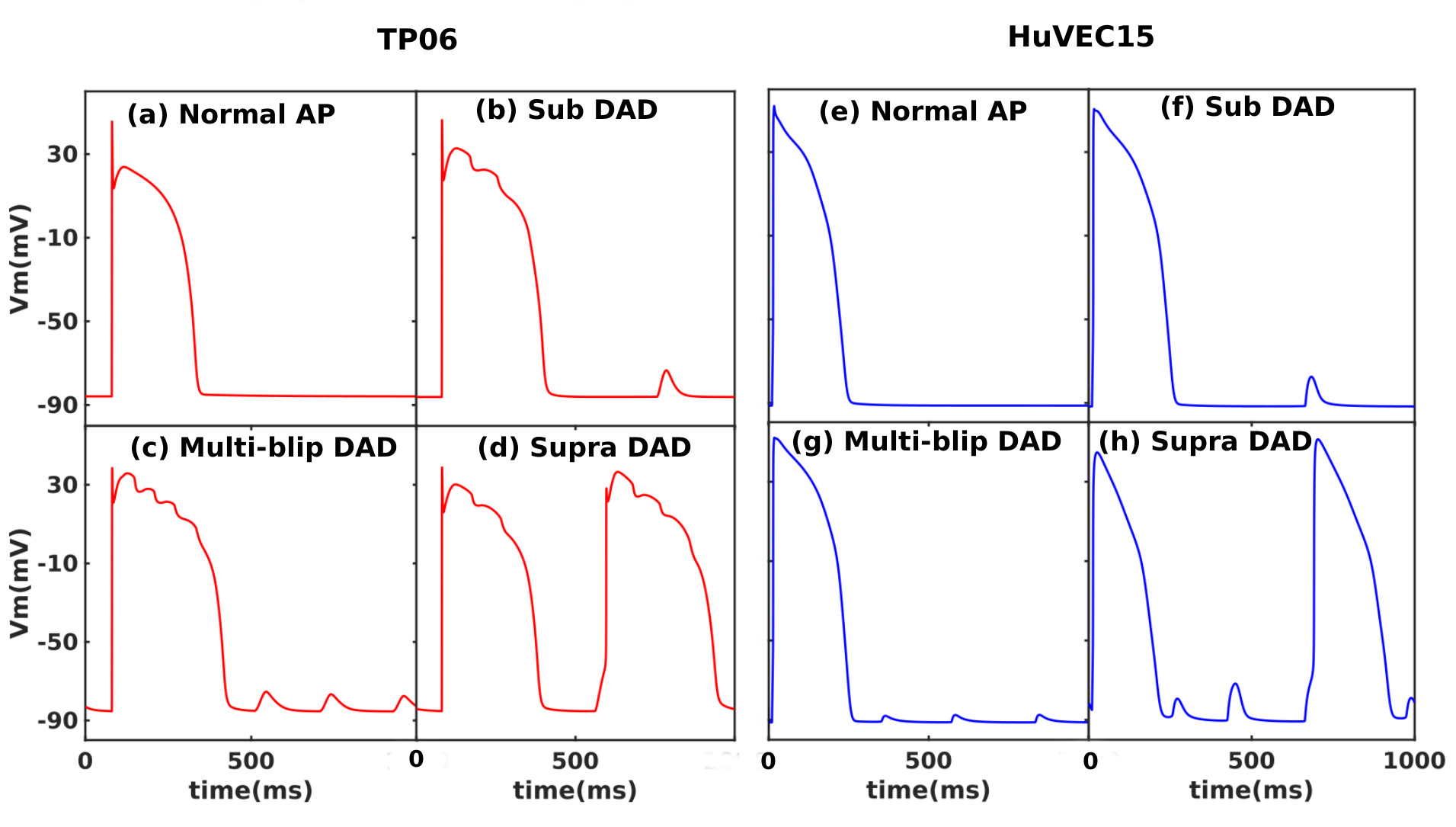}
      \caption{ (Color online) \textbf{Types of DADs:} Plots of the transmembrane potential showing the normal AP and different
      types of DADs (a)-(d) the TP06 model in red and (e)-(h) the HuVEC15 model in blue; (a),(e): normal AP; (b),(f): sub-threshold DADs; (c),(g): multi-blip DADs; (d),(h): supra-threshold DADs. Note that the TP06 model shows [(b)-(d)] both SCRs and LCRS, whereas the HuVEC15 model shows only LCRs [cf., Fig.~\ref{fig:SI:sparks_DAD_EAD}].}
      \label{fig:dadtypes}
\end{figure}

To observe DADs in the complete TP06 and HuVEC15 models, we increase $I_{\rm{CaL}}$ as we have discussed in Subsection~\ref{subsec:Caprotocol}. Then, by tuning the values of parameters such as $V_{\rm{rel}}$, $G_{\rm{CaL}}$, $K_{\rm{NaCa}}$, and $V_{\rm{maxup}}$, we observe three types of the DADs for which we present illustrative plots, along
with a normal AP for comparison, in Fig.~\ref{fig:dadtypes}. In addition to the usual sub-threshold [Figs.~\ref{fig:dadtypes} (b) and (f)] and supra-threshold [Figs.~\ref{fig:dadtypes} (d) and (h)] types of DADS, we uncover a third type,
which we call multi-blip DADs [Fig.~\ref{fig:dadtypes} (c) and (g)]; these are multiple subthreshold DADs, which do not reach the activation threshold of the fast $Na^{+}$-channel, between two successive APs; morphologically similar DADs have been shown in Refs.~\cite{shah2019delayed},
~\cite{zygmunt1998naca}, and ~\cite{catanzaro2006mechanisms}. 
The amplitude of the DAD is defined as the peak $V_{\rm{m}}$ of the DAD relative to the minimum potential during the diastolic interval; and its frequency is the number of DADs per second. This amplitude and frequency are hard to define unambiguously in the supra-threshold DAD regime. In Subsection~\ref{subsec:phases}, we explore the parameter regions in which these types of DADs occur.

\subsection{Sensitivity analysis: crucial parameters for the DAD frequency and amplitude}
\label{subsec:crucial-params}

\begin{figure}
     \centering
      \includegraphics[width=0.45\textwidth]{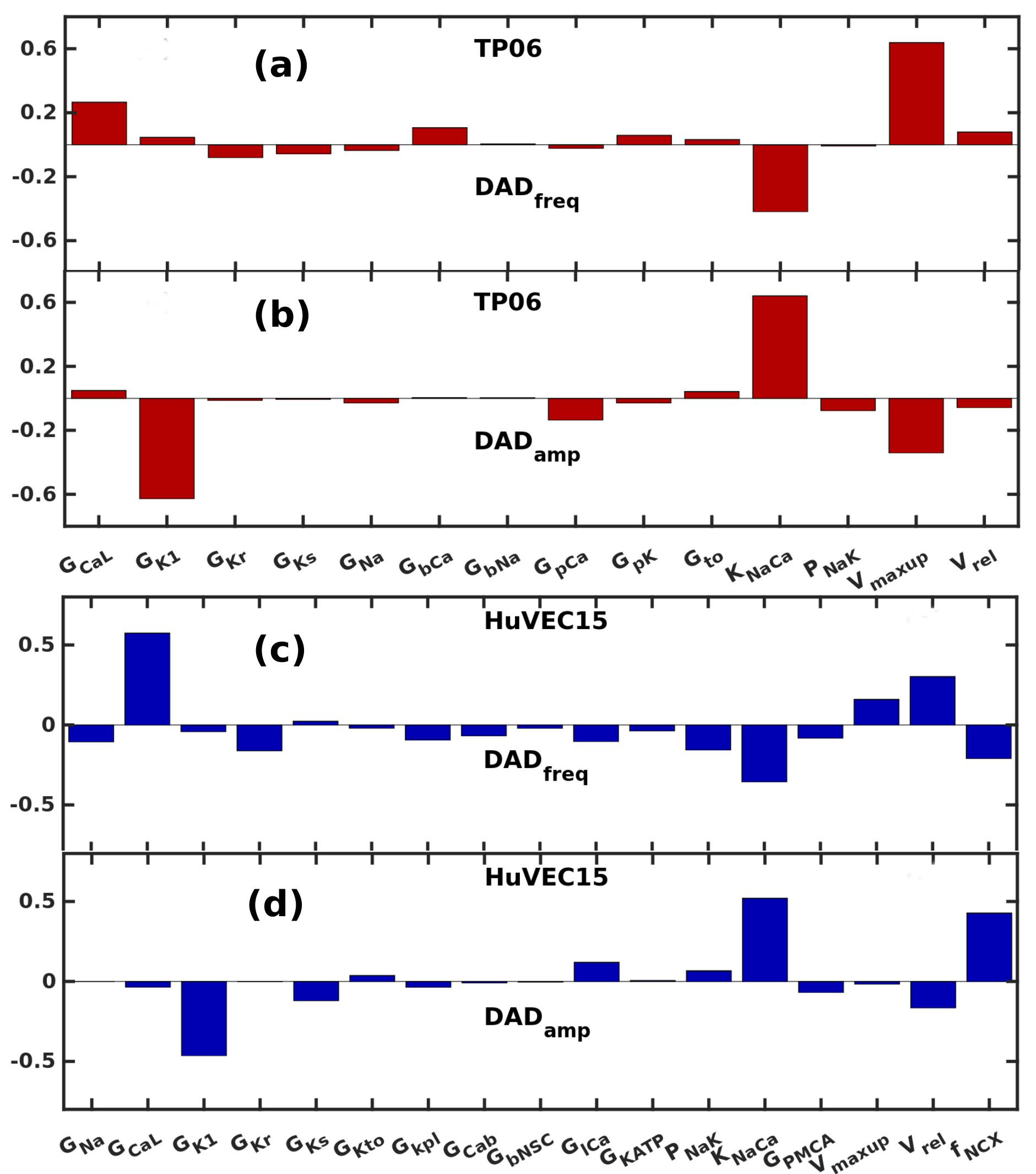}
      \caption{ (Color online) \textbf{Sensitivity analysis:} The columns of the regression-coefficient matrix $\mathbf{B_{PLS}}$ indicate how the scaling of the maximal conductances and fluxes affect DAD\textsubscript{amp} and DAD\textsubscript{freq}, in red for the TP06 model and in blue for the HuVEC15 model. Sensitivity plots for: the TP06 model (a) DAD\textsubscript{freq} and (b) DAD\textsubscript{amp}; the HuVEC15 model (c) DAD\textsubscript{freq} and (d) DAD\textsubscript{amp}. The positive and negative values of the coefficients indicate whether an increase in the parameter increases or decreases the corresponding outputs.}
      
      \label{fig:tp06_sensitivity}
\end{figure}

We have shown different types of DADs and characterised them by using the frequency and amplitude
of the DAD, DAD\textsubscript{amp} and DAD\textsubscript{freq}, respectively. We now perform a parameter-sensitivity analysis to determine the critical parameters for these DAD characteristics. 
To calculate the DAD frequency of the model, for each set of parameters, we stimulate the myocyte models for 500 pacings and record the last $10$ action potentials. As we have noted above, the DAD amplitude and frequency are hard to 
define unambiguously in the supra-threshold DAD regime, so we do not include it in this Subsection. For each simulation, we calculate the average frequency and amplitude of DADs. 
The parameters we choose for our sensitivity analysis are the maximal conductances of all the available transmembrane ionic currents, exchangers, the maximal SERCA pump uptake rate $V_{\rm{maxup}}$, and the maximal release rate of calcium from RyRs, viz., $V_{\rm{rel}}$. This analysis reveals that $V_{\rm{maxup}}$ is the most sensitive parameter for the DAD frequency in the
TP06 model [Fig.\ref{fig:tp06_sensitivity}~(a)]; in contrast, for the HuVEC15 model, $V_{\rm{rel}}$ is the most sensitive parameter for the DAD frequency [Fig.\ref{fig:tp06_sensitivity}~(c)]. For the DAD amplitude, in both TP06 and HuVEC15 models,
we find that $K_{\rm{NaCa}}$ and $G_{\rm{K1}}$ are the most sensitive parameters [Figs.~\ref{fig:tp06_sensitivity}(b) and (d)]. We use a pacing frequency of $1$ Hz here. In the Appendix we show that 
the results of our sensitivity analysis are robust insofar as they are not altered when we change the pacing frequency.

\begin{figure}
     \centering
      \includegraphics[width=0.45\textwidth]{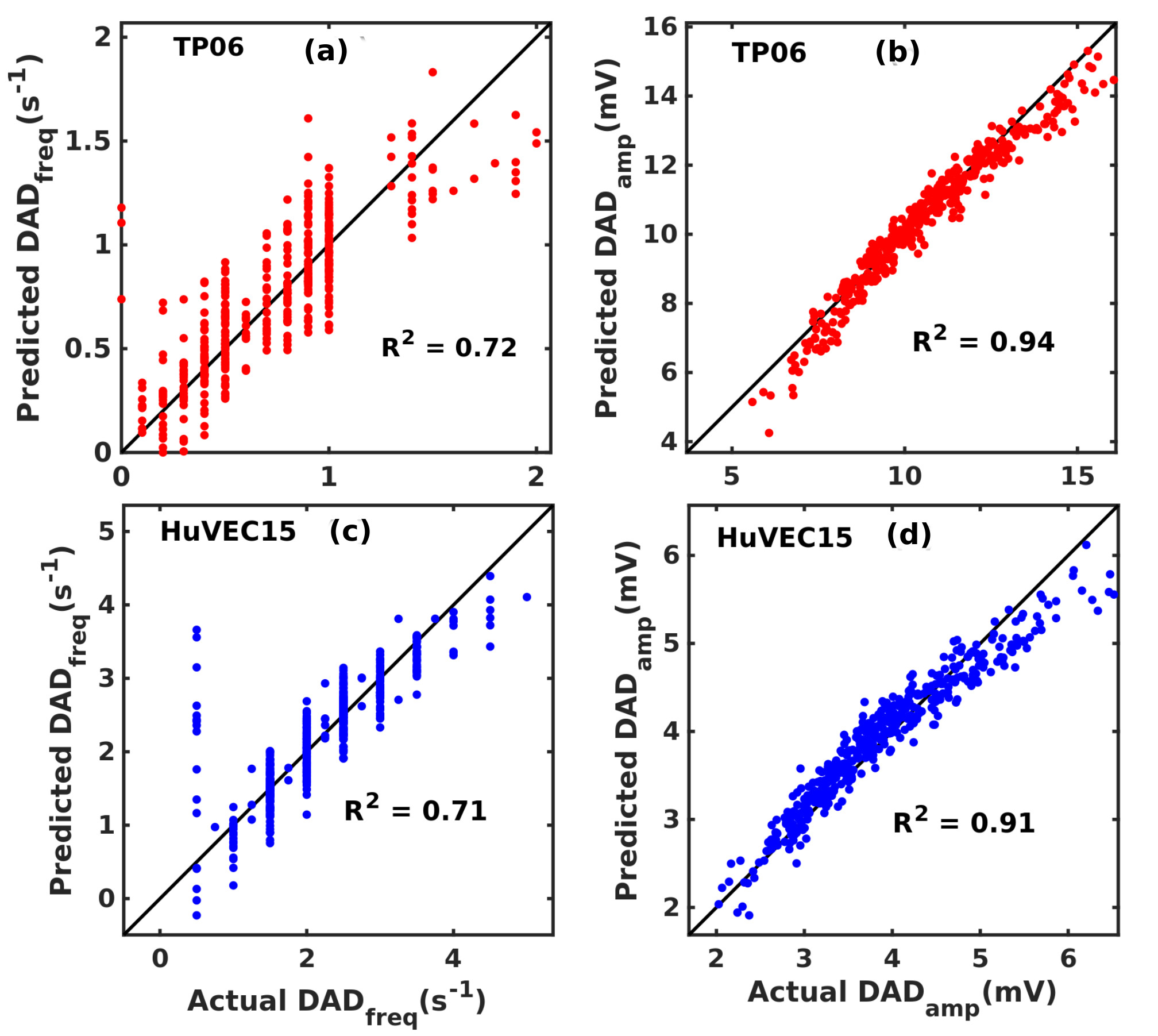}
      \caption{ (Color online) \textbf{The quality of PLS-regression predictions ($R^2$-value)}: Scatter plots for the two outputs, DAD\textsubscript{freq} and  DAD\textsubscript{amp}, in red and blue for TP06 and HuVEC15 models, respectively, with the values computed by our numerical simulations of  equations on the horizontal axis and the values estimated by the PLS regression model on the vertical axis: TP06 model (a) DAD\textsubscript{freq} and (b) DAD\textsubscript{amp}; HuVEC15 model (c) DAD\textsubscript{freq} and (d) DAD\textsubscript{amp}. We perform our regression analysis on a simulated data set that contains approximately $1000$ samples; to obtain the value of $R^2$ value, we use $400$ data points.}
      \label{tp06_rsquare}
\end{figure}

\subsection{DAD Phase Diagrams}{\label{subsec:phases}}

\begin{figure}
     \centering
      \includegraphics[width=0.45\textwidth]{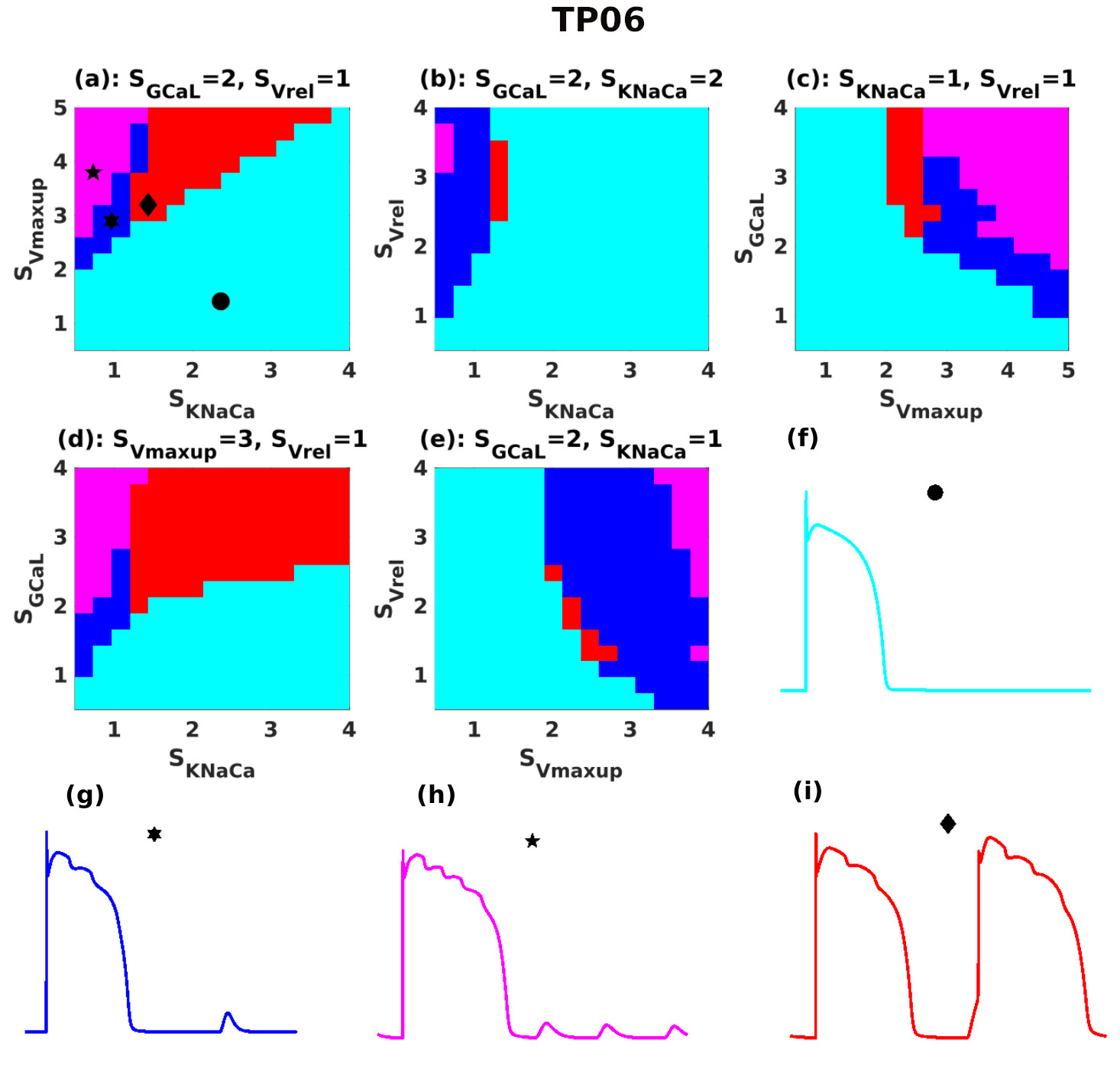}
      \caption{ (Color online) \textbf{TP06-model DAD phase diagrams:} (a)-(e) Phase diagrams, for various combinations of parameter
      values; and (f)-(i) normal APs and those with subthreshold, multi-blip, and suprathreshold DADs are drawn, respectively, in cyan, blue, magenta, and red; the stability regions in the phase diagrams follow the same color
    scheme.
      }
      \label{fig:phase_dia_tp06}
\end{figure}

\begin{figure}
     \centering
      \includegraphics[width=0.45\textwidth]{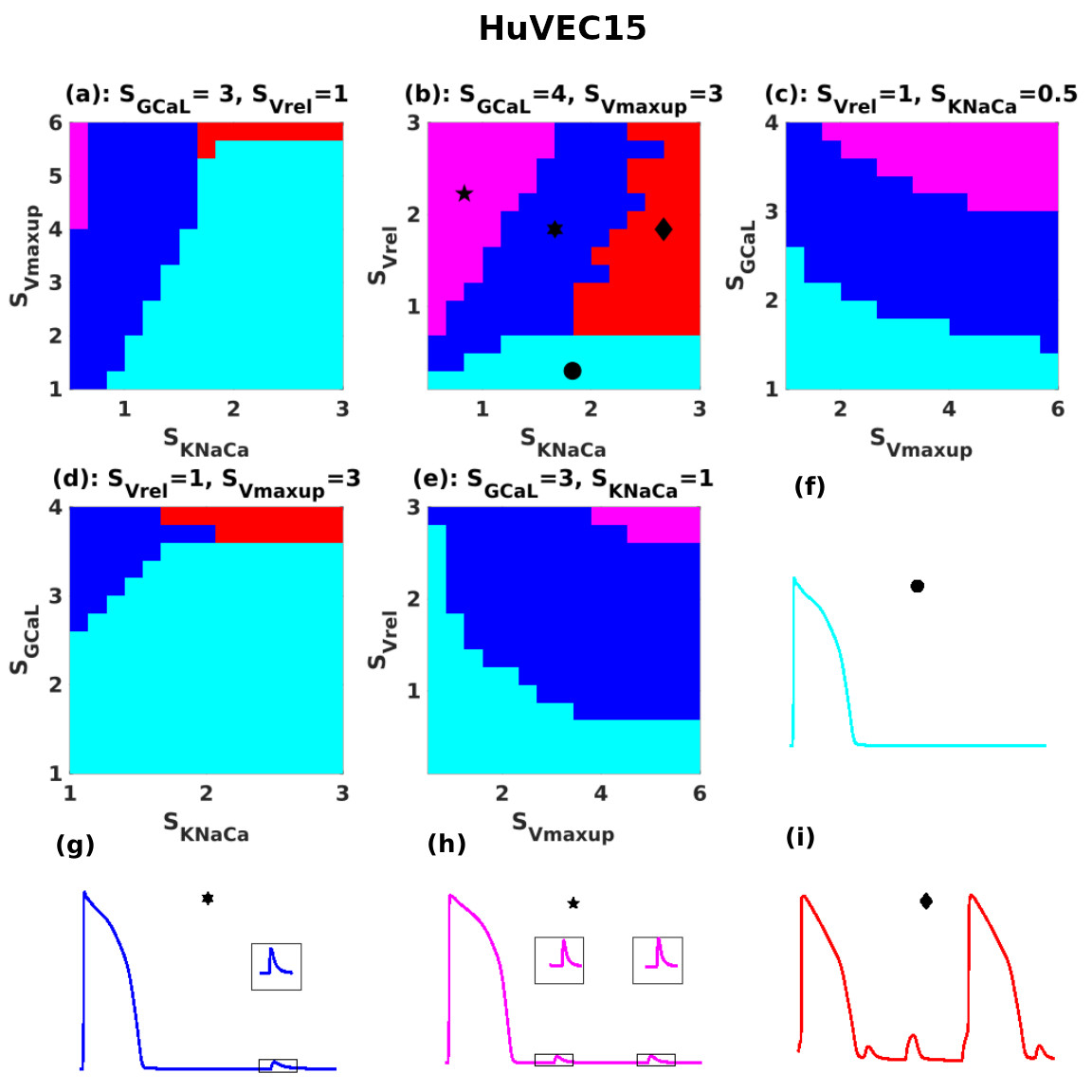}
      \caption{ (Color online) \textbf{HuVEC15-model DAD phase diagrams:}
      (a)-(e) Phase diagrams, for various combinations of parameter
      values; and (f)-(i) normal APs and those with subthreshold, multi-blip, and suprathreshold DADs are drawn, respectively, in cyan, blue, magenta, and red; the stability regions in the phase diagrams follow the same color scheme. The insets in (g) and (h) show enlarged versions of the DADs (that are highlighted with rectangles).
      }
      \label{fig:phase_dia_asakura}
\end{figure}

Now that we have determined the parameters that affect, most sensitively, DAD amplitudes and frequencies, we are in a position to present phase diagrams (or stability diagrams). These indicate the regions of stability for different types of DADs in representative sections through the parameter spaces of the TP06 and HuVEC15 models in Figs.~\ref{fig:phase_dia_tp06} (a)-(e) and Figs.~\ref{fig:phase_dia_asakura} (a)-(e), respectively.
Normal APs and those with subthreshold, multi-blip, and suprathreshold DADs are drawn, respectively, in cyan, blue, magenta, and red; the stability regions in Figs.~\ref{fig:phase_dia_tp06} and \ref{fig:phase_dia_asakura} follow the same color scheme.  We note that, at each point in these DAD phase diagrams, for a given value of $S_{\rm{GCaL}}$, we use the value of $S_{\rm{GKr}}$ that is required to maintain the APD [see Figs.~\ref{fig:compare_APs_L_and_kr}  (c) and (d) and the Appendix]. There is an important difference between the DAD phase diagrams for the TP06 and HuVEC15 models: In the former, an initial increase in $S_{\rm{KNaCa}}$ (see Figs.~\ref{fig:phase_dia_tp06} (a), (b), and (d)) leads to supra-threshold DADs; however, an additional increase in $S_{\rm{KNaCa}}$ brings the system back to the phase with normal APs without DADs; in the DAD phase diagram for the HuVEC15 model (see Figs.~\ref{fig:phase_dia_asakura} (a), (b), and (d)) an
increase in $S_{\rm{KNaCa}}$ leads to the phase with supra-threshold DADs, which is not destabilised by an additional increase in $S_{\rm{KNaCa}}$. 
We explore the consequences of this difference in Subsection~\ref{subsec:protective}.

\subsection{Interplay of Sub-cellular Ca\textsuperscript{2+} Components}{\label{subsec:interplay}}

During Ca\textsuperscript{2+}-overload, the SCRs occur via the opening of RyRs, which increase the cytosolic Ca\textsuperscript{2+} concentrations; the NCX acts in the forward mode and extrudes excess Ca\textsuperscript{2+} outside of the myocyte from the cytosol, whereas the SERCA pump unloads the cytosol by pumping the Ca\textsuperscript{2+} back into the SR store. 
We quantify this interplay of crucial parameters of the Ca{\textsuperscript{2+}}-subsystem that control 
the frequency and amplitude of DADs. From the phase diagrams of Figs.~\ref{fig:phase_dia_tp06} (a)-(b) and Figs.~\ref{fig:phase_dia_asakura} (a)-(b), we note that suprathreshold DADs occur in the region where $S_{\rm{KNaCa}}$ is large; as we reduce $S_{\rm{KNaCa}}$, these systems move to regions in which subthreshold or multi-blip DADs occur.

\subsubsection{The TP06 Model}{\label{subsubsec:TP06}}
	
\begin{figure}
     \centering
      \includegraphics[width=0.47\textwidth]{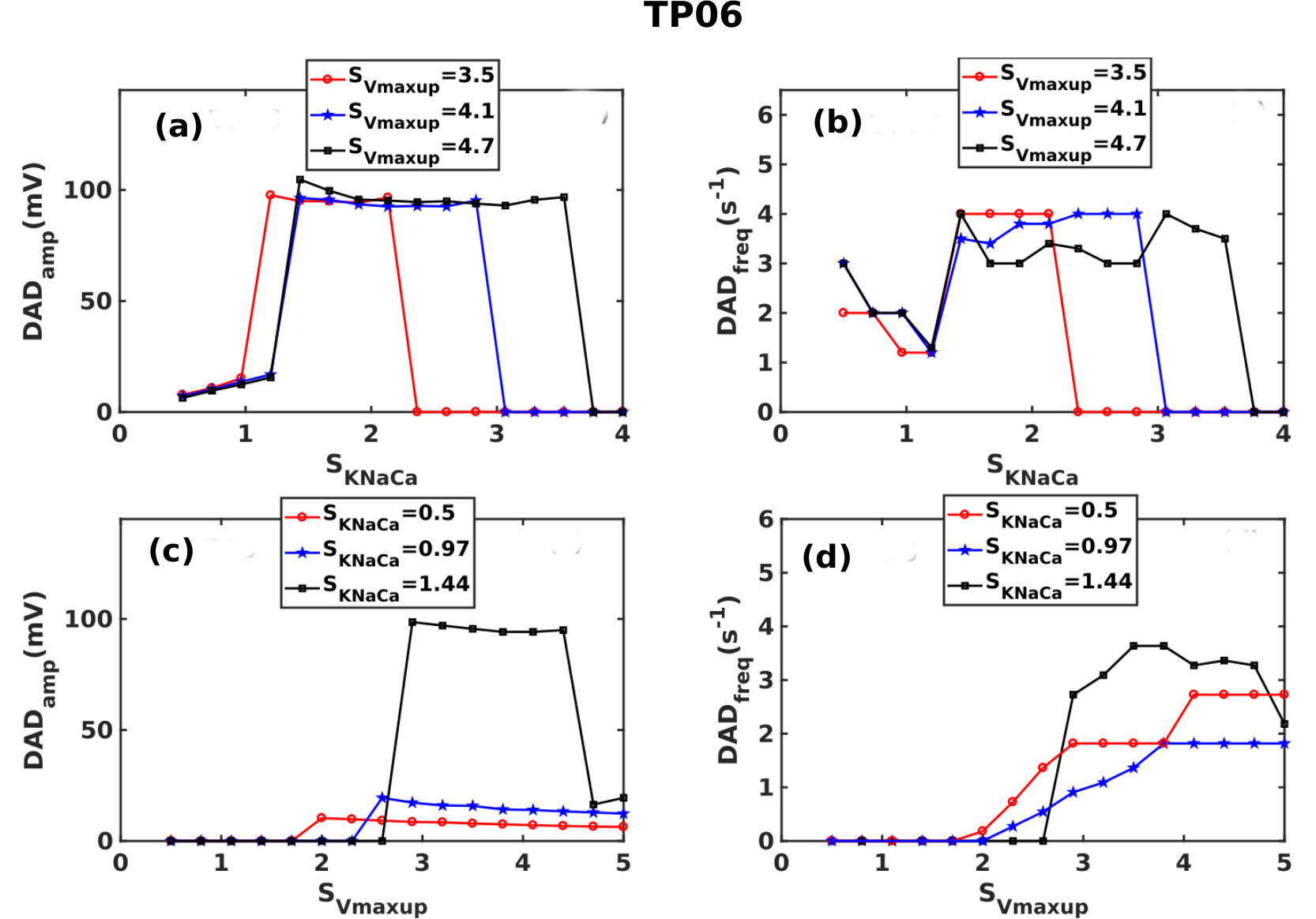}
      \caption{(Color online) \textbf{Effects of the SERCA pump uptake rate $V_{\rm{maxup}}$ and the NCX control parameter $K_{\rm{NaCa}}$ on DADs in the TP06 model:} Plots of: DAD\textsubscript{amp}: (a)  versus $S_{\rm{KNaCa}}$ for different values of $S_{\mathrm{Vmaxup}}$ and (c) versus $S_{\mathrm{Vmaxup}}$ for different values of $S_{\rm{KNaCa}}$;
      DAD\textsubscript{freq}: (b) versus $S_{\rm{KNaCa}}$ for different values of $S_{\mathrm{Vmaxup}}$ and (d) versus $S_{\mathrm{Vmaxup}}$ for different values of $S_{\rm{KNaCa}}$.}
      \label{fig:ncxvsserca}
\end{figure}

We now examine the roles of (a) the SERCA-pump uptake-rate $V_{\rm{maxup}}$ and (b) the NCX control parameter $K_{\rm{NaCa}}$ [see Eq.~\ref{eq:naca} in the Appendix] in controlling the DAD amplitudes and frequencies in the TP06 model. In Fig.~\ref{fig:ncxvsserca}(a) we plot 
the amplitude DAD\textsubscript{amp} versus $S_{\rm{KNaCa}}$ to demonstrate that an increase in $S_{\rm{KNaCa}}$ aids the 
subthreshold DADs to reach the threshold for triggered activity. For each value of $S_{\rm{Vmaxup}}$, there is a window of values of $S_{\rm{KNaCa}}$ in which
we obtain suprathreshold DADs; beyond this window, an additional increase of $S_{\rm{KNaCa}}$ terminates DADs; the width of this window increases with 
$S_{\rm{Vmaxup}}$. In Fig.~\ref{fig:ncxvsserca}(b) we plot the DAD\textsubscript{freq} versus $S_{\rm{KNaCa}}$ to show that, initially, an increase in $S_{\rm{KNaCa}}$ decreases DAD\textsubscript{freq}; an additional increase in $S_{\rm{KNaCa}}$ leads to a jump in DAD\textsubscript{freq}, which arises from a sudden appearance of suprathreshold DADs; an
additional increase in $S_{\mathrm{KNaCa}}$ leads to DAD termination and hence DAD\textsubscript{freq}$=0$.\\
The SERCA pump also has a critical influence on the DADs: Figures~\ref{fig:ncxvsserca}(c) and
(d) show that non-zero DAD\textsubscript{amp} and DAD\textsubscript{freq} appear only after $S_{\mathrm{Vmaxup}}$ crosses a threshold value. Figure~\ref{fig:ncxvsserca}(d) demonstrates that the increase in the $V_{\mathrm{maxup}}$ increases the DAD\textsubscript{freq}. $V_{\mathrm{maxup}}$ has a small effect on DAD\textsubscript{amp}. However, for $S_{\mathrm{KNaCa}}=1.44$, an increase in $S_{\mathrm{Vmaxup}}$ leads to sudden jumps in  DAD\textsubscript{amp} and DAD\textsubscript{freq}; an additional increase in $S_{\mathrm{Vmaxup}}$ reduces both of these. Thus, $S_{\mathrm{Vmaxup}}$ and $S_{\mathrm{KNaCa}}$ play critical roles in the formation of DADs.

\subsubsection{The HuVEC15 Model}{\label{subsubsec:HuVEC15}}

\begin{figure}
     \centering
      \includegraphics[width=0.45\textwidth]{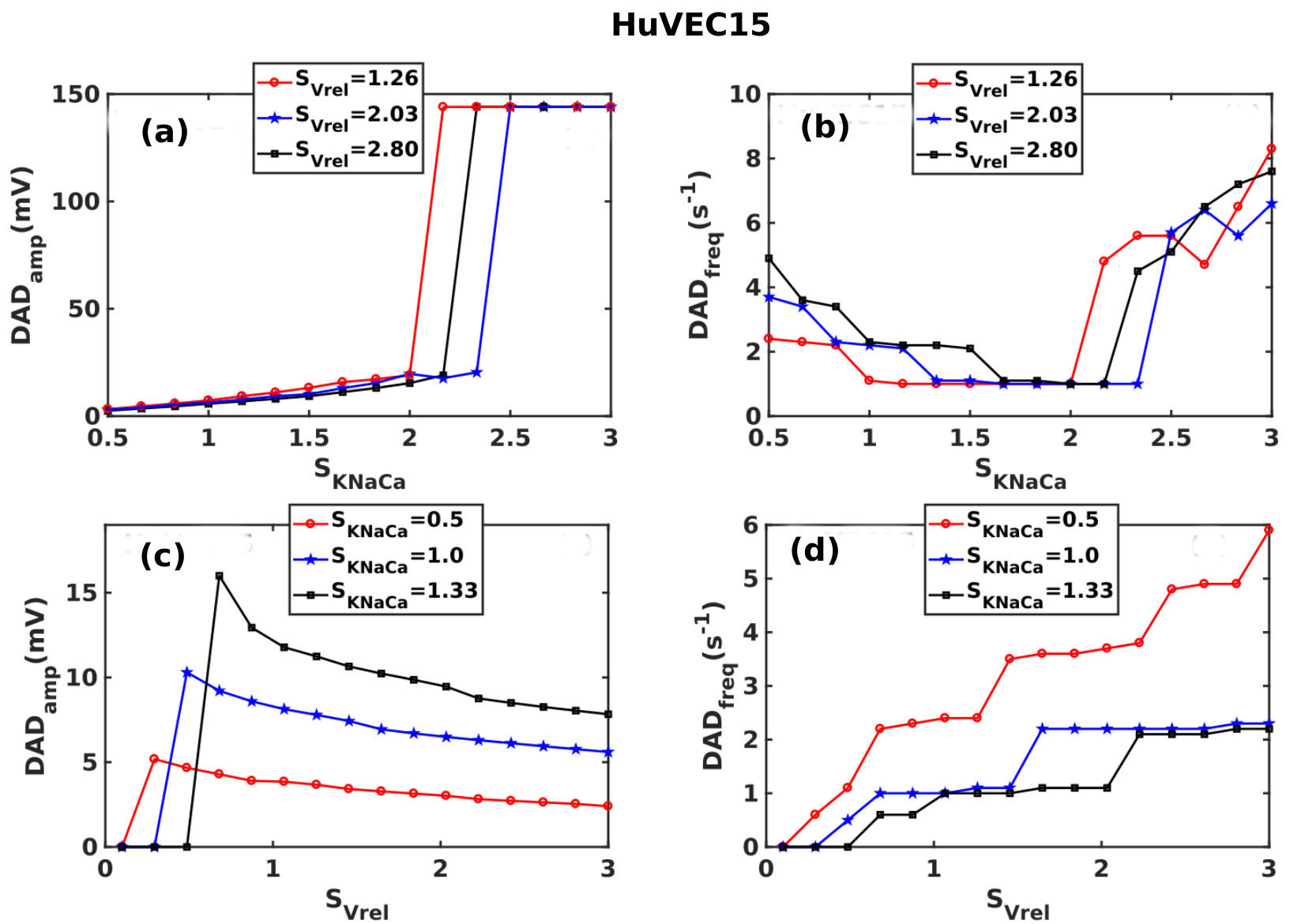}
      \caption{ (Color online) \textbf{The Role of RyR release rate $V_{\rm{rel}}$ and $K_{\rm{NaCa}}$
      on DADs in the HuVEC15 model:} Plots of: DAD\textsubscript{amp}: (a)  versus $S_{\rm{KNaCa}}$ for different values of $S_{\mathrm{Vrel}}$ and (c) versus $S_{\mathrm{Vrel}}$ for different values of $S_{\rm{KNaCa}}$;
      DAD\textsubscript{freq}: (b) versus $S_{\rm{KNaCa}}$ for different values of $S_{\mathrm{Vrel}}$ and (d) versus $S_{\mathrm{Vrel}}$ for different values of $S_{\rm{KNaCa}}$.}
      \label{fig:ncx_vrel}  
\end{figure}

In the HuVEC15 model, the two parameters that influence the DAD amplitude and frequency most significantly are $K_{\rm{NaCa}}$ and $V_{\rm{rel}}$. Our sensitivity analysis in Fig.~\ref{fig:tp06_sensitivity} shows that $K_{\rm{NaCa}}$ reduces the frequency of DADs and also increases the DAD amplitude; this is confirmed by the plots of DAD\textsubscript{amp} and DAD\textsubscript{freq} versus $S_{\rm{KNaCa}}$ in Figs.~\ref{fig:ncx_vrel} (a) and (b), respectively. Similarly, the plots of DAD\textsubscript{amp} and DAD\textsubscript{freq} versus $S_{\rm{Vrel}}$ in Figs.~\ref{fig:ncx_vrel} (c) and (d), respectively, are also in
consonance with our sensitivity analysis, which has demonstrated that $V_{\rm{rel}}$ is the parameter that influences the frequency of DADs most sensitively and which also reduces the DAD amplitudes.

\subsection{NCX protects against DADs in the TP06 model}
\begin{figure}
     \centering
      \includegraphics[width=0.45\textwidth]{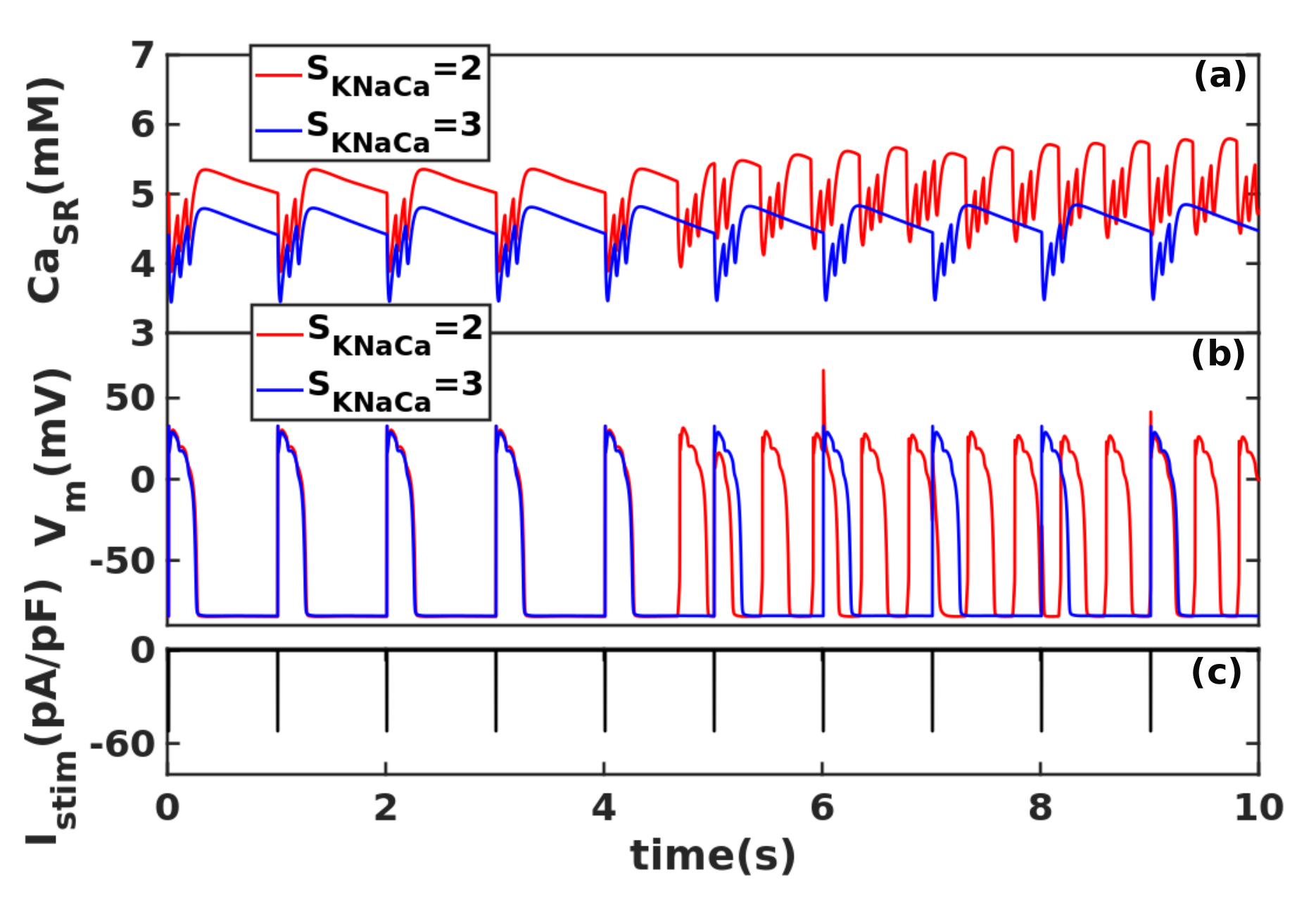}
      \caption{ (Color online) \textbf{The Role of $K_{\rm{NaCa}}$ in terminating DADs in TP06 model:} Plots versus time $t$ of (a) $Ca_{\rm{SR}}$, (b) $V_{\rm{m}}$, and (c) the current $I_{\rm{stim}}$ that provides the pacing stimuli, for two different values of $S_{\rm{KNaCa}}$ [with $S_{\rm{GCaL}}=2$, $S_{\rm{GKr}}=2.8$, $S_{\rm{Vrel}}=1$ and $S_{\rm{Vmaxup}}=4$]. In (a)  $Ca_{\rm{SR}}$ is lower for $S_{\rm{KNaCa}}=3$ than for $S_{\rm{KNaCa}}=2$; in (b) suprathreshold DADs are triggered if $S_{\rm{KNaCa}}=2$, but are absent if $S_{\rm{KNaCa}}=3$.}
      \label{fig:ncx_term_DADs}
\end{figure}

{\label{subsec:protective}}
The usual consensus is that $K_{\rm{NaCa}}$ (NCX) enhances DAD amplitudes and is, therefore, arrhythmogenic. However, our
DAD phase diagrams for the TP06 model, Figs.~\ref{fig:phase_dia_tp06}(a), (b), and (d), and the plots of DAD\textsubscript{amp} versus $S_{\rm{KNaCa}}$ [Fig.~\ref{fig:ncxvsserca} (a)] show that a critical value of $S_{\rm{KNaCa}}$ must be reached before subthreshold DADs undergo a transition to suprathreshold DADs; but too large an increase in $S_{\rm{KNaCa}}$ completely terminates the DADs. This protective mechanism is only present in the TP06 model; we do not find it in the HuVEC15 model. To elucidate this mechanism, we plot in Figs.~\ref{fig:ncx_term_DADs} (a), (b), and (c)
the time series of $Ca_{\rm{SR}}$, $V_{\rm{m}}$, and the current $I_{\rm{stim}}$ that provides the pacing stimuli; we compare 
the time series of $Ca_{\rm{SR}}$ and $V_{\rm{m}}$ for the representative values $S_{\rm{KNaCa}}=2$ and $S_{\rm{KNaCa}}=3$, after $120$ pacings. The plots with the low NCX scaling factors (e.g., $S_{\rm{KNaCa}}=2$) show suprathreshold DADs; in contrast, the plots with high NCX scaling factors (e.g., $S_{\rm{KNaCa}}=3$) yield low values of $Ca_{\rm{SR}}$, so DADs do not appear. Therefore, in the TP06
model, NCX reduces the $Ca_{\rm{SR}}$ load because of late Ca\textsuperscript{2+} release (LCR) (see Fig.~\ref{fig:SI:sparks_DAD_EAD} in the Appendix), which increases the cytosolic calcium concentration during the plateau phase of the AP. This changes the direction of NCX to the forward mode, in which NCX takes one Ca\textsuperscript{2+} ion out of the myocyte and puts three Na\textsuperscript{+} ions back into the myocyte. We recall that, in the backward or reverse mode, NCX expels three Na\textsuperscript{+} ions from the myocyte in exchange for one Ca\textsuperscript{2+} ion; during the plateau region of the AP, with the normal physiological value of Na\textsuperscript{+}, NCX operates in the reverse mode; a sudden increase in cytosolic Ca\textsuperscript{2+} forces the NCX to operate in the forward mode. Such LCRs do not occur in the HuVEC15 model, so NCX does not protect against DADs in this model. [This is related to the important difference [cf. Subsection~\ref{subsec:phases} ] between the DAD phase diagrams of the
TP06 and HuVEC15 models.] We give a representative plot for the direction of NCX and SCRs 
(Fig.~\ref{fig:SI:sparks_DAD_EAD} in the Appendix).

\subsection{Cable and Tissue Simulations}
{\label{subsec:tissue}}

We have presented our results for isolated myocytes; we now describe our results from representative simulations, for both TP06 and HuVEC15 models, in cable and tissue domains with DAD clumps [see Subsection~\ref{subsec:deint}]. In these domains, myocytes are electrotonically coupled.
When DADs occur in the clump, $V_{\rm{m}}$ rises above its value in the resting state, and the electrotonic currents start flowing to the other resting myocytes. The competition between the rising DAD of a certain duration and diffusion processes in cardiac-tissue models leads to the emergence of a length scale~\cite{xie2010so}, which depends on
the parameters of the model; if the linear size of the DAD clump exceeds this length scale, then the clump of DAD myocytes can fire focal excitations. In a DAD clump, myocytes are synchronized by the pacings; therefore, they fire synchronously. We illustrate this in our tissue simulations.

\begin{figure}
	\centering
      \includegraphics[width=.45\textwidth]{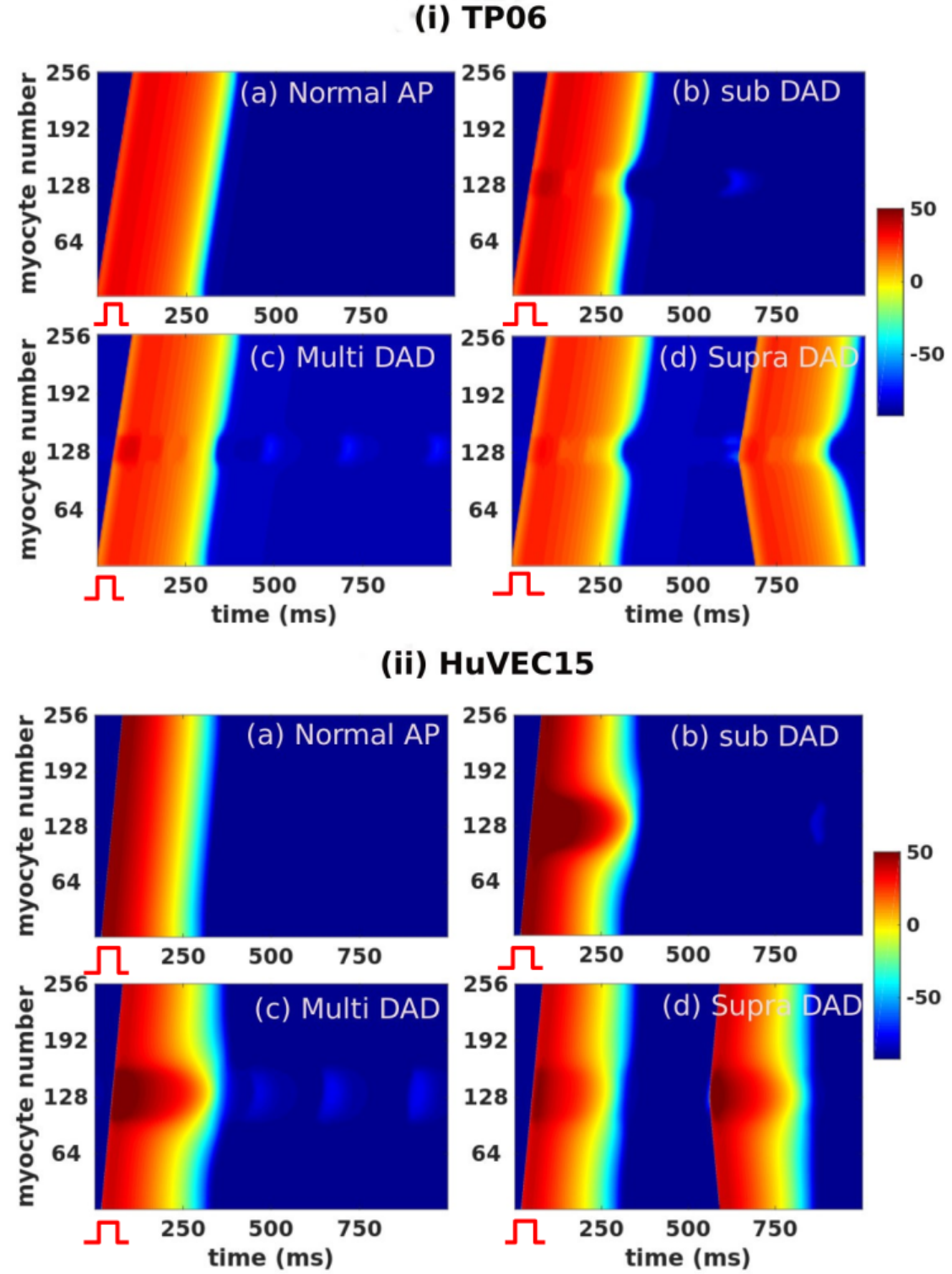}
      \caption{ (Color online) \textbf{Cable Simulations:} Pseudocolor space-time plots of $V_{\rm{m}}$(mV) along our 1D cable domain;  DAD myocytes occupy the middle region of the cable (30 and 60 grid points for the TP06 and HuVEC15 models, respectively). The subplots (i) and (ii) show: (a) a normal AP; (b) subthreshold DADs; (c) multi-blip DADs; and (d) suprathreshold DADs. The stimulation frequency we choose is 1 Hz and the stimulus is applied to the first grid point.}
      \label{fig:cable}
\end{figure}

In Fig.~\ref{fig:cable} we present pseudocolor space-time plots of $V_{\rm{m}}$ in a cable with DAD clumps for both the TP06 (panel (i)) and HuVEC15 (panel (ii)) models; in subfigures (a), (b), (c), and (d) the DAD clumps have a normal AP, subthreshold DADs,  multi-blip DADs, and suprathreshold DADs, respectively [cf. Figs.~\ref{fig:phase_dia_tp06} and ~\ref{fig:phase_dia_asakura}]. We use a 1Hz current stimulus for the first myocyte in the cable and track the signal as it propagates to the other end.  
Figure~\ref{fig:cable}(a) shows the propagation of a normal signal from the first myocyte to the last myocyte of the cable; Fig.~\ref{fig:cable}(b) displays the emergence of sub-threshold depolarization at the center of the cable, following a normal excitation; Fig.~\ref{fig:cable}(c) exhibits a series of 
subthreshold depolarizations (multi-blips) at the center of the cable, following a normal excitation; and Fig.~\ref{fig:cable}(d) depicts the emergence of PVCs, following a normal excitation. We find that, in these cable domains, DAD clumps with $30$ and $60$ grid points representing suprathreshold myocytes are sufficient for triggering PVCs in the TP06 and HuVEC15 models, respectively. 

\begin{figure}
      \includegraphics[width=0.45\textwidth]{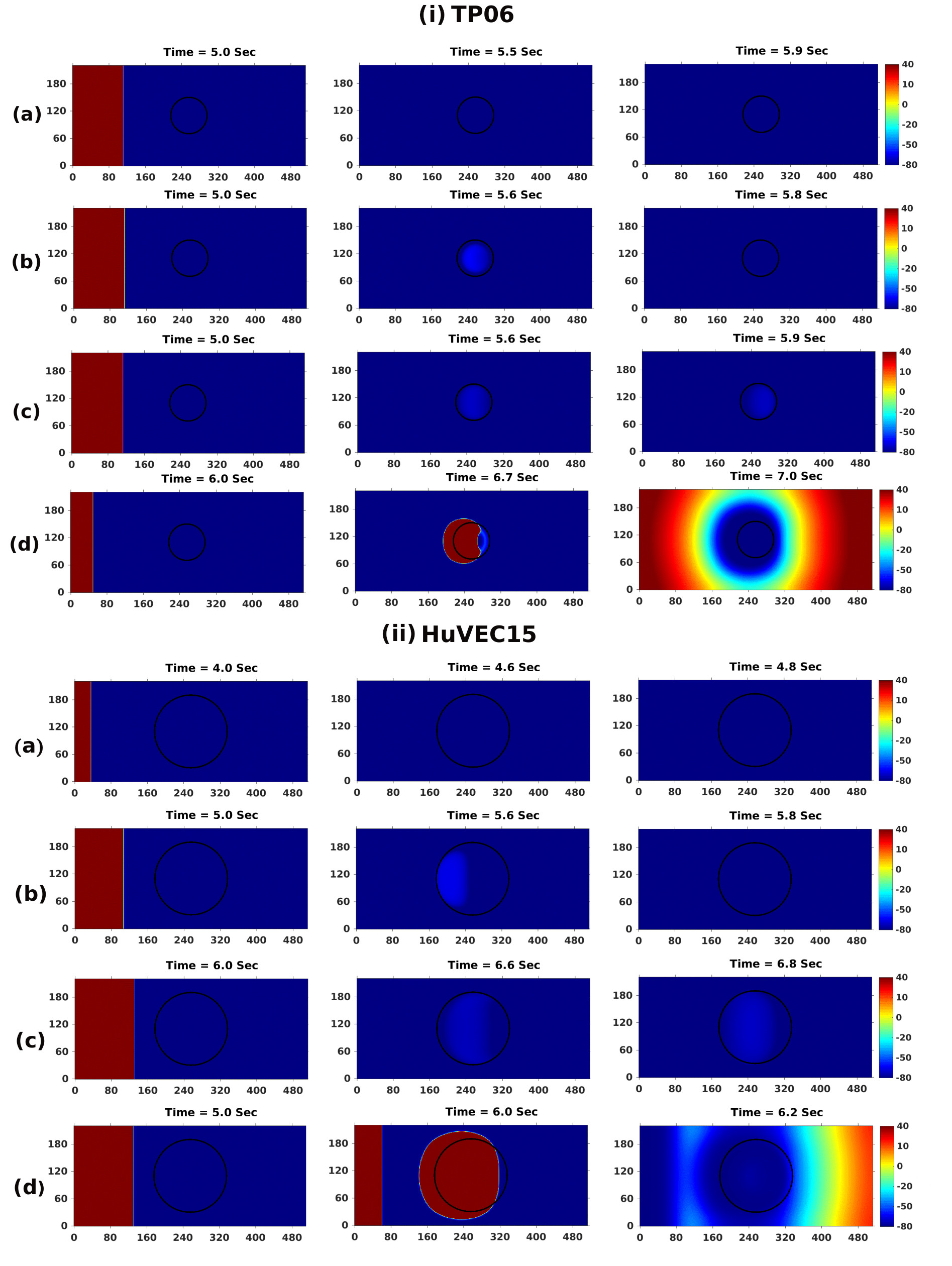}
      \caption{ (Color online) \textbf{2D simulations:} Pseudocolor plots of $V_{m}$(mV) for the TP06 (panel (i)) and HuVEC15 (panel (ii)) models illustrating the effects of pacing on DAD clumps in cardiac tissue:
      (a) myocytes with normal APs (no DADs observed); (b): subthreshold DAD myocytes (after a few pacings the clump fires a PVC, which does not reach the threshold for full excitation and, therefore, is localised in a limited region); (c): multi-blip DAD myocytes (multiple PVCs emerge, but they do not reach the full excitation threshold and, therefore, are localised in a limited region); (d): suprathreshold DAD myocytes (a full-strength PVC emerges and excites the entire domain). For the complete spatiotemporal evolution of $V_{\rm{m}}$ see Videos [\ref{app:2d_tp_plane_wave}-\ref{app:2d_tp_supra_DAD}] (for the TP06 model) and Videos [\ref{app:2d_ask_plane_wave}-\ref{app:2d_ask_supra_DAD}] (for the HuVEC15 model) in the Appendix.}
      \label{fig:2d_tp}
\end{figure}

In Fig.~\ref{fig:2d_tp} we present pseudocolor plots of $V_{m}$ from our simulations for 2D domains for the TP06 (panel (i)) and HuVEC15 (panel (ii)) models. These plots illustrate the effects of circular DAD clumps  [see Subsection~\ref{subsec:deint}] on the propagation of plane waves of electrical activation through these domains, as we pace the tissue at its left boundary. If the clump comprises myocytes with normal APs, then we find normal excitation propagation [subfigures (a)]; if the clump consists of sub-threshold DAD myocytes, we observe the emergence of sub-threshold excitation in [subfigures (b)]; for a clump with multi-blip DAD myocytes, we find multiple subthreshold excitations, with the clump itself containing the subthreshold DADs;  with a clump of supra-threshold DAD myocytes, PVCs emerge after 6 pacings and then propogate through the entire domain. 

\begin{figure}
      \includegraphics[width=0.45\textwidth]{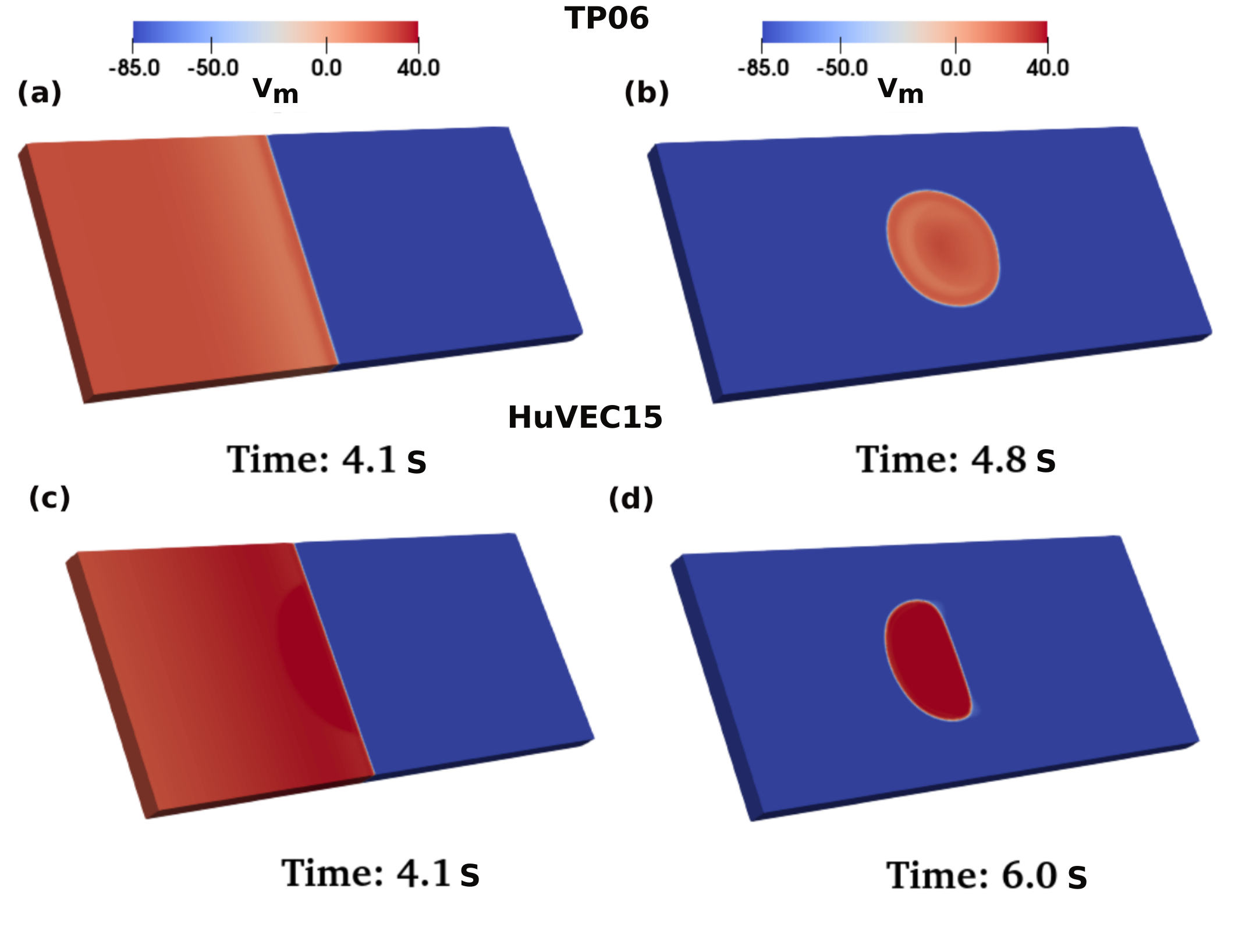}
      \caption{ (Color online) \textbf{3D slabs:} Pseudocolor plots of $V_{\rm{m}}$(mV), for the TP06 and HuVEC15 models, illustrating the spatiotemporal evolution of electrical excitation and the emergence of PVCs from a DAD-myocyte clump, with suprathreshold DADs. TP06 model: (a) pacing-induced plane-wave propagation; (b) PVC emerging after 4 pacings.  HuVEC15 model: (c) pacing-induced plane-wave propagation; (d): PVC emerging after 5 pacings. For the complete spatiotemporal evolution of $V_{\rm{m}}$, see Video~\ref{app:3d_tp_slab} (for the TP06 model) and Video~\ref{app:3d ask slab} (for the HuVEC15 model)] in the Appendix.}
      \label{fig:3d}
\end{figure}

In Fig.~\ref{fig:3d} we present representative simulations in 3D square-cuboid domains, for the TP06 and HuVEC15 cardiac-tissue models, with a cylindrical DAD clump. Our results are qualitatively similar to those we have presented in 2D. We find, e.g., that PVCs emerge from this clump after $4$ and $6$ pacings, respectively, in the TP06 and HuVEC15 models. 

\begin{figure}
      \includegraphics[width=0.45\textwidth]{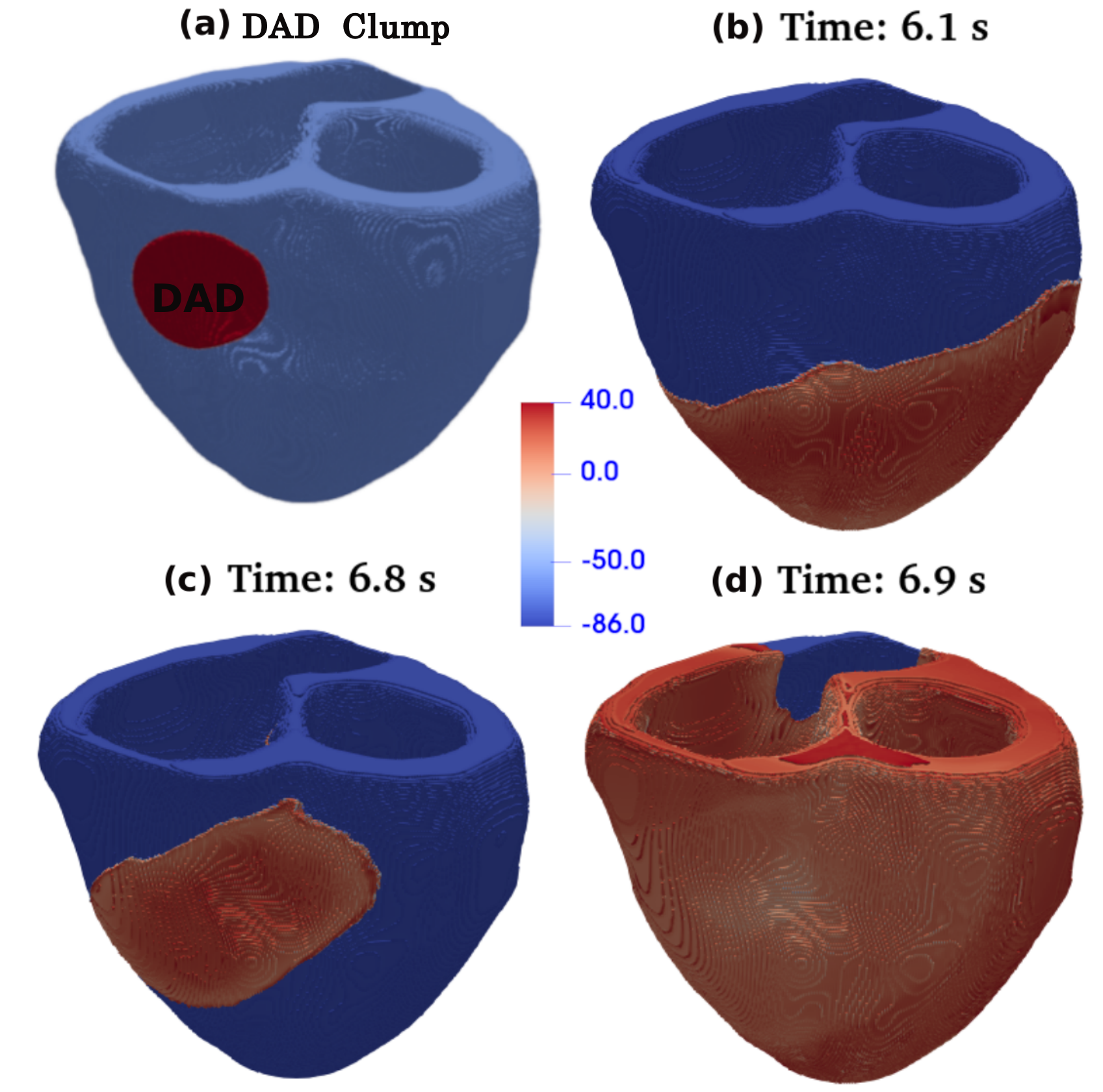}
      \caption{ (Color online) (a) Pseudocolor plot showing a representative clump of DAD myocytes (red) embedded in a human biventricular geometry (blue). Pseudocolor plots of $V_{\rm{m}}$(mV) depicting (b) normal excitation propagation after stimulation of the bi-ventricular geometry from its apex, (c) emergence of PVCs from the DAD clump after $6$ pacings ($1$ Hz pacing frequency), and (d) the propagation of PVCs and subsequent scroll-wave excitations to both the ventricles. For the complete spatiotemporal evolution of $V_{\rm{m}}$ see Video~\ref{app:bi_vent_PVC} in the Appendix.}
      \label{fig:whole_heart}
\end{figure}

In our most realistic study, we perform a full-heart simulation by using the phase-field method; we include fiber orientation to account for the anisotropy of cardiac tissue (see, e.g., Refs.~\cite{rajany2021effects,winslow2011cardiovascular}). We include a DAD clump as we have described in Subsection~\ref{subsec:deint}. In Fig.~\ref{fig:whole_heart}(a), we show such a suprathreshold DAD clump for the TP06 model embedded in a human-bi-ventricular geometry, with roughly $775,000$ grid points. In Fig.~\ref{fig:whole_heart}(b), we present a pseudocolor plot of the $V_{\rm{m}}$ in this geometry; this depicts the propagation of a normal stimulus, applied at the apex of the human bi-ventricular geometry. Once this propagating wave of electrical activation encounters the DAD clump, PVCs emerge, as we show
in Fig.~\ref{fig:whole_heart}(c): these PVCs propagate and finally excite both the ventricles as we can see in Fig.~\ref{fig:whole_heart}(d). 
 

\section{Discussion and Conclusions}{\label{sec:disc_and_concl}}
 We have carried out a detailed investigation of DADs in two human-ventricular myocyte models, the TP06 and HuVEC15, and we have compared the DADs in these models at both single-cell and tissue levels.
 First, by reducing the TP06 myocyte model to its Ca\textsuperscript{2+} subsystem, we have analyzed the SR Ca\textsuperscript{2+} and Na\textsuperscript{+} load requirements for DADs to occur. Then, by using the $Na_{\rm{i}}$ overload in the TP06 Ca\textsuperscript{2+} subsystem, we have demonstrated that, above a threshold value of $Na_{\rm{i}}$, the TP06 Ca\textsuperscript{2+} subsystem does indeed show a transition, via a Hopf-bifurcation, from an asymptotically stable value of the calcium concentration to sustained Ca\textsuperscript{2+} oscillations. By using various values of $V_{\rm{maxup}}$, we have shown that an increase in the SERCA pump uptake rate enhances the threshold $Ca_{\rm{SR}}$ and reduces the threshold $Na_{\rm{i}}$ requirements for Ca\textsuperscript{2+} oscillations in the reduced TP06 model. Furthermore,
 we have examined the RyR leak and the NCX to reveal that (a) the former reduces significantly both the $Ca_{\rm{SR}}$ and $Na_{\rm{i}}$ threshold requirements for Ca\textsuperscript{2+} oscillations but (b) the
 latter does not influence the Hopf transition point in the reduced model.

Next, by increasing $G_{\rm{CaL}}$ in the full myocyte models, we have obtained calcium-overload conditions and shown the following.
Under Ca\textsuperscript{2+} overload, the TP06 model shows LCRs, which lead to EAD-type depolarizations without reopening the $I_{\rm{CaL}}$ channel; such LCRs have been reported in Refs.~\cite{song2015calcium},~\cite{fink2010pharmacodynamic},~\cite{fowler2018late},~\cite{shiferaw2012intracellular}, ~\cite{fowler2020arrhythmogenic}; however, the HuVEC15 model does not exhibit these LCRs.

 By tuning the calcium overload and making a few other modifications, such as the introduction of leak currents (in the TP06 model) and changing the distribution of NCX channels (in the HuVEC15 model), we have obtained DADs in both these models. We have then shown that, by using various combinations of parameters (such as $G_{\rm{CaL}}$, $V_{\rm{rel}}$, $K_{\rm{NaCa}}$ and $V_{\rm{maxup}}$) three different types of DADs occur: (a) subthreshold, (b) suprathreshold, and (c) multi-blip (or multiple sub-threshold DADs between two subsequent APs) DADs. To distinguish between these DAD types, we have identified two essential characteristics, namely,
 DAD\textsubscript{amp}  and DAD\textsubscript{freq}, whose parameter-sensitivity we have examined.\\
 Our parameter-sensitivity analyses have shown that, in the TP06 model, $K_{\rm{NaCa}}$, $G_{\rm{K1}}$, and $V_{\rm{maxup}}$ and, in the HuVEC15 model, $K_{\rm{NaCa}}$, $f_{\rm{NCX}}$ and $G_{\rm{K1}}$
 affect the amplitude of the DADs most significantly. The importance of $K_{\rm{NaCa}}$ and $G_{\rm{K1}}$ has been reported in the earlier literature ~\cite{jost2013orm,voigt2012enhanced,bogeholz2015suppression,myles2015decreased,xu2007ik1}. However, the negative effect of $V_{\rm{maxup}}$ appears in the sensitivity analysis of DAD\textsubscript{amp}, in the TP06 model, because $V_{\rm{maxup}}$ removes excess calcium from the cytosol. Therefore, $V_{\rm{maxup}}$ directly competes with electrogenic $K_{\rm{NaCa}}$ and thus reduces the rate of increase of DAD\textsubscript{amp}. The parameter $f_{\rm{NCX}}$, in the HuVEC15 model, represents the fraction of the Na\textsuperscript{+}-Ca{\textsuperscript{2+}} exchangers in the vicinity of the intermediate zone (iz in Fig. \ref{fig:myocyte_schema}), which has a higher calcium concentration than the cytosol; therefore, thermodynamic or electrochemical forces, which appear in the equations for the sodium-calcium exchanger current (see Ref.~\cite{himeno2015human}), make $f_{\rm{NCX}}$ evacuate calcium at an enhanced rate and contribute positively to DAD\textsubscript{amp}.

We have shown that DAD\textsubscript{freq} is affected principally by $G_{\rm{CaL}}$, $K_{\rm{NaCa}}$ and $V_{\rm{maxup}}$, in the TP06 model, and by $G_{\rm{CaL}}$, $K_{\rm{NaCa}}$, $f_{\rm{NCX}}$ and $V_{\rm{rel}}$, in the HuVEC15 model. The sensitivity of these conductances and fluxes can be explained as follows: (a) $G_{\rm{CaL}}$ is involved in calcium overloading, which is necessary for DADs; (b) $K_{\rm{NaCa}}$ and $f_{\rm{NCX}}$ (the latter in the HuVEC15 model) impact DAD\textsubscript{freq} negatively, because they lead to calcium unloading in the forward mode; (c) an increase in $V_{\rm{maxup}}$ enhances the refilling of the SR and, therefore, for the release of calcium; (d) in the HuVEC15 model, $V_{\rm{rel}}$ increases DAD\textsubscript{freq}. 
The differences in the parameter sensitivity of DAD\textsubscript{freq} in the TP06 and HuVEC15 models arises principally because of the different ways in which these models account for RyRs (see Sec.~\ref{subsec:Models}).
The DAD phase diagrams in Figs.~\ref{fig:phase_dia_tp06} and \ref{fig:phase_dia_asakura}, which 
we present in Sec.~\ref{subsec:phases}, show the types of DADs that occur as we change parameters in the TP06 and HuVEC15 models.
 Although the formulation of calcium dynamics is different in the TP06 and HuVEC15 models, our results for both these models agree on the factors that promote SCRs and DADs. A few earlier studies [e.g., Refs.~\cite{sato2021increasing} and \cite{salazar2016role}] show that an increase in $V_{\rm{maxup}}$ promotes DADs as we also find in our study; in contrast, some other studies claim that the SERCA pump reduces the incidence DADs [e.g., Refs.~\cite{fink2011ca2+} and \cite{davia2001serca2a}]. These contrasting differences in these claims may arise from the differences in (a) the sensitivity of the RyR activation to the calcium in the SR and SS or (b) the methods used to generate the Ca\textsuperscript{2+} overload. For example, Ref.~\cite{fink2011ca2+} uses $Na_{\rm{i}}$ overload to increase the cytosolic and SR Ca\textsuperscript{2+} in the myocyte; by contrast, we use increases in $I_{\rm{CaL}}$ and the RyR leak to trigger DADs. The cytosolic- and SR-compartment ratios also play roles in deciding the $V_{\rm{maxup}}$-dependence of DADs. Moreover, the steepness of the opening of the RyR can influence the dependence of DADs on $V_{\rm{rel}}$.
   The multi-blip DADs that we have observed in both TP06 and HuVEC15 models are the consequence of fast Ca\textsuperscript{2+} uptake by the SERCA pump, which refills the SR and prepares it for the subsequent firing. Such DADs are similar to diastolic-membrane-potential oscillations reported in the sinoatrial node (SAN)~\cite{catanzaro2006mechanisms}, a Purkinje-cell model~\cite{shah2019delayed}, and in myocardial myocytes~\cite{zygmunt1998naca}. 
  
An increase in the value of $K_{\rm{NaCa}}$ usually increases the DAD amplitude and promotes suprathreshold DADs. However, in the TP06 model, we have demonstrated the termination of suprathreshold DADs beyond a threshold value of $K_{\rm{NaCa}}$. This termination occurs because of the LCRs in the TP06 model: LCRs increase the cytosolic calcium, during the late phases of AP; this drives the NCX into the forward mode; a prominent $K_{\rm{NaCa}}$ can unload the excess Ca\textsuperscript{2+} from the myocyte and thus eliminate SCRs and DADs.
  
 Subthreshold DADs are known to inactivate the $I_{\rm{Na}}$ channels~\cite{singer1967interrelationships} and can have consequences at the tissue scale~\cite{liu2015delayed}. We have shown that multi-blip and subthreshold DADs can partially inactivate Na\textsuperscript{+} channels [see Fig.~\ref{fig:SI:hj_gates} in the Appendix]. In particular, multi-blip DADs can inactivate Na\textsuperscript{+} channels multiple times, thereby increasing the chances of conduction blocks in cardiac tissue (cf., Ref.~\cite{liu2015delayed}). Moreover, by reducing $G_{\rm{K1}}$, we can facilitate the formation of multi-blip DADs [see Fig.~\ref{fig:SI:Ik1_effect} in the Appendix].

In Subsection~\ref{subsec:tissue} we have discussed, for both TP06 and HuVEC15 models, the results of our studies  with DAD clumps, with the three types of DAD myocytes.  We have examined wave dynamics in 1D cable, 2D tissue, and 3D anatomically realistic bi-ventricular domains. We have demonstrated that, if the linear size of the DAD clump is above a parameter-dependent threshold value, this clump fires PVCs. Our simulations have shown, in particular, how a normal stimulus, applied at the apex of the human bi-ventricular geometry, leads to the propagation of electrical activation that encounters the DAD clump from which PVCs emerge [Figs.~\ref{fig:whole_heart}(c) and (d)]. 

Thus, we have shown that both TP06 and HuVEC15 models are useful for studying PVCs, induced by different types of DAD clumps in tissue and bi-ventricular domains. During calcium overload, LCRs, which frequently accompany diastolic SCRs and DADs, can play a crucial role in the dynamics of $V_{\rm{m}}$ and Ca\textsuperscript{2+} overload. We have shown that both LCRs and SCRs can occur in the TP06 model [Figs.~\ref{fig:dadtypes}(b)-(d) and, in the Appendix, Fig.~\ref{fig:SI:sparks_DAD_EAD}]. Therefore, the TP06 model is a natural candidate for the examination of DAD-induced PVCs; we investigate this in detail in Paper II. 

\section{Limitations of Our Study}
\label{subsec:limitations}
  The origin of DADs is related to the sub-cellular phenomena of calcium sparks and calcium waves; our study does not discuss the latter in detail.
  The dependencies of the widths and durations of DADs on sub-cellular-scale parameters are not addressed here.
\acknowledgments
We thank Mahesh Mulimani and Soling Zimik for valuable discussions.  

\begin{widetext}
\appendix
\section*{Appendix}
\label{supp}
In this Appendix, we give details of the following:
\begin{itemize}
\item Videos illustrating the spatiotemporal evolution of our tissue simulations [Subsections~\ref{app:2d_tp_plane_wave} - \ref{app:bi_vent_PVC}].
\item $Na_{\rm{i}}$ overload and the Ca\textsuperscript{2+}-subsystem, in Subsection~\ref{app:subsystem}.
\item $S_{\rm{GKr}}$ values required for $S_{\rm{GCaL}}$, in Subsection~\ref{app:linear_fit}.
\item The TP06 and HuVEC15 models with Ca\textsuperscript{2+} overload, in Subsection~\ref{app:overload}.
\item Changes introduced in the TP06 and HuVEC15 myocyte models, in Subsection~\ref{app:changes}.
\item Subthreshold DADs and $I_{\rm{Na}}$ inactivation, in Subsection~\ref{app:na_avail}.
\item Robustness of our parameter-sensitivity results, in Subsection~\ref{app:robust}.
\item The effect of the $I_{\rm{K1}}$ conductance on the DAD amplitude, in Subsection~\ref{app:Ik1}.
\end{itemize}

\subsection{Video 1}
\label{app:2d_tp_plane_wave}
Animation of the pseudocolor plots of $V_{\rm{m}}$(mV), for the TP06 model, illustrating the spatiotemporal evolution of plane wave pacing in cardiac tissue as in Fig.~\ref{fig:2d_tp}(i)(a). The parameter set we use is: $S_{\rm{GCaL}} = 1.0$, $S_{\rm{Vmaxup}} = 1.0$, $S_{\rm{KNaCa}} = 1.0$, 
$S_{\rm{Vrel}} = 1.0$,  $S_{\rm{GKr}} = 1.0$ . For the video, we use $30$ frames per second with each frame separated from the succeeding frame by $20$ms in real-time. See video here: \url{https://youtu.be/qkWOesjLJxg}.

\subsection{Video 2}
\label{app:2d_tp_sub_DAD}
Animation of the pseudocolor plots of $V_{\rm{m}}$(mV), for the TP06 model, illustrating the spatiotemporal evolution of plane-wave pacing in cardiac tissue and the emergence of PVCs, from the subthreshold-DAD clump, as in Fig.~\ref{fig:2d_tp}(i)(b). The parameter set we use is: $S_{\rm{GCaL}} = 2.0$, $S_{\rm{Vmaxup}} = 3.0$, $S_{\rm{KNaCa}} = 1.0$, 
$S_{\rm{Vrel}} = 1.0$,  $S_{\rm{GKr}} = 1.0$. For the video, we use $30$ frames per second (fps), with an inter-frame separation (ifs) of $20$ ms in real-time. See video here
\url{https://youtu.be/CsPXpn7yTlk}.

\subsection{Video 3}
\label{app:2d_tp_multi_DAD}
Animation of the pseudocolor plots of $V_{\rm{m}}$(mV), for the TP06 model, illustrating the spatiotemporal evolution of plane-wave pacing in cardiac tissue and the emergence of PVCs, from the multiblip-DAD clump, as in Fig~\ref{fig:2d_tp}(i)(c). The parameter set we use is: $S_{\rm{GCaL}} = 2.0$, $S_{\rm{Vmaxup}} = 4.5$, $S_{\rm{KNaCa}} = 0.8$, $S_{\rm{Vrel}} = 1.0$,  $S_{\rm{GKr}} = 1.0$ . For the video, we use fps$=30$ and ifs$=20$ms in real-time. See video here: \url{https://youtu.be/JqEOs0vYSFM}.

\subsection{Video 4}
\label{app:2d_tp_supra_DAD}
Animation of the pseudocolor plots of $V_{\rm{m}}$(mV), for the TP06 model, illustrating the spatiotemporal evolution of plane-wave pacing in cardiac tissue and the emergence of PVCs, from the suprathreshold-DAD clump, as in Fig~\ref{fig:2d_tp}(i)(d). The parameter set we use is: $S_{\rm{GCaL}} = 2.0$, $S_{\rm{Vmaxup}} = 3.0$, $S_{\rm{KNaCa}} = 2.5$, $S_{\rm{Vrel}} = 1.0$,  $S_{\rm{GKr}} = 1.0$. For the video, we use fps $=30$ and ifs $=20$ms in real-time. See video here: \url{https://youtu.be/E6pScuUPS9E}.

\subsection{Video 5}
\label{app:2d_ask_plane_wave}
Animation of the pseudocolor plots of $V_{\rm{m}}$(mV), for the HuVEC15 model, illustrating the spatiotemporal evolution of plane-wave pacing in cardiac tissue, as in Fig~\ref{fig:2d_tp}(ii)(a). The parameter set we use is: $S_{\rm{GCaL}} = 1.0$, $S_{\rm{Vmaxup}} = 1.0$, $S_{\rm{KNaCa}} = 1.0$, $S_{\rm{Vrel}} = 1.0$,  $S_{\rm{GKr}} = 1.0$. For the video, we use fps $=30$ and ifs $=20$ms in real-time. See video here: \url{https://youtu.be/T_P9_NqDnVY}.

\subsection{Video 6}
\label{app:2d_ask_sub_DAD}
Animation of the pseudocolor plots of $V_{\rm{m}}$(mV), for the HuVEC15 model, illustrating the spatiotemporal evolution of plane-wave pacing in cardiac tissue and the emergence of PVCs, from the subthreshold-DAD clump, as in Fig~\ref{fig:2d_tp}(ii)(b). The parameter set we use is: $S_{\rm{GCaL}} = 4.0$, $S_{\rm{Vmaxup}} = 3.0$, $S_{\rm{KNaCa}} = 1.0$, 
$S_{\rm{Vrel}} = 1.2$,  $S_{\rm{GKr}} = 4.7$, $S_{\rm{GK1}} = 1$. For the video, we use fps $=30$ frames per second with each frame separated from the succeeding frame by ifs $=20$ms in real-time. See video here: \url{https://youtu.be/stxNdr0bqv8/}.
\subsection{Video 7}
\label{app:2d_ask_multiDAD}
Animation of the pseudocolor plots of $V_{\rm{m}}$(mV), for the HuVEC15 model, illustrating the spatiotemporal evolution of plane-wave pacing in cardiac tissue and the emergence of subthreshold PVCs (multiple times between two successive pacings), from the multiblip-DAD clump, as in Fig~\ref{fig:2d_tp}(ii)(c). The parameter set we use is: $S_{\rm{GCaL}} = 4.0$, $S_{\rm{Vmaxup}} = 3.0$, $S_{\rm{KNaCa}} = .7$, 
$S_{\rm{Vrel}} = 2.2$,  $S_{\rm{GKr}} = 4.7$, $S_{\rm{GK1}} = 0.35$. For the video, we use fps $=30$ and ifs $=20$ms in real-time. See video here: \url{https://youtu.be/K_uyKwcfTh8}.

\subsection{Video 8}
\label{app:2d_ask_supra_DAD}
Animation of the pseudocolor plots of $V_{\rm{m}}$(mV), for the HuVEC15 model, illustrating the spatiotemporal evolution of plane-wave pacing in cardiac tissue and the emergence of suprathreshold PVCs, from the DAD clump, as in Fig~\ref{fig:2d_tp}(ii)(d). The parameter set we use is: $S_{\rm{GCaL}} = 4.0$, $S_{\rm{Vmaxup}} = 3.0$, $S_{\rm{KNaCa}} = 1.0$, 
$S_{\rm{Vrel}} = 1.0$, $S_{\rm{GKr}} = 5.2$, $S_{\rm{GK1}} = 1.0$. For the video, we use fps $=30$ and ifs $=20$ms in real-time. See video here: \url{https://youtu.be/nmaMdlUCLMY}.

\subsection{Video 9}
\label{app:3d_tp_slab}
Animation of the pseudocolor plots of $V_{\rm{m}}$(mV), for the TP06 model, illustrating the spatiotemporal evolution of plane-wave pacing in cuboidal cardiac tissue, with a disc-shaped suprathreshold-DAD clump embedded in it, and the emergence of PVCs from the clump, as in Fig~\ref{fig:3d}(a)-(b). The parameter set we use is: $S_{\rm{GCaL}} = 2.5$, $S_{\rm{Vmaxup}} = 4.5$, $S_{\rm{KNaCa}} = 2.5$, 
$S_{\rm{Vrel}} = 1.0$, $S_{\rm{GKr}} = 2.8$. For the video, we use fps $=30$ and ifs $=20$ms in real-time. See video here: \url{https://youtube.com/shorts/D9bI7WA9XkE}.

\subsection{Video 10}
\label{app:3d ask slab}
Animation of the pseudocolor plots of $V_{\rm{m}}$(mV), for the HuVEC15 model, illustrating the spatiotemporal evolution of plane-wave pacing in cuboidal cardiac tissue, with a disc-shaped suprathreshold-DAD clump embedded in it, and the emergence of PVCs from the clump, as in Fig~\ref{fig:3d}(c)-(d). The parameter set we use is: $S_{\rm{GCaL}} = 3.0$, $S_{\rm{Vmaxup}} = 3.0$, $S_{\rm{KNaCa}} = 2.7$, 
$S_{\rm{Vrel}} = 1.0$, $S_{\rm{GKr}} = 5.2$, $S_{\rm{GK1}} = 1.0$. For the video, we use fps $=30$ and ifs $=20$ms in real-time. See video here: \url{https://youtube.com/shorts/3Qc-Av-jg64}.

\subsection{Video 11}
\label{app:bi_vent_PVC}
Animation of the pseudocolor plots of $V_{\rm{m}}$(mV), for the TP06 model, illustrating the spatiotemporal evolution of $V_{\rm{m}}$(mV) in a human bi-ventricular geometry, with a suprathreshold-DAD clump embedded in it, and the emergence of PVCs from the clump, as in Fig~\ref{fig:whole_heart}(b)-(d). The parameter set we use is: $S_{\rm{GCaL}} = 2.5$, $S_{\rm{Vmaxup}} = 4.5$, $S_{\rm{KNaCa}} = 2.5$, $S_{\rm{Vrel}} = 1.0$, $S_{\rm{GKr}} = 2.8$. For the video, we use fps $=30$ and ifs $=20$ms in real-time. See video here: \url{https://youtu.be/oRqnPLYqTGE}.

\subsection{$Na_{\rm{i}}$ overload and the Ca\textsuperscript{2+}-subsystem}
\label{app:subsystem}
The Na{\textsuperscript{+}}-Ca\textsuperscript{2+} exchanger (NCX) functions in both the 
forward and backward directions. During Na{\textsuperscript{+}} overload, it removes $3$ Na{\textsuperscript{+}} outside of the cell in exchange for a single Ca{\textsuperscript{2+}} ion (backward mode); however, an increase in intracellular Ca\textsuperscript{2+} can force the NCX in the opposite direction (forward mode), in which NCX removes $1$ Ca\textsuperscript{2+} outside in exchange of $3$ Na{\textsuperscript{+}} inside (forward mode). In the reduced model (in the absence of $I_{\rm{CaL}}$ channels)  we use intracellular Na{\textsuperscript{+}} ($Na_{\rm{i}}$) overload, which forces the Na{\textsuperscript{+}}-Ca{\textsuperscript{2+}} exchanger to the backward mode that overloads the Ca\textsuperscript{2+}-subsystem. The NCX is modeled by the following equation:
\begin{equation}{\label{eq:naca}}
I_{\rm{NaCa}} = {K_{\rm{NaCa}}}{\frac {{\exp{\left (\frac{{\gamma}{V_{\rm{m}}F}}{{{RT}}}\right)}}
{{{Na_{\rm{i}}}}^3}{Ca_{\rm{o}}}-{\exp{\left (\frac{(\gamma -1){V_{\rm{m}}F}}{{{RT}}}\right )}}
{{{Na_{\rm{o}}}}^3}{Ca_{\rm{i}}}{\alpha}} {({{K_{\rm{mNai}}}^3}+{{{Na_{\rm{o}}}}^3})({K_{{mCa}}}+{{Ca_{\rm{o}}}})
{\left (1+K_{\rm{sat}}{\exp{\left (\frac{(\gamma -1){V_{\rm{m}}F}}{{{RT}}}\right )}}\right)}}}\,, \\
\end{equation}
where $V_{\rm{m}}$ is the membrane potential, $Na_{\rm{i}}$  the intracellular Na\textsuperscript{+} concentration, $Ca_{\rm{i}}$ the intracellular Ca\textsuperscript{2+}  concentration, and
\begin{eqnarray*}
\text{constant}~ \gamma &=& 0.35 \,;\\
\text{parameter}~ K_{\rm{NaCa}} &=& 1000.0 ~\dfrac{pA}{pF}\,; \\
\text{Faraday constant}~ F &=& 96.485 ~\dfrac{C}{m-mole}\,; \\
\text{Gas constant}~ R &=& 8.314 ~\dfrac{joule}{mole-K}\,; \\
\text{Temperature}~ T &=& 310 ~K \,;\\
\text{Extracellular calcium concentration}~ Ca_{\rm{o}} &=& 2 ~mM \,;\\
\text{Extracellular Na concentration}~ Na_{\rm{o}} &=& 140 ~mM \,;\\
\text{constant}~ K_{\rm{mNai}} &=& 87.5 ~mM \,;\\
\text{constant}~ K_{\rm{mCa}} ~ &=& 1.38 ~mM \,;\\
\text{constant}~ K_{\rm{sat}} ~ &=& ~0.1 \,.
\end{eqnarray*}
For the full description of the Ca\textsuperscript{2+}-subsystem of TP06 model we require, in addition, the following set of equations:  
\begin{eqnarray*}
\text{constant}~ Buf_{\rm{c}} &=& 0.2  ~mM \,;\\
\text{constant}~ Buf_{\rm{sr}} &=& 10.0 ~mM \,;\\
\text{constant}~ Buf_{\rm{ss}} &=& 0.4 ~mM \,;\\
\text{constant}~ EC &=& 1.5   ~mM \,;\\ 
\text{constant}~ K_{\rm{buf_c}} &=& 0.001  ~mM \,;\\
\text{constant}~ K_{\rm{buf_{sr}}} &=& 0.3   ~mM \,;\\
\text{constant}~ K_{\rm{buf_{ss}}} &=& 0.00025  ~mM \,;\\
\text{constant}~ K_{\rm{up}} &=& 0.00025  ~mM \,;\\ 
\text{constant}~ V_{\rm{{leak}}} &=& 0.00036 ~s^{-1} \,;\\ 
\text{constant}~ V_{\rm{rel}} &=& 0.102  ~s^{-1} \,;\\  
\text{constant}~ V_{\rm{sr}} &=& 0.001094 ~\mu M^3 \,;\\  
\text{constant}~ V_{\rm{ss}} &=& 0.00005468 ~\mu M^3 \,;\\  
\text{constant}~ V_{\rm{xfer}} &=& 0.0038  ~ms^{-1} \,;\\ 
\text{constant}~ V_{\rm{maxup}} &=& 0.006375   \dfrac{mM}{ms} \,;\\
\text{constant}~ k_{\rm{1}}^{\prime} &=& 0.15 ~{(M^2-ms)}^{-1} \,;\\ 
\text{constant}~ k_{\rm{2}}^{\prime} &=& 0.045 ~{(mM-ms)}^{-1} \,;\\ 
\text{constant}~ k_{\rm{3}} &=& 0.06  \,;\\
\text{constant}~ k_{\rm{4}} &=& 0.005 \,;  \\
\text{constant}~ max_{\rm{sr}} &=& 2. \,;  \\
\text{constant}~ min_{\rm{sr}} &=& 1.0 \,;  \\
\text{constant}~ C_{\rm{m}} &=& 0.185 ~\mu F \,; \\
\text{constant}~ V &=& -85.23 ~mV \,; \\
\text{constant}~ V_{\rm{c}} &=& 0.016404  ~\mu m^{3}  \,; \\
\text{constant}~ \alpha &=& 2.5 \,;\\
\end{eqnarray*}
\begin{eqnarray}
I_{\rm{leak}} &=&
\begin{cases}
     V_{\rm{leak}}\left({Ca_{\rm{SR}}-{Ca_{\rm{i}}}}\right),& \text{if } \text{isolated myocyte} \,;\\
    0,              & \text{tissue}\,;
\end{cases} \\
I_{\rm{up}} &=& \dfrac{V_{\rm{maxup}}}{1+\left({\frac {K_{\rm{up}}}{{Ca_{\rm{i}}}}}\right)^2} \,;\\
I_{\rm{rel}} &=& \left(V_{\rm{rel}}.O + V_{\rm{RyRL}}\right) \left(Ca_{\rm{SS}}- Ca_{\rm{SS}}\right) \,;\\
V_{\rm{RyRL}} &=& 0.00018 \, \nonumber ;\\
I_{\rm{xfer}} &=& V_{\rm{xfer}} \left(Ca_{\rm{SS}}- Ca_{\rm{i}}\right) \,;\\
O &=& \dfrac{k_{\rm{1}} {Ca_{\rm{SS}}}^2 \bar R}{{k_{\rm{3}}}+ {k_{\rm{1}} {Ca_{\rm{SS}}}^2}}
\end{eqnarray}
\begin{eqnarray}
\dfrac{d\bar R}{dt} &=& -{k_{\rm{2}}} Ca_{\rm{SS}} \bar R+ {k_{\rm{4}}}\left(1- \bar R\right) \,;\\
Ca_{\rm{i_{bufc}}} &=& \dfrac{1.0}{
1.0+ \dfrac{Buf_c.K_{\rm{buf_c}}}{\left(Ca_{\rm{i}}+K_{\rm{buf_c}}\right)^2}} \,;\\
Ca_{\rm{sr_{bufsr}}} &=& \dfrac{1.0}{1.0+ \dfrac{Buf_{sr} K_{\rm{buf_{sr}}}}{\left(Ca_{\rm{SR}}+K_{\rm{buf_{sr}}}\right)^2}} \,;\\
Ca_{\rm{ss_{bufss}}} &=& \dfrac{1.0}{1.0+\dfrac{Buf_{\rm{ss}}K_{\rm{buf_{ss}}}}{\left(Ca_{\rm{SS}}+K_{\rm{buf_{ss}}}\right)^2}} \,;\\
\dfrac{dCa_{\rm{i}}}{dt} &=& Ca_{\rm{i_{bufc}}}\left(\left(I_{\rm{leak}}-I_{\rm{up}}\right)\dfrac{V_{\rm{sr}}}{V_{\rm{c}}}+I_{\rm{xfer}}-2.I_{\rm{NaCa}}.C_{\rm{m}}/\left(2V_{\rm{c}}F\right)\right) \,;\\
\dfrac{dCa_{\rm{SR}}}{dt} &=& Ca_{\rm{sr_{bufsr}}}\left(I_{\rm{up}}-\left(I_{\rm{rel}}+I_{\rm{leak}}\right)\right) \,;\\
\dfrac{dCa_{\rm{SS}}}{dt} &=& Ca_{\rm{ss_{bufss}}}\left(I_{\rm{rel}}\dfrac{V_{\rm{sr}}}{V_{\rm{ss}}}-I_{\rm{xfer}}\dfrac{V_{\rm{c}}}{V_{\rm{ss}}}\right) \,;\\
k_1 &=& \dfrac{k_{\rm{1}}^{\prime}}{k_{\rm{casr}}} \,;\\
k_2 &=& {k_{\rm{2}}^{\prime}}{k_{\rm{casr}}} \,;\\
k_{\rm{casr}} &=& max_{\rm{sr}}- \dfrac{max_{\rm{sr}}- min_{\rm{sr}}}{1+ \left(\dfrac{EC}{Ca_{\rm{SR}}}\right)^2} \,.
\end{eqnarray}

\subsection{ $S_{\rm{GKr}}$ values required for $S_{\rm{GCaL}}$}
\label{app:linear_fit}
In the main paper, we use a range of values for $S_{\rm{GCaL}}$; however, to counter the change in the action potential duration (APD), because of the change in $S_{\rm{GCaL}}$, we must adjust the factor $S_{\rm{GKr}}$, so we determine the values of $S_{\rm{GKr}}$ that are required for $4$ values of $S_{\rm{GCaL}}$, which we use for a straight-line fit (Fig.~\ref{fig:SI:fit}), whence we obtain other values of $S_{\rm{GKr}}$ that should be used for a given value of $S_{\rm{GCaL}}$.

\begin{figure}
     \centering
      \includegraphics[width=0.8\textwidth]{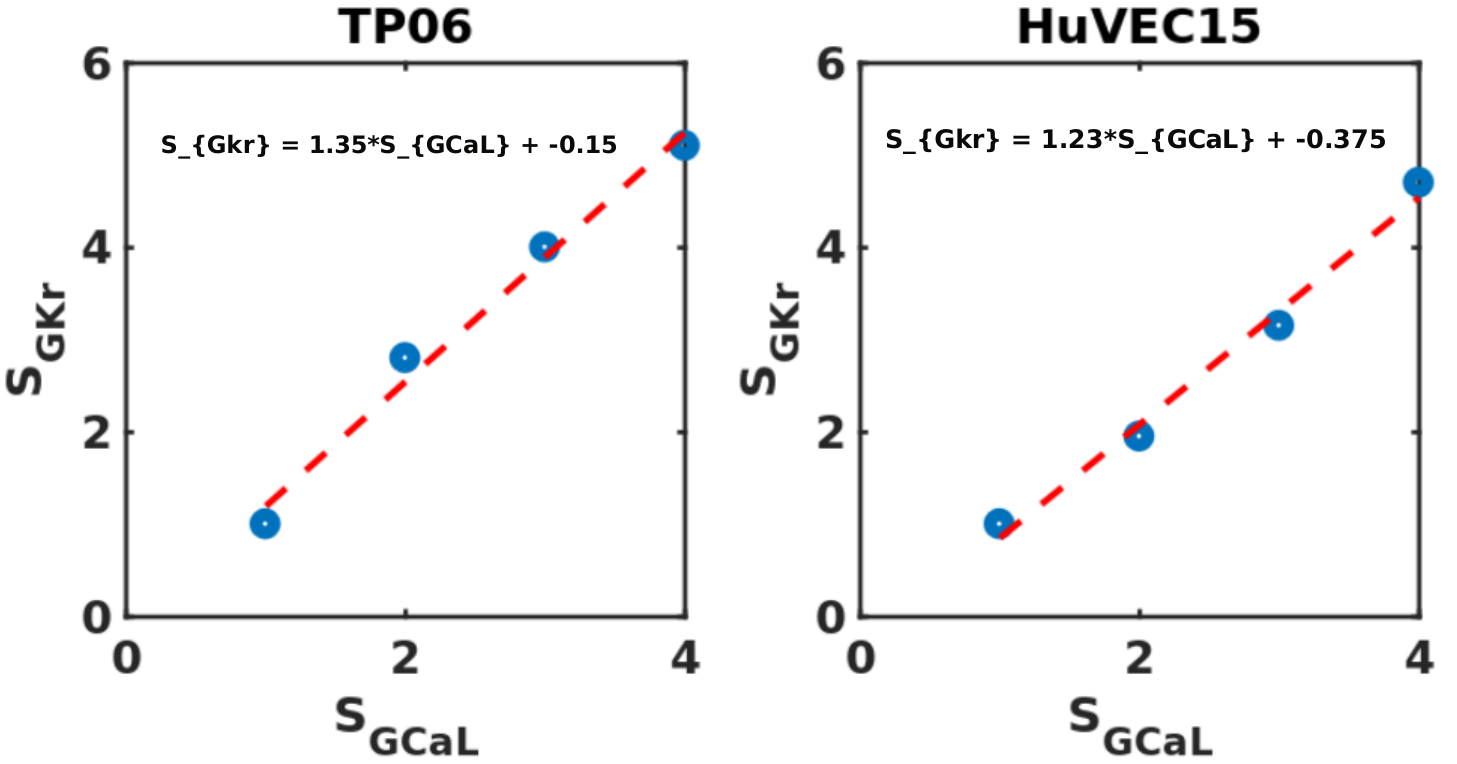}
      \caption{ (Color online) \textbf{Linear fits for the $S_{\rm{GKr}}$ required for a given $S_{\rm{GCaL}}$ (see text):} for the TP06 (left panel) and HuVEC15 (right panel) models.}
      \label{fig:SI:fit}
\end{figure}

\subsection{The TP06 and HuVEC15 models with Ca\textsuperscript{2+} overload}
\label{app:overload}
We increase the calcium load of the TP06 and HuVEC15 models by using $S_{\rm{GCaL}}=2$ (and the corresponding values of $S_{\rm{GKr}}$ values discussed in the previous Section). We stimulate the myocyte, in both these models, for 500 AP (1 Hz). In this Ca\textsuperscript{2+}-overload condition, the TP06 model triggers an extra systolic calcium spark or late calcium release (LCR) during the AP (see Fig. \ref{fig:SI:sparks_DAD_EAD}(a)), whereas the HuVEC15 model shows a spontaneous calcium release (SCR) in the diastolic interval (Fig.\ref{fig:SI:sparks_DAD_EAD}(b)); these two types of calcium releases force the NCX to the forward mode and, therefore, increase $V_{\rm{m}}$. The LCRs leads to $V_{\rm{m}}$ depolarizations that are similar to EADs, whereas the SCRs lead to DADs.

\begin{figure}
     \centering
    \includegraphics[width=\textwidth]{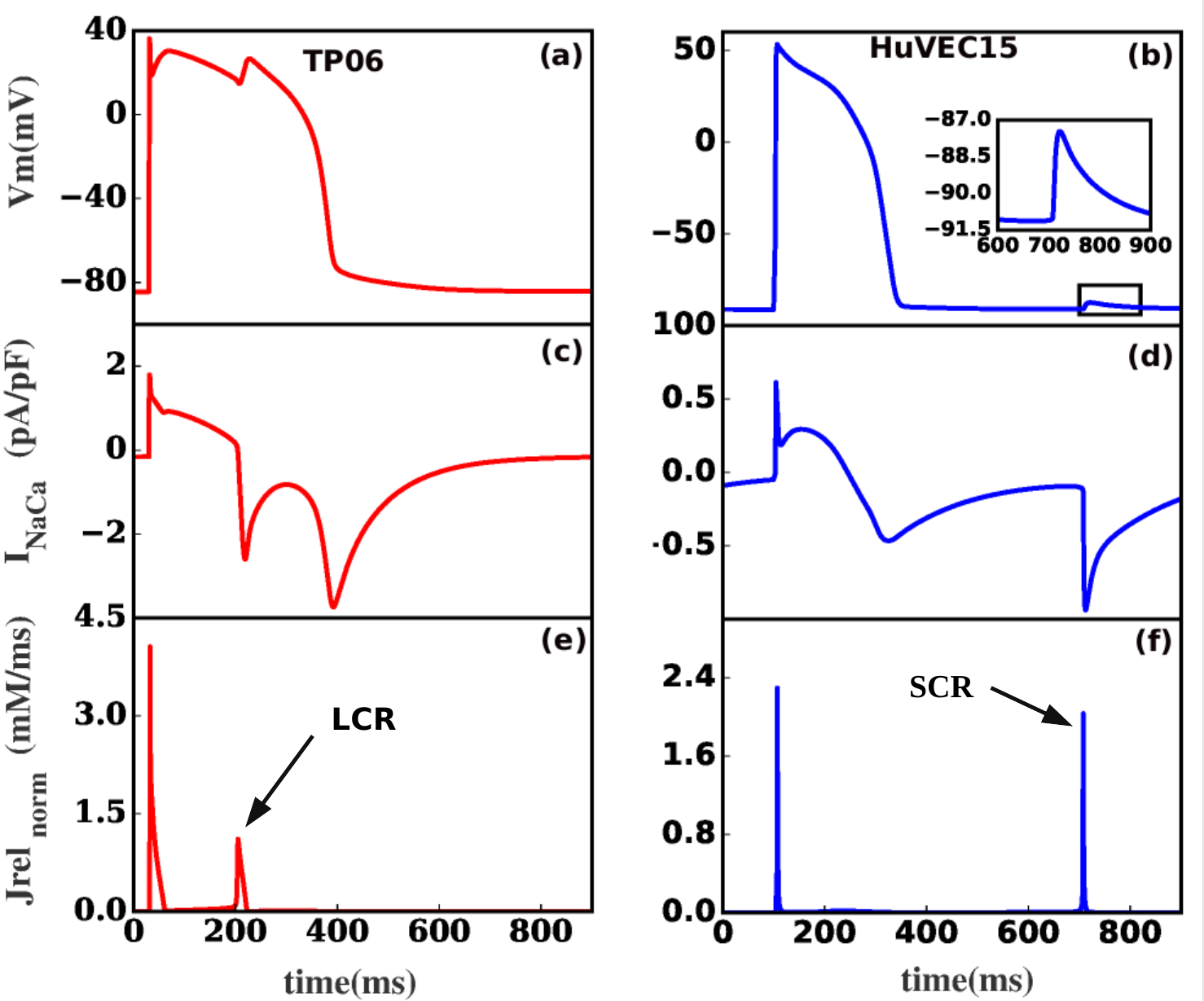}
         \caption{ (Color online) \textbf{Ca\textsuperscript{2+} Sparks and Afterdepolarizations, with Ca\textsuperscript{2+} overload, in the TP06 (left) and HuVEC15 (right) models:} (a) Ca\textsuperscript{2+} sparks causing EADs in the TP06 model; (b DADs in the HuVEC15 model; (c) and (d): plots of $I_{\rm{NaCa}}$ depicting how Ca\textsuperscript{2+} sparks change the direction of NCX from outward to inward; (e) and (f): normalised calcium-release flux from the SR compartment to the subspace (in the TP06 model) or the junctional space (in the 
         HuVEC15 model). We use $S_{\rm{GCaL}}=2$ (and the corresponding values of $S_{\rm{GKr}}$ values discussed in the previous Subsection).}
         \label{fig:SI:sparks_DAD_EAD}
\end{figure}

\subsection{Changes introduced in the TP06 and HuVEC15 myocyte models}
\label{app:changes}
We have discussed above how we treat Ca\textsuperscript{2+} overload in both the TP06 and HuVEC15 models; and we have shown that the resulting Ca\textsuperscript{2+} overload is enough to trigger DADs in the HuVEC15 model, but not in the TP06 model; furthermore, we find that the amplitude of the DAD in the HuVEC15 is only $2$ mV. Therefore, we introduce the following changes in these two models: (a) we introduce a leak of Ca\textsuperscript{2+} ions, through the RyR, in the TP06 model to trigger DADs; (b) while preserving the overall density of NCX in the various compartments, we increase the fraction of 
NCX in the intermediate zone of the HuVEC15 model.

\subsection{Role of the RyR Leak Current in the TP06 and HuVEC15 Models}
\label{app:leak}
In both these models, the opening of the RyRs is modeled via the Ca-induced-Ca release (CICR). The calcium availability inside the SR stores ($Ca_{\rm{SR}}$ and $Ca_{\rm{SRrl}}$ for TP06 and HuVEC15
models, respectively) and the subspaces outside the RyR, also called triggers ($Ca_{\rm{SS}}$ and $Ca_{\rm{jnc}}$ for TP06 and HuVEC15 models, respectively) both modulate the opening probability of 
the RyR. If the trigger is not enough during the diastole, then Ca\textsuperscript{2+} sparks and 
DADs do not occur. Therefore, a small calcium leak, through the closed RyRs, suffices for $[Ca^{2+}]_{\rm{SS}}$ to reach the threshold trigger levels required for to SCRs and DADs.
 Therefore, we add a small calcium leak (background SR Ca\textsuperscript{2+} release that is independent of the RyR opening probability) through the RyR channel, in the TP06 model, as follows:
\begin{align}
\label{eq:SI:TP06leak}
\begin{split}
 I_{\rm{rel}} =& ~(V_{\rm{rel}}.O + V_{\rm{RyRL}})([Ca^{2+}]_{\rm{SR}} - [Ca^{2+}]_{\rm{SS}})\,;
\\
 V_{\rm{rel}} =& ~0.102 ~\text{ms}^{-1}\,.
\end{split}
\end{align}
$I_{\rm{rel}}$ is the molar calcium-induced calcium release (CICR) current, $O$ the opening probability of the RyR, $V_{\rm{RyRL}} = 0.00018$ms$^{-1}$ the rate constant of calcium leak through RyR, $V_{\rm{rel}}$ the rate constant of the calcium release through RyR, and $[Ca^{2+}]_{\rm{SR}}$ and $[Ca^{2+}]_{\rm{SS}}$ are 
the SR and subspace molar calcium concentrations, respectively. This kind of RyR leak is already present in the HuVEC15 model as follows:

\begin{align}
\label{eq:SI:Asakuraleak}
\begin{split}
 I_{\rm{rel}} =& ~\dfrac{J_{\rm{rel}}}{V_{\rm{SRrl}}}\,; V_{\rm{SRrl}} = ~300 ~ \text{fL}\,; 
\\
V_{\rm{rel}} =& ~\frac{P_{\rm{RyR}}}{V_{\rm{SRrl}}}\,; P_{\rm{RyR}} = ~5191 ~ \text{fL/mS}\,;\\
 J_{\rm{rel}} =&  ~P_{\rm{RyR}}(p_{\rm{ORyR}} + 0.000075) ([Ca^{2+}]_{\rm{SRrl}} - [Ca^{2+}]_{\rm{jnc}}) 
\end{split}
\end{align}

$J_{\rm{rel}}$ is the CICR current , $P_{\rm{RyR}}$ the rate constant, $ V_{\rm{rel}} \times 0.000075$ms$^{-1}$
is the molar RyR leak current, $p_{\rm{ORyR}}$ is the RyR opening probability, $[Ca^{2+}]_{\rm{SRrl}}$ and $[Ca^{2+}]_{\rm{jnc}}$ are, respectively, the molar calcium concentrations in the junctional space and release compartments of the SR. The RyR leak plays a crucial role in triggering SCRs and DADs, in both TP06 and HuVEC15 models.
\begin{figure}
     \centering
      \includegraphics[width=0.9\textwidth]{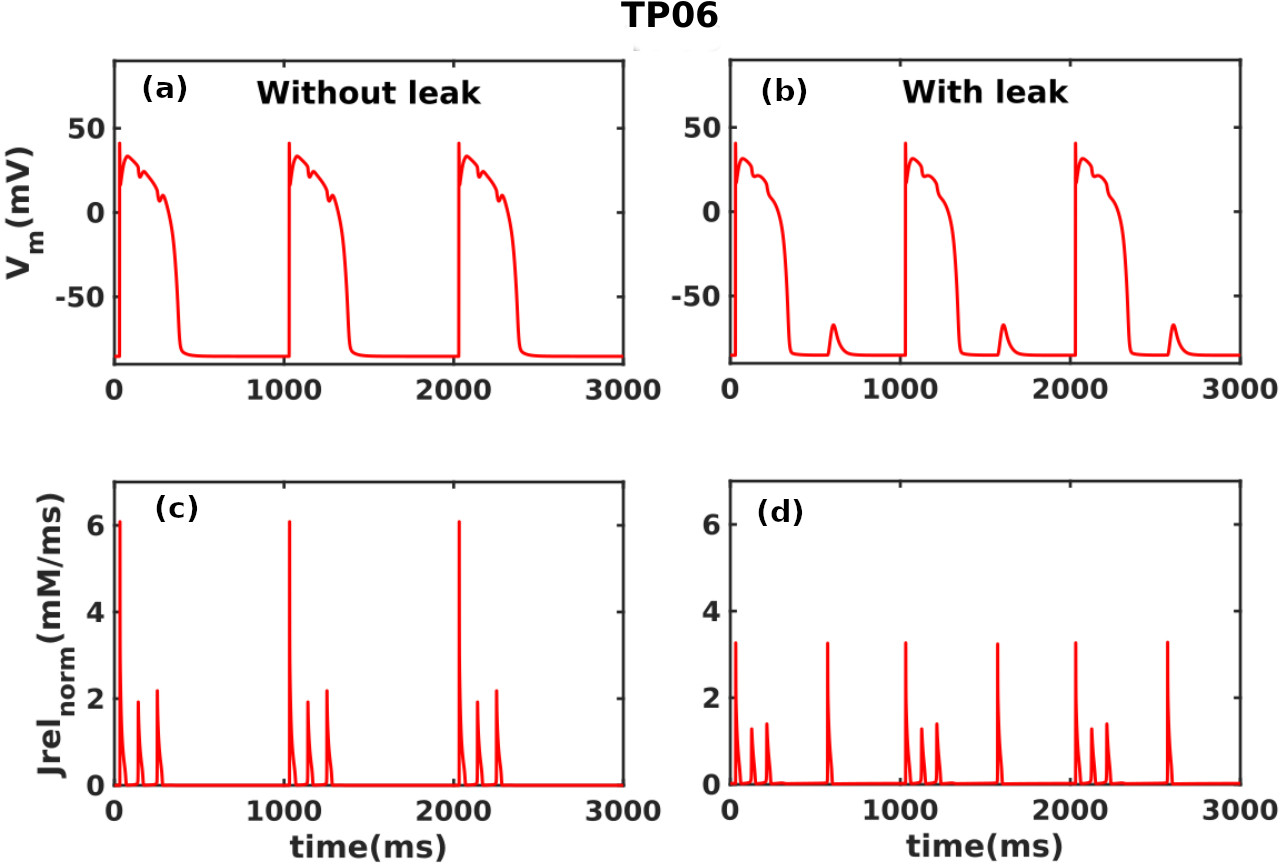}
      \caption{\textbf{The role of the RyR leak in triggering DADs in the TP06 model:} (a): No DADs are observed without the RyR leak; (b) after modifying the equations to introduce an RyR leak (see text) this model triggers DADs.}
      \label{fig:SI:leak_nonleak}
\end{figure}
 
\subsection{Distribution of NCX Channels}
\label{app:ncx_dist}
Cardiac myocytes have multiple compartments with different calcium concentrations; this is usually called calcium compartmentalization ~\cite{langer1994myocardial}. The distribution of ion channels in each compartment is different. The HuVEC15 model has 90\% of ion channels in the bulk cytosol region, and the remaining 10\% are in the intermediate zone (iz near RyRs in Fig. 1 in the main paper). However, the fraction of NCX ion channels near the RyRs may be up to 45\% Ref.~\cite{chu2016modeling}. Therefore, in the HUVEC15 model, we increase the fraction of NCX channels in the iz to 25\% from 10\%. This modification in the NCX distribution increases the calcium-to-voltage coupling gain~\cite{maruyama2010diastolic} of the myocyte and, thereby, increases the DAD amplitude for a given SCR.

\subsection{Subthreshold DADs and $I_{\rm{Na}}$ inactivation}
\label{app:na_avail}
We demonstrate the role of subthreshold and multi-blip DADs in the inactivation of the Na\textsuperscript{+} channel. The study of Ref.~\cite{liu2015delayed} has suggested that subthreshold DADs can inactivate the Na\textsuperscript{+} channel and can act as a substrate for promoting conduction block. In particular, we compare the effects of multi-blip and subthreshold DADs on the inactivation gates of the $I_{\rm{Na}}$. In Fig.~\ref{fig:SI:hj_gates}(a) we show that the subthreshold DADs can inactivate the $I_{\rm{Na}}$ gates, i.e., $h$ (fast-inactivation) and $j$ (slow-inactivation) gates in the TP06 model, as shown by the  plots of the product $h*j$ of these gates (Fig.~\ref{fig:SI:hj_gates}(c)). Similarly, multi-blip DADs (Fig.~\ref{fig:SI:hj_gates}(b)) inactivate fast $I_{\rm{Na}}$ on multiple occasions (see Fig.~\ref{fig:SI:hj_gates}(d)).

\begin{figure}
     \centering
      \includegraphics[width=\textwidth]{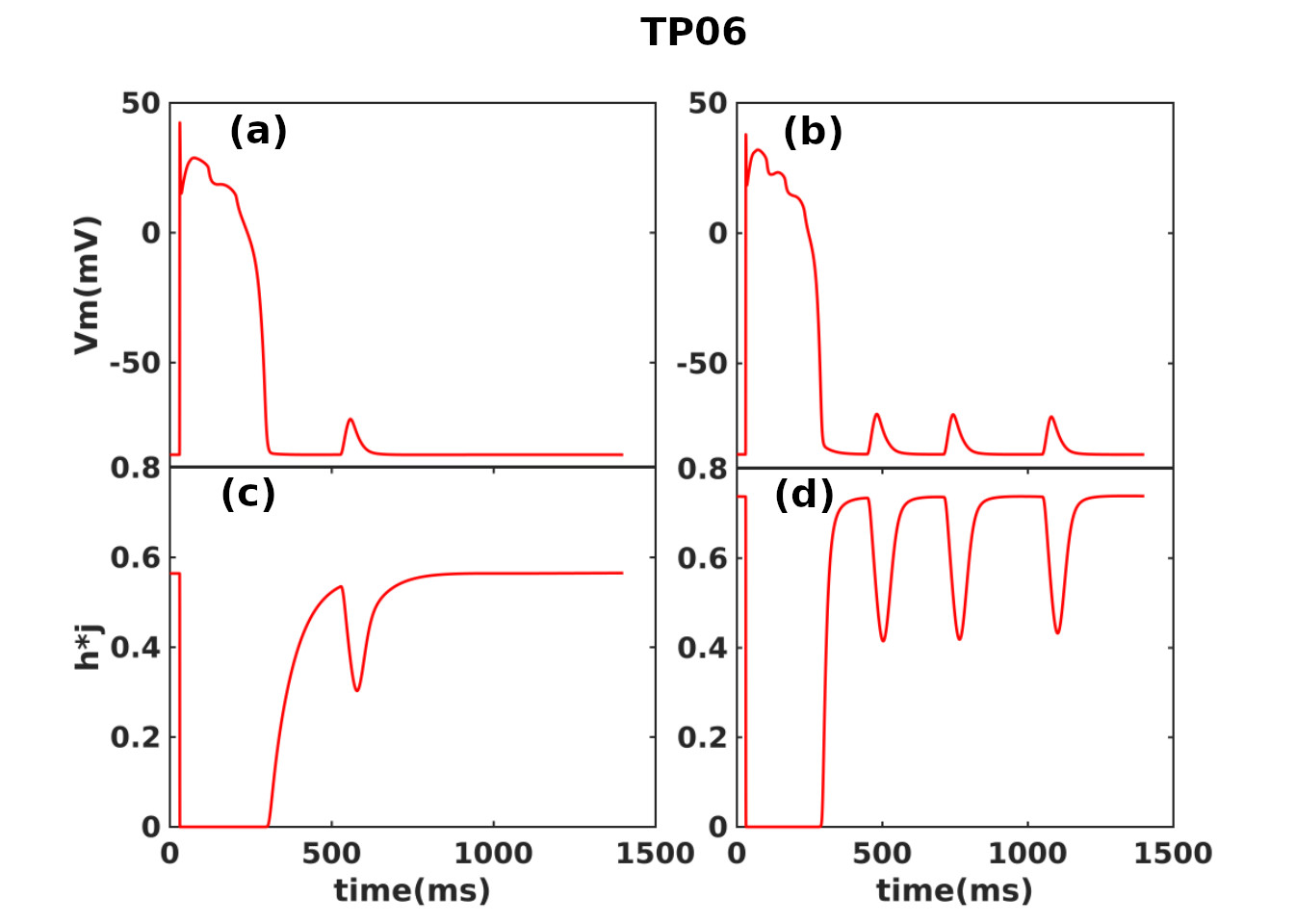}
      \caption{\textbf{Comparison of subthreshold and multi-blip DADs based on $I_{\rm{Na}}$ channel inactivation:}. The top panels show TP06-model membrane potentials for subthreshold DADs (left) and multi-blip DADs (right). The bottom panels show the product of $I_{\rm{Na}}$ inactivation gates [$h*j$ (see text)] for these DADs: (a) Subthreshold DADs; (b) multi-blip DADs; the product of inactivation gates ($h*j$) for (c) subthreshold DADs and (d) multi-blip DADs.}
      \label{fig:SI:hj_gates}  
\end{figure}

\subsection{Robustness of our parameter-sensitivity results}
\label{app:robust}
We demonstrate that our parameter-sensitivity analyses are robust for a range of pacing frequencies in the sense that they
yield results similar to those in the main paper, even if we use a different pacing frequency for stimulating the myocyte [see Figs.~\ref{fig:SI:huvec_sens} and \ref{fig:SI:tp06_sens}].

\begin{figure}
     \centering
      \includegraphics[width=0.9\textwidth]{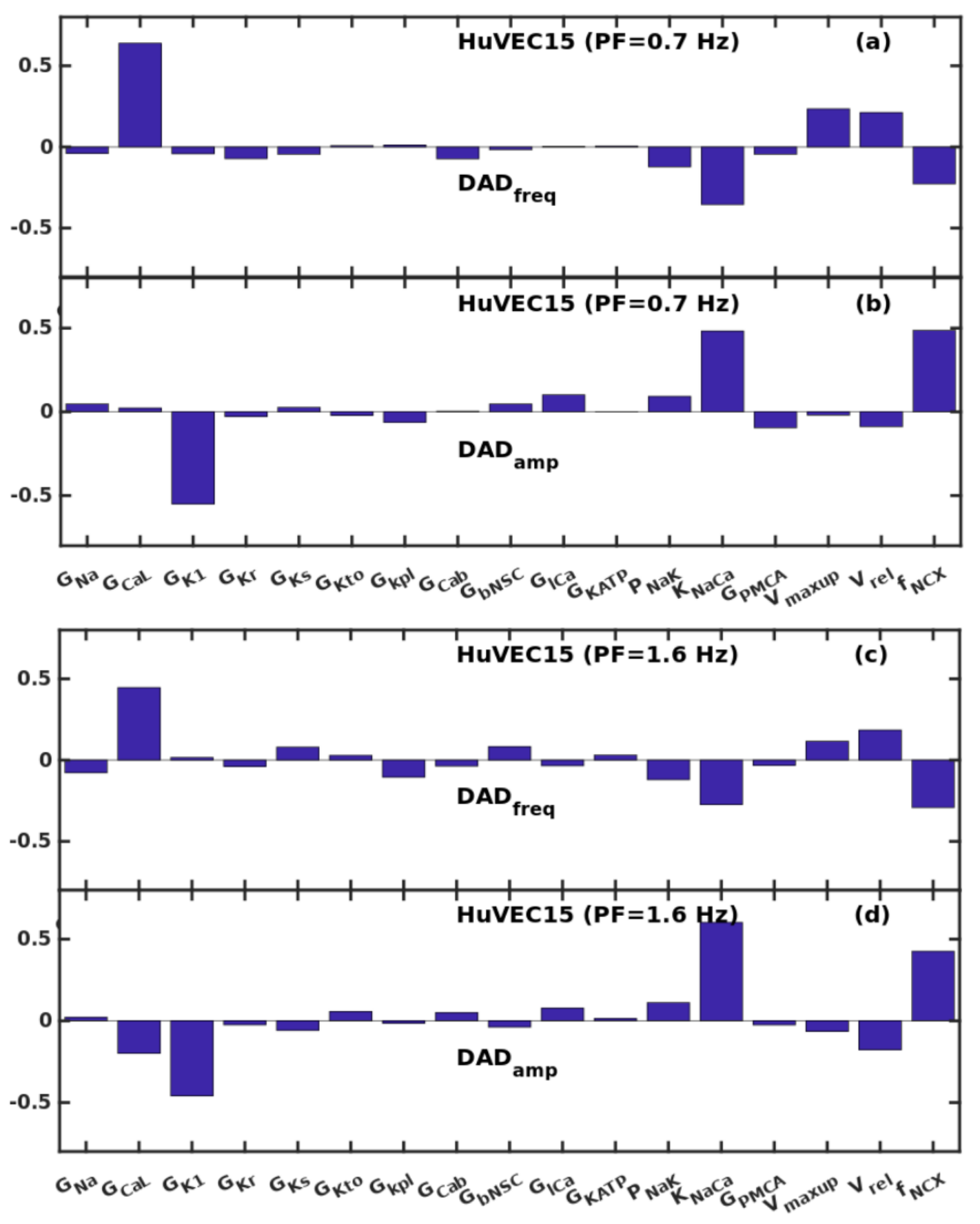}
      \caption{\textbf{Parameter sensitivity at 0.7 Hz and 1.6 Hz pacing frequencies (HuVEC15 model).} (a) DAD-frequency sensitivity plot at 0.7 Hz; (b) DAD-amplitude sensitivity plot at 0.7 Hz; (c) DAD-frequency sensitivity plot at 1.6 Hz; (b) DAD amplitude sensitivity plot at 1.6 Hz.}
      \label{fig:SI:huvec_sens}
\end{figure}

\begin{figure}
     \centering
      \includegraphics[width=\textwidth]{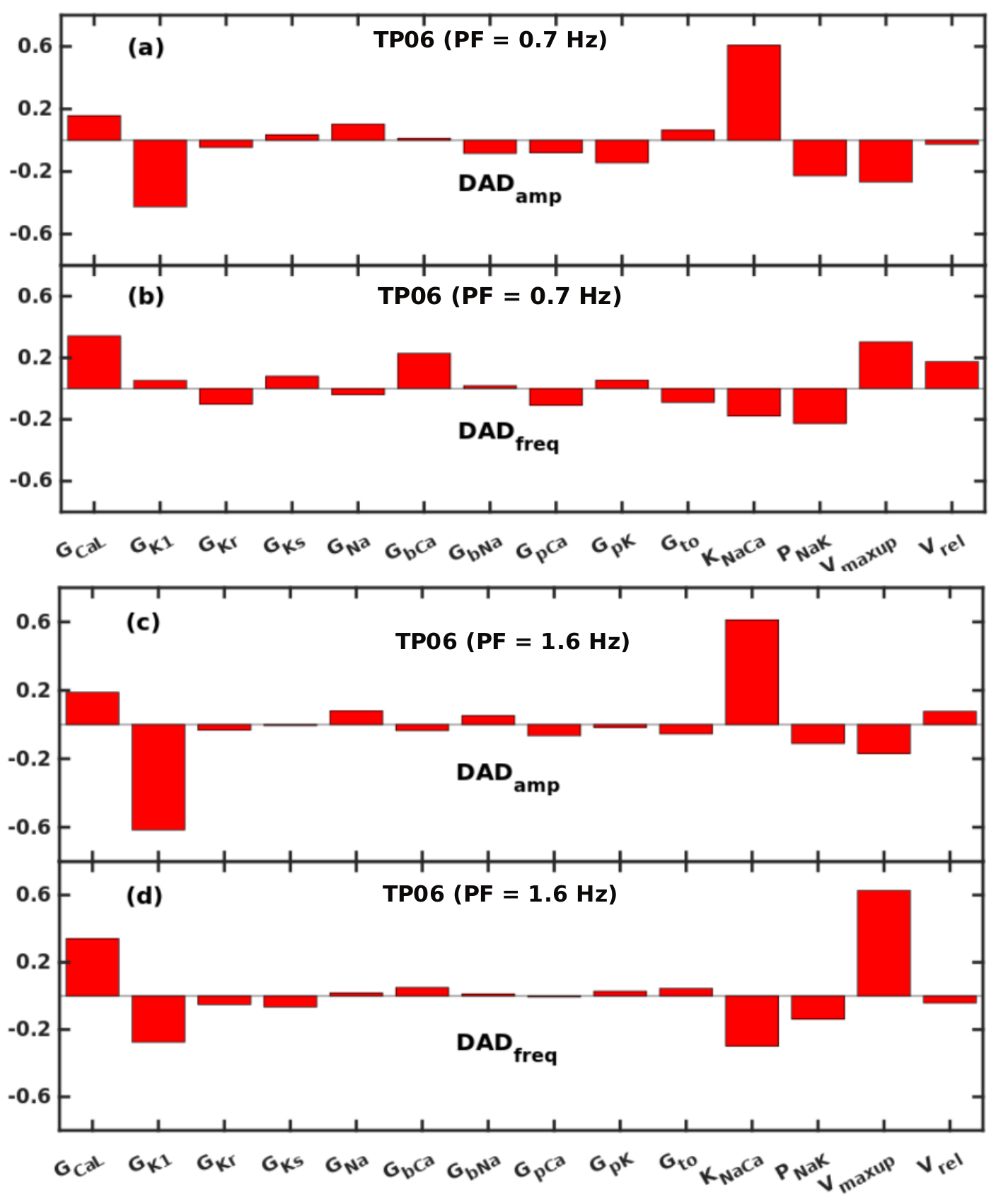}
      \caption{\textbf{Parameter sensitivity at 0.7 Hz and 1.6 Hz pacing frequencies (TP06 model).} 
      (a) DAD-frequency sensitivity plot at 0.7 Hz; (b) DAD-amplitude sensitivity plot at 0.7 Hz; (c) DAD-frequency sensitivity plot at 1.6 Hz; (b) DAD amplitude sensitivity plot at 1.6 Hz.}
      \label{fig:SI:tp06_sens}
\end{figure}

\subsection{Role of $I_{\rm{K1}}$ conductance in DAD amplitude}\label{app:Ik1}
In the main paper, we have discussed the three types of DADs that can occur in the TP06 and HuVEC15 models and the parameters that control the incidence frequencies and amplitudes of these DADs. We know that NCX increases $V_{\rm{m}}$ in response to CICR; the ratio of the rise in $V_{\rm{m}}$, in response to the SCR (or $Ca_{\rm{i}}$) amplitude, is known as calcium-voltage coupling gain~\cite{maruyama2010diastolic}. In the context of DADs, given the amplitudes of SCRs, the calcium-voltage coupling gain, during the diastolic interval, depends on two currents, namely, $I_{\rm{NaCa}}$ and $I_{\rm{K1}}$; the former 
competes against the latter. Therefore, the interplay of the parameters $S_{\rm{KNaCa}}$ and  $S_{\rm{GK1}}$ controls the maximum amplitude that a DAD reaches. A loss-of-function of the $I_{\rm{K1}}$ channel has been identified in some conditions such as Anderson's syndrome~\cite{verkerk2001ionic}; therefore, we use a reduced value of $G_{\rm{K1}}$ to simulate this effect on the DAD amplitude. 
In Fig.~\ref{fig:SI:Ik1_effect} we show that a reduction in $G_{\rm{K1}}$ can amplify subthreshold DADs so that they become suprathreshold DADs. We also know that $V_{\rm{maxup}}$ can reduce the coupling interval between DADs and the previous AP. Therefore, a reduction in $G_{\rm{K1}}$ can lead to a sustained incidence of triggered activity if this reduction is combined with multi-blip DADs as the coupling interval between DADs and the AP is very small in the case of multi-blip DADs. 

\begin{figure}
     \centering
      \includegraphics[width=\textwidth]{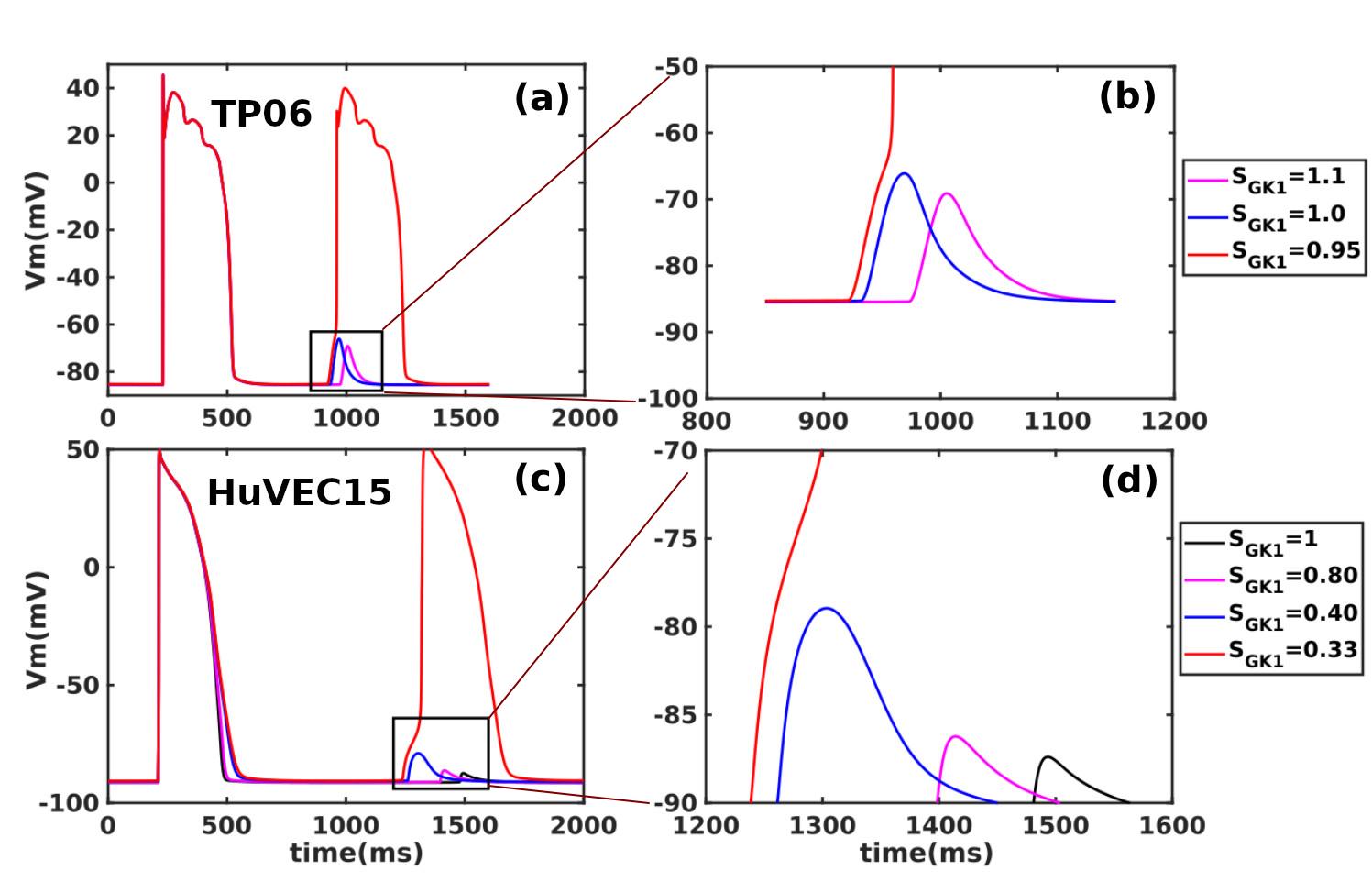}
      \caption{\textbf{Effect of $G_{\rm{K1}}$ on the DAD\textsubscript{amp} (see text):} (a) for the TP06 model; (b) expanded version of the inset in (a); (c) for the HuVEC15 model; (d) expanded version of the inset in (c). In both these models a reduction in $G_{\rm{K1}}$ converts subthreshold DADs to suprathreshold amplitude.}
      \label{fig:SI:Ik1_effect}
\end{figure}

\end{widetext}

\bibliography{main} BibTeX.

\begin{thebibliography}{73}%
\makeatletter
\providecommand \@ifxundefined [1]{%
 \@ifx{#1\undefined}
}%
\providecommand \@ifnum [1]{%
 \ifnum #1\expandafter \@firstoftwo
 \else \expandafter \@secondoftwo
 \fi
}%
\providecommand \@ifx [1]{%
 \ifx #1\expandafter \@firstoftwo
 \else \expandafter \@secondoftwo
 \fi
}%
\providecommand \natexlab [1]{#1}%
\providecommand \enquote  [1]{``#1''}%
\providecommand \bibnamefont  [1]{#1}%
\providecommand \bibfnamefont [1]{#1}%
\providecommand \citenamefont [1]{#1}%
\providecommand \href@noop [0]{\@secondoftwo}%
\providecommand \href [0]{\begingroup \@sanitize@url \@href}%
\providecommand \@href[1]{\@@startlink{#1}\@@href}%
\providecommand \@@href[1]{\endgroup#1\@@endlink}%
\providecommand \@sanitize@url [0]{\catcode `\\12\catcode `\$12\catcode
  `\&12\catcode `\#12\catcode `\^12\catcode `\_12\catcode `\%12\relax}%
\providecommand \@@startlink[1]{}%
\providecommand \@@endlink[0]{}%
\providecommand \url  [0]{\begingroup\@sanitize@url \@url }%
\providecommand \@url [1]{\endgroup\@href {#1}{\urlprefix }}%
\providecommand \urlprefix  [0]{URL }%
\providecommand \Eprint [0]{\href }%
\providecommand \doibase [0]{https://doi.org/}%
\providecommand \selectlanguage [0]{\@gobble}%
\providecommand \bibinfo  [0]{\@secondoftwo}%
\providecommand \bibfield  [0]{\@secondoftwo}%
\providecommand \translation [1]{[#1]}%
\providecommand \BibitemOpen [0]{}%
\providecommand \bibitemStop [0]{}%
\providecommand \bibitemNoStop [0]{.\EOS\space}%
\providecommand \EOS [0]{\spacefactor3000\relax}%
\providecommand \BibitemShut  [1]{\csname bibitem#1\endcsname}%
\let\auto@bib@innerbib\@empty
\bibitem [{\citenamefont {Ten~Tusscher}\ and\ \citenamefont
  {Panfilov}(2006)}]{ten2006alternans}%
  \BibitemOpen
  \bibfield  {author} {\bibinfo {author} {\bibfnamefont {K.~H.}\ \bibnamefont
  {Ten~Tusscher}}\ and\ \bibinfo {author} {\bibfnamefont {A.~V.}\ \bibnamefont
  {Panfilov}},\ }\bibfield  {title} {\bibinfo {title} {Alternans and spiral
  breakup in a human ventricular tissue model},\ }\href@noop {} {\bibfield
  {journal} {\bibinfo  {journal} {American Journal of Physiology-Heart and
  Circulatory Physiology}\ }\textbf {\bibinfo {volume} {291}},\ \bibinfo
  {pages} {H1088} (\bibinfo {year} {2006})}\BibitemShut {NoStop}%
\bibitem [{\citenamefont {Himeno}\ \emph {et~al.}(2015)\citenamefont {Himeno},
  \citenamefont {Asakura}, \citenamefont {Cha}, \citenamefont {Memida},
  \citenamefont {Powell}, \citenamefont {Amano},\ and\ \citenamefont
  {Noma}}]{himeno2015human}%
  \BibitemOpen
  \bibfield  {author} {\bibinfo {author} {\bibfnamefont {Y.}~\bibnamefont
  {Himeno}}, \bibinfo {author} {\bibfnamefont {K.}~\bibnamefont {Asakura}},
  \bibinfo {author} {\bibfnamefont {C.~Y.}\ \bibnamefont {Cha}}, \bibinfo
  {author} {\bibfnamefont {H.}~\bibnamefont {Memida}}, \bibinfo {author}
  {\bibfnamefont {T.}~\bibnamefont {Powell}}, \bibinfo {author} {\bibfnamefont
  {A.}~\bibnamefont {Amano}},\ and\ \bibinfo {author} {\bibfnamefont
  {A.}~\bibnamefont {Noma}},\ }\bibfield  {title} {\bibinfo {title} {A human
  ventricular myocyte model with a refined representation of
  excitation-contraction coupling},\ }\href@noop {} {\bibfield  {journal}
  {\bibinfo  {journal} {Biophysical journal}\ }\textbf {\bibinfo {volume}
  {109}},\ \bibinfo {pages} {415} (\bibinfo {year} {2015})}\BibitemShut
  {NoStop}%
\bibitem [{\citenamefont {Nowbar}\ \emph {et~al.}(2019)\citenamefont {Nowbar},
  \citenamefont {Gitto}, \citenamefont {Howard}, \citenamefont {Francis},\ and\
  \citenamefont {Al-Lamee}}]{nowbar2019mortality}%
  \BibitemOpen
  \bibfield  {author} {\bibinfo {author} {\bibfnamefont {A.~N.}\ \bibnamefont
  {Nowbar}}, \bibinfo {author} {\bibfnamefont {M.}~\bibnamefont {Gitto}},
  \bibinfo {author} {\bibfnamefont {J.~P.}\ \bibnamefont {Howard}}, \bibinfo
  {author} {\bibfnamefont {D.~P.}\ \bibnamefont {Francis}},\ and\ \bibinfo
  {author} {\bibfnamefont {R.}~\bibnamefont {Al-Lamee}},\ }\bibfield  {title}
  {\bibinfo {title} {Mortality from ischemic heart disease: Analysis of data
  from the world health organization and coronary artery disease risk factors
  from ncd risk factor collaboration},\ }\href@noop {} {\bibfield  {journal}
  {\bibinfo  {journal} {Circulation: cardiovascular quality and outcomes}\
  }\textbf {\bibinfo {volume} {12}},\ \bibinfo {pages} {e005375} (\bibinfo
  {year} {2019})}\BibitemShut {NoStop}%
\bibitem [{\citenamefont {Zimik}\ \emph {et~al.}(2015)\citenamefont {Zimik},
  \citenamefont {Vandersickel}, \citenamefont {Nayak}, \citenamefont
  {Panfilov},\ and\ \citenamefont {Pandit}}]{zimik2015comparative}%
  \BibitemOpen
  \bibfield  {author} {\bibinfo {author} {\bibfnamefont {S.}~\bibnamefont
  {Zimik}}, \bibinfo {author} {\bibfnamefont {N.}~\bibnamefont {Vandersickel}},
  \bibinfo {author} {\bibfnamefont {A.~R.}\ \bibnamefont {Nayak}}, \bibinfo
  {author} {\bibfnamefont {A.~V.}\ \bibnamefont {Panfilov}},\ and\ \bibinfo
  {author} {\bibfnamefont {R.}~\bibnamefont {Pandit}},\ }\bibfield  {title}
  {\bibinfo {title} {A comparative study of early afterdepolarization-mediated
  fibrillation in two mathematical models for human ventricular cells},\
  }\href@noop {} {\bibfield  {journal} {\bibinfo  {journal} {PloS one}\
  }\textbf {\bibinfo {volume} {10}},\ \bibinfo {pages} {e0130632} (\bibinfo
  {year} {2015})}\BibitemShut {NoStop}%
\bibitem [{\citenamefont {Volders}\ \emph {et~al.}(1997)\citenamefont
  {Volders}, \citenamefont {Kulcs{\'a}r}, \citenamefont {Vos}, \citenamefont
  {Sipido}, \citenamefont {Wellens}, \citenamefont {Lazzara},\ and\
  \citenamefont {Szabo}}]{volders1997similarities}%
  \BibitemOpen
  \bibfield  {author} {\bibinfo {author} {\bibfnamefont {P.~G.}\ \bibnamefont
  {Volders}}, \bibinfo {author} {\bibfnamefont {A.}~\bibnamefont
  {Kulcs{\'a}r}}, \bibinfo {author} {\bibfnamefont {M.~A.}\ \bibnamefont
  {Vos}}, \bibinfo {author} {\bibfnamefont {K.~R.}\ \bibnamefont {Sipido}},
  \bibinfo {author} {\bibfnamefont {H.~J.}\ \bibnamefont {Wellens}}, \bibinfo
  {author} {\bibfnamefont {R.}~\bibnamefont {Lazzara}},\ and\ \bibinfo {author}
  {\bibfnamefont {B.}~\bibnamefont {Szabo}},\ }\bibfield  {title} {\bibinfo
  {title} {Similarities between early and delayed afterdepolarizations induced
  by isoproterenol in canine ventricular myocytes},\ }\href@noop {} {\bibfield
  {journal} {\bibinfo  {journal} {Cardiovascular research}\ }\textbf {\bibinfo
  {volume} {34}},\ \bibinfo {pages} {348} (\bibinfo {year} {1997})}\BibitemShut
  {NoStop}%
\bibitem [{\citenamefont {Vandersickel}\ \emph {et~al.}(2014)\citenamefont
  {Vandersickel}, \citenamefont {Kazbanov}, \citenamefont {Nuitermans},
  \citenamefont {Weise}, \citenamefont {Pandit},\ and\ \citenamefont
  {Panfilov}}]{vandersickel2014study}%
  \BibitemOpen
  \bibfield  {author} {\bibinfo {author} {\bibfnamefont {N.}~\bibnamefont
  {Vandersickel}}, \bibinfo {author} {\bibfnamefont {I.~V.}\ \bibnamefont
  {Kazbanov}}, \bibinfo {author} {\bibfnamefont {A.}~\bibnamefont
  {Nuitermans}}, \bibinfo {author} {\bibfnamefont {L.~D.}\ \bibnamefont
  {Weise}}, \bibinfo {author} {\bibfnamefont {R.}~\bibnamefont {Pandit}},\ and\
  \bibinfo {author} {\bibfnamefont {A.~V.}\ \bibnamefont {Panfilov}},\
  }\bibfield  {title} {\bibinfo {title} {A study of early afterdepolarizations
  in a model for human ventricular tissue},\ }\href@noop {} {\bibfield
  {journal} {\bibinfo  {journal} {PloS one}\ }\textbf {\bibinfo {volume} {9}},\
  \bibinfo {pages} {e84595} (\bibinfo {year} {2014})}\BibitemShut {NoStop}%
\bibitem [{\citenamefont {Verkerk}\ \emph {et~al.}(2001)\citenamefont
  {Verkerk}, \citenamefont {Veldkamp}, \citenamefont {Baartscheer},
  \citenamefont {Schumacher}, \citenamefont {Kl{\"o}pping}, \citenamefont {van
  Ginneken},\ and\ \citenamefont {Ravesloot}}]{verkerk2001ionic}%
  \BibitemOpen
  \bibfield  {author} {\bibinfo {author} {\bibfnamefont {A.~O.}\ \bibnamefont
  {Verkerk}}, \bibinfo {author} {\bibfnamefont {M.~W.}\ \bibnamefont
  {Veldkamp}}, \bibinfo {author} {\bibfnamefont {A.}~\bibnamefont
  {Baartscheer}}, \bibinfo {author} {\bibfnamefont {C.~A.}\ \bibnamefont
  {Schumacher}}, \bibinfo {author} {\bibfnamefont {C.}~\bibnamefont
  {Kl{\"o}pping}}, \bibinfo {author} {\bibfnamefont {A.~C.}\ \bibnamefont {van
  Ginneken}},\ and\ \bibinfo {author} {\bibfnamefont {J.~H.}\ \bibnamefont
  {Ravesloot}},\ }\bibfield  {title} {\bibinfo {title} {Ionic mechanism of
  delayed afterdepolarizations in ventricular cells isolated from human
  end-stage failing hearts},\ }\href@noop {} {\bibfield  {journal} {\bibinfo
  {journal} {Circulation}\ }\textbf {\bibinfo {volume} {104}},\ \bibinfo
  {pages} {2728} (\bibinfo {year} {2001})}\BibitemShut {NoStop}%
\bibitem [{\citenamefont {Kass}\ and\ \citenamefont
  {Tsien}(1982)}]{kass1982fluctuations}%
  \BibitemOpen
  \bibfield  {author} {\bibinfo {author} {\bibfnamefont {R.~S.}\ \bibnamefont
  {Kass}}\ and\ \bibinfo {author} {\bibfnamefont {R.~W.}\ \bibnamefont
  {Tsien}},\ }\bibfield  {title} {\bibinfo {title} {Fluctuations in membrane
  current driven by intracellular calcium in cardiac purkinje fibers},\
  }\href@noop {} {\bibfield  {journal} {\bibinfo  {journal} {Biophysical
  journal}\ }\textbf {\bibinfo {volume} {38}},\ \bibinfo {pages} {259}
  (\bibinfo {year} {1982})}\BibitemShut {NoStop}%
\bibitem [{\citenamefont {Marban}\ \emph {et~al.}(1986)\citenamefont {Marban},
  \citenamefont {Robinson}, \citenamefont {Wier} \emph
  {et~al.}}]{marban1986mechanisms}%
  \BibitemOpen
  \bibfield  {author} {\bibinfo {author} {\bibfnamefont {E.}~\bibnamefont
  {Marban}}, \bibinfo {author} {\bibfnamefont {S.~W.}\ \bibnamefont
  {Robinson}}, \bibinfo {author} {\bibfnamefont {W.~G.}\ \bibnamefont {Wier}},
  \emph {et~al.},\ }\bibfield  {title} {\bibinfo {title} {Mechanisms of
  arrhythmogenic delayed and early afterdepolarizations in ferret ventricular
  muscle.},\ }\href@noop {} {\bibfield  {journal} {\bibinfo  {journal} {The
  Journal of clinical investigation}\ }\textbf {\bibinfo {volume} {78}},\
  \bibinfo {pages} {1185} (\bibinfo {year} {1986})}\BibitemShut {NoStop}%
\bibitem [{\citenamefont {Rizzi}\ \emph {et~al.}(2008)\citenamefont {Rizzi},
  \citenamefont {Liu}, \citenamefont {Napolitano}, \citenamefont {Nori},
  \citenamefont {Turcato}, \citenamefont {Colombi}, \citenamefont {Bicciato},
  \citenamefont {Arcelli}, \citenamefont {Spedito}, \citenamefont {Scelsi}
  \emph {et~al.}}]{rizzi2008unexpected}%
  \BibitemOpen
  \bibfield  {author} {\bibinfo {author} {\bibfnamefont {N.}~\bibnamefont
  {Rizzi}}, \bibinfo {author} {\bibfnamefont {N.}~\bibnamefont {Liu}}, \bibinfo
  {author} {\bibfnamefont {C.}~\bibnamefont {Napolitano}}, \bibinfo {author}
  {\bibfnamefont {A.}~\bibnamefont {Nori}}, \bibinfo {author} {\bibfnamefont
  {F.}~\bibnamefont {Turcato}}, \bibinfo {author} {\bibfnamefont
  {B.}~\bibnamefont {Colombi}}, \bibinfo {author} {\bibfnamefont
  {S.}~\bibnamefont {Bicciato}}, \bibinfo {author} {\bibfnamefont
  {D.}~\bibnamefont {Arcelli}}, \bibinfo {author} {\bibfnamefont
  {A.}~\bibnamefont {Spedito}}, \bibinfo {author} {\bibfnamefont
  {M.}~\bibnamefont {Scelsi}}, \emph {et~al.},\ }\bibfield  {title} {\bibinfo
  {title} {Unexpected structural and functional consequences of the r33q
  homozygous mutation in cardiac calsequestrin: a complex arrhythmogenic
  cascade in a knock in mouse model},\ }\href@noop {} {\bibfield  {journal}
  {\bibinfo  {journal} {Circulation research}\ }\textbf {\bibinfo {volume}
  {103}},\ \bibinfo {pages} {298} (\bibinfo {year} {2008})}\BibitemShut
  {NoStop}%
\bibitem [{\citenamefont {Song}\ \emph {et~al.}(2008)\citenamefont {Song},
  \citenamefont {Shryock},\ and\ \citenamefont
  {Belardinelli}}]{song2008increase}%
  \BibitemOpen
  \bibfield  {author} {\bibinfo {author} {\bibfnamefont {Y.}~\bibnamefont
  {Song}}, \bibinfo {author} {\bibfnamefont {J.~C.}\ \bibnamefont {Shryock}},\
  and\ \bibinfo {author} {\bibfnamefont {L.}~\bibnamefont {Belardinelli}},\
  }\bibfield  {title} {\bibinfo {title} {An increase of late sodium current
  induces delayed afterdepolarizations and sustained triggered activity in
  atrial myocytes},\ }\href@noop {} {\bibfield  {journal} {\bibinfo  {journal}
  {American Journal of Physiology-Heart and Circulatory Physiology}\ }\textbf
  {\bibinfo {volume} {294}},\ \bibinfo {pages} {H2031} (\bibinfo {year}
  {2008})}\BibitemShut {NoStop}%
\bibitem [{\citenamefont {Stambler}\ \emph {et~al.}(2003)\citenamefont
  {Stambler}, \citenamefont {Fenelon}, \citenamefont {Shepard}, \citenamefont
  {Clemo},\ and\ \citenamefont {Guiraudon}}]{stambler2003characterization}%
  \BibitemOpen
  \bibfield  {author} {\bibinfo {author} {\bibfnamefont {B.~S.}\ \bibnamefont
  {Stambler}}, \bibinfo {author} {\bibfnamefont {G.}~\bibnamefont {Fenelon}},
  \bibinfo {author} {\bibfnamefont {R.~K.}\ \bibnamefont {Shepard}}, \bibinfo
  {author} {\bibfnamefont {H.~F.}\ \bibnamefont {Clemo}},\ and\ \bibinfo
  {author} {\bibfnamefont {C.~M.}\ \bibnamefont {Guiraudon}},\ }\bibfield
  {title} {\bibinfo {title} {Characterization of sustained atrial tachycardia
  in dogs with rapid ventricular pacing-induced heart failure},\ }\href@noop {}
  {\bibfield  {journal} {\bibinfo  {journal} {Journal of cardiovascular
  electrophysiology}\ }\textbf {\bibinfo {volume} {14}},\ \bibinfo {pages}
  {499} (\bibinfo {year} {2003})}\BibitemShut {NoStop}%
\bibitem [{\citenamefont {Wongcharoen}\ \emph {et~al.}(2007)\citenamefont
  {Wongcharoen}, \citenamefont {CHEN}, \citenamefont {CHEN}, \citenamefont
  {LIN},\ and\ \citenamefont {CHEN}}]{wongcharoen2007effects}%
  \BibitemOpen
  \bibfield  {author} {\bibinfo {author} {\bibfnamefont {W.}~\bibnamefont
  {Wongcharoen}}, \bibinfo {author} {\bibfnamefont {Y.-C.}\ \bibnamefont
  {CHEN}}, \bibinfo {author} {\bibfnamefont {Y.-J.}\ \bibnamefont {CHEN}},
  \bibinfo {author} {\bibfnamefont {C.-I.}\ \bibnamefont {LIN}},\ and\ \bibinfo
  {author} {\bibfnamefont {S.-A.}\ \bibnamefont {CHEN}},\ }\bibfield  {title}
  {\bibinfo {title} {Effects of aging and ouabain on left atrial
  arrhythmogenicity},\ }\href@noop {} {\bibfield  {journal} {\bibinfo
  {journal} {Journal of cardiovascular electrophysiology}\ }\textbf {\bibinfo
  {volume} {18}},\ \bibinfo {pages} {526} (\bibinfo {year} {2007})}\BibitemShut
  {NoStop}%
\bibitem [{\citenamefont {Xie}\ \emph {et~al.}(2006)\citenamefont {Xie},
  \citenamefont {Walker},\ and\ \citenamefont {Wang}}]{xie2006dioxin}%
  \BibitemOpen
  \bibfield  {author} {\bibinfo {author} {\bibfnamefont {A.}~\bibnamefont
  {Xie}}, \bibinfo {author} {\bibfnamefont {N.~J.}\ \bibnamefont {Walker}},\
  and\ \bibinfo {author} {\bibfnamefont {D.}~\bibnamefont {Wang}},\ }\bibfield
  {title} {\bibinfo {title} {Dioxin (2, 3, 7, 8-tetrachlorodibenzo-p-doxin)
  enhances triggered afterdepolarizations in rat ventricular myocytes},\
  }\href@noop {} {\bibfield  {journal} {\bibinfo  {journal} {Cardiovascular
  toxicology}\ }\textbf {\bibinfo {volume} {6}},\ \bibinfo {pages} {99}
  (\bibinfo {year} {2006})}\BibitemShut {NoStop}%
\bibitem [{\citenamefont {Pogwizd}\ and\ \citenamefont
  {Bers}(2004)}]{pogwizd2004cellular}%
  \BibitemOpen
  \bibfield  {author} {\bibinfo {author} {\bibfnamefont {S.~M.}\ \bibnamefont
  {Pogwizd}}\ and\ \bibinfo {author} {\bibfnamefont {D.~M.}\ \bibnamefont
  {Bers}},\ }\bibfield  {title} {\bibinfo {title} {Cellular basis of triggered
  arrhythmias in heart failure},\ }\href@noop {} {\bibfield  {journal}
  {\bibinfo  {journal} {Trends in cardiovascular medicine}\ }\textbf {\bibinfo
  {volume} {14}},\ \bibinfo {pages} {61} (\bibinfo {year} {2004})}\BibitemShut
  {NoStop}%
\bibitem [{\citenamefont {Leenhardt}\ \emph {et~al.}(2012)\citenamefont
  {Leenhardt}, \citenamefont {Denjoy},\ and\ \citenamefont
  {Guicheney}}]{leenhardt2012catecholaminergic}%
  \BibitemOpen
  \bibfield  {author} {\bibinfo {author} {\bibfnamefont {A.}~\bibnamefont
  {Leenhardt}}, \bibinfo {author} {\bibfnamefont {I.}~\bibnamefont {Denjoy}},\
  and\ \bibinfo {author} {\bibfnamefont {P.}~\bibnamefont {Guicheney}},\
  }\bibfield  {title} {\bibinfo {title} {Catecholaminergic polymorphic
  ventricular tachycardia},\ }\href@noop {} {\bibfield  {journal} {\bibinfo
  {journal} {Circulation: Arrhythmia and Electrophysiology}\ }\textbf {\bibinfo
  {volume} {5}},\ \bibinfo {pages} {1044} (\bibinfo {year} {2012})}\BibitemShut
  {NoStop}%
\bibitem [{\citenamefont {Lazzara}\ \emph {et~al.}(1973)\citenamefont
  {Lazzara}, \citenamefont {El-Sherif},\ and\ \citenamefont
  {Scherlag}}]{lazzara1973electrophysiological}%
  \BibitemOpen
  \bibfield  {author} {\bibinfo {author} {\bibfnamefont {R.}~\bibnamefont
  {Lazzara}}, \bibinfo {author} {\bibfnamefont {N.}~\bibnamefont {El-Sherif}},\
  and\ \bibinfo {author} {\bibfnamefont {B.~J.}\ \bibnamefont {Scherlag}},\
  }\bibfield  {title} {\bibinfo {title} {Electrophysiological properties of
  canine purkinje cells in one-day-old myocardial infarction},\ }\href@noop {}
  {\bibfield  {journal} {\bibinfo  {journal} {Circulation research}\ }\textbf
  {\bibinfo {volume} {33}},\ \bibinfo {pages} {722} (\bibinfo {year}
  {1973})}\BibitemShut {NoStop}%
\bibitem [{\citenamefont {Orchard}\ and\ \citenamefont
  {Cingolani}(1994)}]{orchard1994acidosis}%
  \BibitemOpen
  \bibfield  {author} {\bibinfo {author} {\bibfnamefont {C.~H.}\ \bibnamefont
  {Orchard}}\ and\ \bibinfo {author} {\bibfnamefont {H.~E.}\ \bibnamefont
  {Cingolani}},\ }\bibfield  {title} {\bibinfo {title} {Acidosis and
  arrhythmias in cardiac muscle.},\ }\href@noop {} {\bibfield  {journal}
  {\bibinfo  {journal} {Cardiovascular research}\ }\textbf {\bibinfo {volume}
  {28}},\ \bibinfo {pages} {1312} (\bibinfo {year} {1994})}\BibitemShut
  {NoStop}%
\bibitem [{\citenamefont {Lascano}\ \emph {et~al.}(2013)\citenamefont
  {Lascano}, \citenamefont {Said}, \citenamefont {Vittone}, \citenamefont
  {Mattiazzi}, \citenamefont {Mundi{\~n}a-Weilenmann},\ and\ \citenamefont
  {Negroni}}]{lascano2013role}%
  \BibitemOpen
  \bibfield  {author} {\bibinfo {author} {\bibfnamefont {E.~C.}\ \bibnamefont
  {Lascano}}, \bibinfo {author} {\bibfnamefont {M.}~\bibnamefont {Said}},
  \bibinfo {author} {\bibfnamefont {L.}~\bibnamefont {Vittone}}, \bibinfo
  {author} {\bibfnamefont {A.}~\bibnamefont {Mattiazzi}}, \bibinfo {author}
  {\bibfnamefont {C.}~\bibnamefont {Mundi{\~n}a-Weilenmann}},\ and\ \bibinfo
  {author} {\bibfnamefont {J.~A.}\ \bibnamefont {Negroni}},\ }\bibfield
  {title} {\bibinfo {title} {Role of camkii in post acidosis arrhythmias: a
  simulation study using a human myocyte model},\ }\href@noop {} {\bibfield
  {journal} {\bibinfo  {journal} {Journal of molecular and cellular
  cardiology}\ }\textbf {\bibinfo {volume} {60}},\ \bibinfo {pages} {172}
  (\bibinfo {year} {2013})}\BibitemShut {NoStop}%
\bibitem [{\citenamefont {Vermeulen}\ \emph {et~al.}(1994)\citenamefont
  {Vermeulen}, \citenamefont {Mcguire}, \citenamefont {Opthof}, \citenamefont
  {Coronel}, \citenamefont {De~Bakker}, \citenamefont {Kl{\"o}pping},\ and\
  \citenamefont {Janse}}]{vermeulen1994triggered}%
  \BibitemOpen
  \bibfield  {author} {\bibinfo {author} {\bibfnamefont {J.~T.}\ \bibnamefont
  {Vermeulen}}, \bibinfo {author} {\bibfnamefont {M.~A.}\ \bibnamefont
  {Mcguire}}, \bibinfo {author} {\bibfnamefont {T.}~\bibnamefont {Opthof}},
  \bibinfo {author} {\bibfnamefont {R.}~\bibnamefont {Coronel}}, \bibinfo
  {author} {\bibfnamefont {J.~M.}\ \bibnamefont {De~Bakker}}, \bibinfo {author}
  {\bibfnamefont {C.}~\bibnamefont {Kl{\"o}pping}},\ and\ \bibinfo {author}
  {\bibfnamefont {M.~J.}\ \bibnamefont {Janse}},\ }\bibfield  {title} {\bibinfo
  {title} {Triggered activity and automaticity in ventricular trabeculae of
  failing human and rabbit hearts},\ }\href@noop {} {\bibfield  {journal}
  {\bibinfo  {journal} {Cardiovascular research}\ }\textbf {\bibinfo {volume}
  {28}},\ \bibinfo {pages} {1547} (\bibinfo {year} {1994})}\BibitemShut
  {NoStop}%
\bibitem [{\citenamefont {Ferrier}\ \emph {et~al.}(1973)\citenamefont
  {Ferrier}, \citenamefont {SOUNDERS},\ and\ \citenamefont
  {Mendez}}]{ferrier1973cellular}%
  \BibitemOpen
  \bibfield  {author} {\bibinfo {author} {\bibfnamefont {G.~R.}\ \bibnamefont
  {Ferrier}}, \bibinfo {author} {\bibfnamefont {J.~H.}\ \bibnamefont
  {SOUNDERS}},\ and\ \bibinfo {author} {\bibfnamefont {C.}~\bibnamefont
  {Mendez}},\ }\bibfield  {title} {\bibinfo {title} {A cellular mechanism for
  the generation of ventricular arrhythmias by acetylstrophanthidin},\
  }\href@noop {} {\bibfield  {journal} {\bibinfo  {journal} {Circulation
  research}\ }\textbf {\bibinfo {volume} {32}},\ \bibinfo {pages} {600}
  (\bibinfo {year} {1973})}\BibitemShut {NoStop}%
\bibitem [{\citenamefont {Rosen}\ \emph {et~al.}(1973)\citenamefont {Rosen},
  \citenamefont {Gelband}, \citenamefont {Merker},\ and\ \citenamefont
  {Hoffman}}]{rosen1973mechanisms}%
  \BibitemOpen
  \bibfield  {author} {\bibinfo {author} {\bibfnamefont {M.~R.}\ \bibnamefont
  {Rosen}}, \bibinfo {author} {\bibfnamefont {H.}~\bibnamefont {Gelband}},
  \bibinfo {author} {\bibfnamefont {C.}~\bibnamefont {Merker}},\ and\ \bibinfo
  {author} {\bibfnamefont {B.~F.}\ \bibnamefont {Hoffman}},\ }\bibfield
  {title} {\bibinfo {title} {Mechanisms of digitalis toxicity: effects of
  ouabain on phase four of canine purkinje fiber transmembrane potentials},\
  }\href@noop {} {\bibfield  {journal} {\bibinfo  {journal} {Circulation}\
  }\textbf {\bibinfo {volume} {47}},\ \bibinfo {pages} {681} (\bibinfo {year}
  {1973})}\BibitemShut {NoStop}%
\bibitem [{\citenamefont {Priori}\ and\ \citenamefont
  {Corr}(1990)}]{priori1990mechanisms}%
  \BibitemOpen
  \bibfield  {author} {\bibinfo {author} {\bibfnamefont {S.~G.}\ \bibnamefont
  {Priori}}\ and\ \bibinfo {author} {\bibfnamefont {P.~B.}\ \bibnamefont
  {Corr}},\ }\bibfield  {title} {\bibinfo {title} {Mechanisms underlying early
  and delayed afterdepolarizations induced by catecholamines},\ }\href@noop {}
  {\bibfield  {journal} {\bibinfo  {journal} {American Journal of
  Physiology-Heart and Circulatory Physiology}\ }\textbf {\bibinfo {volume}
  {258}},\ \bibinfo {pages} {H1796} (\bibinfo {year} {1990})}\BibitemShut
  {NoStop}%
\bibitem [{\citenamefont {Vassalle}\ and\ \citenamefont
  {Lin}(2004)}]{vassalle2004calcium}%
  \BibitemOpen
  \bibfield  {author} {\bibinfo {author} {\bibfnamefont {M.}~\bibnamefont
  {Vassalle}}\ and\ \bibinfo {author} {\bibfnamefont {C.-I.}\ \bibnamefont
  {Lin}},\ }\bibfield  {title} {\bibinfo {title} {Calcium overload and cardiac
  function},\ }\href@noop {} {\bibfield  {journal} {\bibinfo  {journal}
  {Journal of biomedical science}\ }\textbf {\bibinfo {volume} {11}},\ \bibinfo
  {pages} {542} (\bibinfo {year} {2004})}\BibitemShut {NoStop}%
\bibitem [{\citenamefont {Killeen}\ \emph {et~al.}(2007)\citenamefont
  {Killeen}, \citenamefont {Thomas}, \citenamefont {Gurung}, \citenamefont
  {Goddard}, \citenamefont {Fraser}, \citenamefont {Mahaut-Smith},
  \citenamefont {Colledge}, \citenamefont {Grace},\ and\ \citenamefont
  {Huang}}]{killeen2007arrhythmogenic}%
  \BibitemOpen
  \bibfield  {author} {\bibinfo {author} {\bibfnamefont {M.~J.}\ \bibnamefont
  {Killeen}}, \bibinfo {author} {\bibfnamefont {G.}~\bibnamefont {Thomas}},
  \bibinfo {author} {\bibfnamefont {I.}~\bibnamefont {Gurung}}, \bibinfo
  {author} {\bibfnamefont {C.}~\bibnamefont {Goddard}}, \bibinfo {author}
  {\bibfnamefont {J.}~\bibnamefont {Fraser}}, \bibinfo {author} {\bibfnamefont
  {M.}~\bibnamefont {Mahaut-Smith}}, \bibinfo {author} {\bibfnamefont
  {W.}~\bibnamefont {Colledge}}, \bibinfo {author} {\bibfnamefont
  {A.}~\bibnamefont {Grace}},\ and\ \bibinfo {author} {\bibfnamefont {C.-H.}\
  \bibnamefont {Huang}},\ }\bibfield  {title} {\bibinfo {title} {Arrhythmogenic
  mechanisms in the isolated perfused hypokalaemic murine heart},\ }\href@noop
  {} {\bibfield  {journal} {\bibinfo  {journal} {Acta Physiologica}\ }\textbf
  {\bibinfo {volume} {189}},\ \bibinfo {pages} {33} (\bibinfo {year}
  {2007})}\BibitemShut {NoStop}%
\bibitem [{\citenamefont {Wu}\ and\ \citenamefont
  {Corr}(1995)}]{wu1995palmitoylcarnitine}%
  \BibitemOpen
  \bibfield  {author} {\bibinfo {author} {\bibfnamefont {J.}~\bibnamefont
  {Wu}}\ and\ \bibinfo {author} {\bibfnamefont {P.~B.}\ \bibnamefont {Corr}},\
  }\bibfield  {title} {\bibinfo {title} {Palmitoylcarnitine increases [na+] i
  and initiates transient inward current in adult ventricular myocytes},\
  }\href@noop {} {\bibfield  {journal} {\bibinfo  {journal} {American Journal
  of Physiology-Heart and Circulatory Physiology}\ }\textbf {\bibinfo {volume}
  {268}},\ \bibinfo {pages} {H2405} (\bibinfo {year} {1995})}\BibitemShut
  {NoStop}%
\bibitem [{\citenamefont {Verkerk}\ \emph {et~al.}(2000)\citenamefont
  {Verkerk}, \citenamefont {Veldkamp}, \citenamefont {de~Jonge}, \citenamefont
  {Wilders},\ and\ \citenamefont {van Ginneken}}]{verkerk2000injury}%
  \BibitemOpen
  \bibfield  {author} {\bibinfo {author} {\bibfnamefont {A.~O.}\ \bibnamefont
  {Verkerk}}, \bibinfo {author} {\bibfnamefont {M.~W.}\ \bibnamefont
  {Veldkamp}}, \bibinfo {author} {\bibfnamefont {N.}~\bibnamefont {de~Jonge}},
  \bibinfo {author} {\bibfnamefont {R.}~\bibnamefont {Wilders}},\ and\ \bibinfo
  {author} {\bibfnamefont {A.~C.}\ \bibnamefont {van Ginneken}},\ }\bibfield
  {title} {\bibinfo {title} {Injury current modulates afterdepolarizations in
  single human ventricular cells},\ }\href@noop {} {\bibfield  {journal}
  {\bibinfo  {journal} {Cardiovascular research}\ }\textbf {\bibinfo {volume}
  {47}},\ \bibinfo {pages} {124} (\bibinfo {year} {2000})}\BibitemShut
  {NoStop}%
\bibitem [{\citenamefont {Wleklinski}\ \emph {et~al.}(2020)\citenamefont
  {Wleklinski}, \citenamefont {Kannankeril},\ and\ \citenamefont
  {Knollmann}}]{wleklinski2020molecular}%
  \BibitemOpen
  \bibfield  {author} {\bibinfo {author} {\bibfnamefont {M.~J.}\ \bibnamefont
  {Wleklinski}}, \bibinfo {author} {\bibfnamefont {P.~J.}\ \bibnamefont
  {Kannankeril}},\ and\ \bibinfo {author} {\bibfnamefont {B.~C.}\ \bibnamefont
  {Knollmann}},\ }\bibfield  {title} {\bibinfo {title} {Molecular and tissue
  mechanisms of catecholaminergic polymorphic ventricular tachycardia},\
  }\href@noop {} {\bibfield  {journal} {\bibinfo  {journal} {The Journal of
  physiology}\ }\textbf {\bibinfo {volume} {598}},\ \bibinfo {pages} {2817}
  (\bibinfo {year} {2020})}\BibitemShut {NoStop}%
\bibitem [{\citenamefont {Fabiato}(1983)}]{fabiato1983calcium}%
  \BibitemOpen
  \bibfield  {author} {\bibinfo {author} {\bibfnamefont {A.}~\bibnamefont
  {Fabiato}},\ }\bibfield  {title} {\bibinfo {title} {Calcium-induced release
  of calcium from the cardiac sarcoplasmic reticulum},\ }\href@noop {}
  {\bibfield  {journal} {\bibinfo  {journal} {American Journal of
  Physiology-Cell Physiology}\ }\textbf {\bibinfo {volume} {245}},\ \bibinfo
  {pages} {C1} (\bibinfo {year} {1983})}\BibitemShut {NoStop}%
\bibitem [{\citenamefont {Matsuda}\ \emph {et~al.}(1997)\citenamefont
  {Matsuda}, \citenamefont {Takuma},\ and\ \citenamefont
  {Baba}}]{matsuda1997na+}%
  \BibitemOpen
  \bibfield  {author} {\bibinfo {author} {\bibfnamefont {T.}~\bibnamefont
  {Matsuda}}, \bibinfo {author} {\bibfnamefont {K.}~\bibnamefont {Takuma}},\
  and\ \bibinfo {author} {\bibfnamefont {A.}~\bibnamefont {Baba}},\ }\bibfield
  {title} {\bibinfo {title} {Na+-ca2+ exchanger: physiology and pharmacology},\
  }\href@noop {} {\bibfield  {journal} {\bibinfo  {journal} {The Japanese
  Journal of Pharmacology}\ }\textbf {\bibinfo {volume} {74}},\ \bibinfo
  {pages} {1} (\bibinfo {year} {1997})}\BibitemShut {NoStop}%
\bibitem [{\citenamefont {Knollmann}\ \emph {et~al.}(2006)\citenamefont
  {Knollmann}, \citenamefont {Chopra}, \citenamefont {Hlaing}, \citenamefont
  {Akin}, \citenamefont {Yang}, \citenamefont {Ettensohn}, \citenamefont
  {Knollmann}, \citenamefont {Horton}, \citenamefont {Weissman}, \citenamefont
  {Holinstat} \emph {et~al.}}]{knollmann2006casq2}%
  \BibitemOpen
  \bibfield  {author} {\bibinfo {author} {\bibfnamefont {B.~C.}\ \bibnamefont
  {Knollmann}}, \bibinfo {author} {\bibfnamefont {N.}~\bibnamefont {Chopra}},
  \bibinfo {author} {\bibfnamefont {T.}~\bibnamefont {Hlaing}}, \bibinfo
  {author} {\bibfnamefont {B.}~\bibnamefont {Akin}}, \bibinfo {author}
  {\bibfnamefont {T.}~\bibnamefont {Yang}}, \bibinfo {author} {\bibfnamefont
  {K.}~\bibnamefont {Ettensohn}}, \bibinfo {author} {\bibfnamefont {B.~E.}\
  \bibnamefont {Knollmann}}, \bibinfo {author} {\bibfnamefont {K.~D.}\
  \bibnamefont {Horton}}, \bibinfo {author} {\bibfnamefont {N.~J.}\
  \bibnamefont {Weissman}}, \bibinfo {author} {\bibfnamefont {I.}~\bibnamefont
  {Holinstat}}, \emph {et~al.},\ }\bibfield  {title} {\bibinfo {title} {Casq2
  deletion causes sarcoplasmic reticulum volume increase, premature ca 2+
  release, and catecholaminergic polymorphic ventricular tachycardia},\
  }\href@noop {} {\bibfield  {journal} {\bibinfo  {journal} {The Journal of
  clinical investigation}\ }\textbf {\bibinfo {volume} {116}},\ \bibinfo
  {pages} {2510} (\bibinfo {year} {2006})}\BibitemShut {NoStop}%
\bibitem [{\citenamefont {Palade}\ \emph {et~al.}(1983)\citenamefont {Palade},
  \citenamefont {Mitchell},\ and\ \citenamefont
  {Fleischer}}]{palade1983spontaneous}%
  \BibitemOpen
  \bibfield  {author} {\bibinfo {author} {\bibfnamefont {P.}~\bibnamefont
  {Palade}}, \bibinfo {author} {\bibfnamefont {R.~D.}\ \bibnamefont
  {Mitchell}},\ and\ \bibinfo {author} {\bibfnamefont {S.}~\bibnamefont
  {Fleischer}},\ }\bibfield  {title} {\bibinfo {title} {Spontaneous calcium
  release from sarcoplasmic reticulum. general description and effects of
  calcium.},\ }\href@noop {} {\bibfield  {journal} {\bibinfo  {journal}
  {Journal of Biological Chemistry}\ }\textbf {\bibinfo {volume} {258}},\
  \bibinfo {pages} {8098} (\bibinfo {year} {1983})}\BibitemShut {NoStop}%
\bibitem [{\citenamefont {Shiferaw}\ \emph {et~al.}(2003)\citenamefont
  {Shiferaw}, \citenamefont {Watanabe}, \citenamefont {Garfinkel},
  \citenamefont {Weiss},\ and\ \citenamefont {Karma}}]{shiferaw2003model}%
  \BibitemOpen
  \bibfield  {author} {\bibinfo {author} {\bibfnamefont {Y.}~\bibnamefont
  {Shiferaw}}, \bibinfo {author} {\bibfnamefont {M.}~\bibnamefont {Watanabe}},
  \bibinfo {author} {\bibfnamefont {A.}~\bibnamefont {Garfinkel}}, \bibinfo
  {author} {\bibfnamefont {J.}~\bibnamefont {Weiss}},\ and\ \bibinfo {author}
  {\bibfnamefont {A.}~\bibnamefont {Karma}},\ }\bibfield  {title} {\bibinfo
  {title} {Model of intracellular calcium cycling in ventricular myocytes},\
  }\href@noop {} {\bibfield  {journal} {\bibinfo  {journal} {Biophysical
  journal}\ }\textbf {\bibinfo {volume} {85}},\ \bibinfo {pages} {3666}
  (\bibinfo {year} {2003})}\BibitemShut {NoStop}%
\bibitem [{\citenamefont {Colman}(2019)}]{colman2019arrhythmia}%
  \BibitemOpen
  \bibfield  {author} {\bibinfo {author} {\bibfnamefont {M.~A.}\ \bibnamefont
  {Colman}},\ }\bibfield  {title} {\bibinfo {title} {Arrhythmia mechanisms and
  spontaneous calcium release: Bi-directional coupling between re-entrant and
  focal excitation},\ }\href@noop {} {\bibfield  {journal} {\bibinfo  {journal}
  {PLoS computational biology}\ }\textbf {\bibinfo {volume} {15}},\ \bibinfo
  {pages} {e1007260} (\bibinfo {year} {2019})}\BibitemShut {NoStop}%
\bibitem [{\citenamefont {Walker}\ \emph {et~al.}(2017)\citenamefont {Walker},
  \citenamefont {Gurev}, \citenamefont {Rice}, \citenamefont {Greenstein},\
  and\ \citenamefont {Winslow}}]{walker2017estimating}%
  \BibitemOpen
  \bibfield  {author} {\bibinfo {author} {\bibfnamefont {M.~A.}\ \bibnamefont
  {Walker}}, \bibinfo {author} {\bibfnamefont {V.}~\bibnamefont {Gurev}},
  \bibinfo {author} {\bibfnamefont {J.~J.}\ \bibnamefont {Rice}}, \bibinfo
  {author} {\bibfnamefont {J.~L.}\ \bibnamefont {Greenstein}},\ and\ \bibinfo
  {author} {\bibfnamefont {R.~L.}\ \bibnamefont {Winslow}},\ }\bibfield
  {title} {\bibinfo {title} {Estimating the probabilities of rare arrhythmic
  events in multiscale computational models of cardiac cells and tissue},\
  }\href@noop {} {\bibfield  {journal} {\bibinfo  {journal} {PLoS computational
  biology}\ }\textbf {\bibinfo {volume} {13}},\ \bibinfo {pages} {e1005783}
  (\bibinfo {year} {2017})}\BibitemShut {NoStop}%
\bibitem [{\citenamefont {Xie}\ \emph {et~al.}(2010)\citenamefont {Xie},
  \citenamefont {Sato}, \citenamefont {Garfinkel}, \citenamefont {Qu},\ and\
  \citenamefont {Weiss}}]{xie2010so}%
  \BibitemOpen
  \bibfield  {author} {\bibinfo {author} {\bibfnamefont {Y.}~\bibnamefont
  {Xie}}, \bibinfo {author} {\bibfnamefont {D.}~\bibnamefont {Sato}}, \bibinfo
  {author} {\bibfnamefont {A.}~\bibnamefont {Garfinkel}}, \bibinfo {author}
  {\bibfnamefont {Z.}~\bibnamefont {Qu}},\ and\ \bibinfo {author}
  {\bibfnamefont {J.~N.}\ \bibnamefont {Weiss}},\ }\bibfield  {title} {\bibinfo
  {title} {So little source, so much sink: requirements for
  afterdepolarizations to propagate in tissue},\ }\href@noop {} {\bibfield
  {journal} {\bibinfo  {journal} {Biophysical journal}\ }\textbf {\bibinfo
  {volume} {99}},\ \bibinfo {pages} {1408} (\bibinfo {year}
  {2010})}\BibitemShut {NoStop}%
\bibitem [{\citenamefont {Iyer}\ \emph {et~al.}(2007)\citenamefont {Iyer},
  \citenamefont {Hajjar},\ and\ \citenamefont
  {Armoundas}}]{iyer2007mechanisms}%
  \BibitemOpen
  \bibfield  {author} {\bibinfo {author} {\bibfnamefont {V.}~\bibnamefont
  {Iyer}}, \bibinfo {author} {\bibfnamefont {R.~J.}\ \bibnamefont {Hajjar}},\
  and\ \bibinfo {author} {\bibfnamefont {A.~A.}\ \bibnamefont {Armoundas}},\
  }\bibfield  {title} {\bibinfo {title} {Mechanisms of abnormal calcium
  homeostasis in mutations responsible for catecholaminergic polymorphic
  ventricular tachycardia},\ }\href@noop {} {\bibfield  {journal} {\bibinfo
  {journal} {Circulation research}\ }\textbf {\bibinfo {volume} {100}},\
  \bibinfo {pages} {e22} (\bibinfo {year} {2007})}\BibitemShut {NoStop}%
\bibitem [{\citenamefont {Fink}\ \emph {et~al.}(2008)\citenamefont {Fink},
  \citenamefont {Noble}, \citenamefont {Virag}, \citenamefont {Varro},\ and\
  \citenamefont {Giles}}]{fink2008contributions}%
  \BibitemOpen
  \bibfield  {author} {\bibinfo {author} {\bibfnamefont {M.}~\bibnamefont
  {Fink}}, \bibinfo {author} {\bibfnamefont {D.}~\bibnamefont {Noble}},
  \bibinfo {author} {\bibfnamefont {L.}~\bibnamefont {Virag}}, \bibinfo
  {author} {\bibfnamefont {A.}~\bibnamefont {Varro}},\ and\ \bibinfo {author}
  {\bibfnamefont {W.~R.}\ \bibnamefont {Giles}},\ }\bibfield  {title} {\bibinfo
  {title} {Contributions of herg k+ current to repolarization of the human
  ventricular action potential},\ }\href@noop {} {\bibfield  {journal}
  {\bibinfo  {journal} {Progress in biophysics and molecular biology}\ }\textbf
  {\bibinfo {volume} {96}},\ \bibinfo {pages} {357} (\bibinfo {year}
  {2008})}\BibitemShut {NoStop}%
\bibitem [{\citenamefont {Fink}\ \emph {et~al.}(2011)\citenamefont {Fink},
  \citenamefont {Noble},\ and\ \citenamefont {Noble}}]{fink2011ca2+}%
  \BibitemOpen
  \bibfield  {author} {\bibinfo {author} {\bibfnamefont {M.}~\bibnamefont
  {Fink}}, \bibinfo {author} {\bibfnamefont {P.~J.}\ \bibnamefont {Noble}},\
  and\ \bibinfo {author} {\bibfnamefont {D.}~\bibnamefont {Noble}},\ }\bibfield
   {title} {\bibinfo {title} {Ca2+-induced delayed afterdepolarizations are
  triggered by dyadic subspace ca2+ affirming that increasing serca reduces
  aftercontractions},\ }\href@noop {} {\bibfield  {journal} {\bibinfo
  {journal} {American Journal of Physiology-Heart and Circulatory Physiology}\
  }\textbf {\bibinfo {volume} {301}},\ \bibinfo {pages} {H921} (\bibinfo {year}
  {2011})}\BibitemShut {NoStop}%
\bibitem [{\citenamefont {Shannon}\ \emph {et~al.}(2004)\citenamefont
  {Shannon}, \citenamefont {Wang}, \citenamefont {Puglisi}, \citenamefont
  {Weber},\ and\ \citenamefont {Bers}}]{shannon2004mathematical}%
  \BibitemOpen
  \bibfield  {author} {\bibinfo {author} {\bibfnamefont {T.~R.}\ \bibnamefont
  {Shannon}}, \bibinfo {author} {\bibfnamefont {F.}~\bibnamefont {Wang}},
  \bibinfo {author} {\bibfnamefont {J.}~\bibnamefont {Puglisi}}, \bibinfo
  {author} {\bibfnamefont {C.}~\bibnamefont {Weber}},\ and\ \bibinfo {author}
  {\bibfnamefont {D.~M.}\ \bibnamefont {Bers}},\ }\bibfield  {title} {\bibinfo
  {title} {A mathematical treatment of integrated ca dynamics within the
  ventricular myocyte},\ }\href@noop {} {\bibfield  {journal} {\bibinfo
  {journal} {Biophysical journal}\ }\textbf {\bibinfo {volume} {87}},\ \bibinfo
  {pages} {3351} (\bibinfo {year} {2004})}\BibitemShut {NoStop}%
\bibitem [{\citenamefont {Stern}\ \emph {et~al.}(1999)\citenamefont {Stern},
  \citenamefont {Song}, \citenamefont {Cheng}, \citenamefont {Sham},
  \citenamefont {Yang}, \citenamefont {Boheler},\ and\ \citenamefont
  {R{\'\i}os}}]{stern1999local}%
  \BibitemOpen
  \bibfield  {author} {\bibinfo {author} {\bibfnamefont {M.~D.}\ \bibnamefont
  {Stern}}, \bibinfo {author} {\bibfnamefont {L.-S.}\ \bibnamefont {Song}},
  \bibinfo {author} {\bibfnamefont {H.}~\bibnamefont {Cheng}}, \bibinfo
  {author} {\bibfnamefont {J.~S.}\ \bibnamefont {Sham}}, \bibinfo {author}
  {\bibfnamefont {H.~T.}\ \bibnamefont {Yang}}, \bibinfo {author}
  {\bibfnamefont {K.~R.}\ \bibnamefont {Boheler}},\ and\ \bibinfo {author}
  {\bibfnamefont {E.}~\bibnamefont {R{\'\i}os}},\ }\bibfield  {title} {\bibinfo
  {title} {Local control models of cardiac excitation--contraction coupling: a
  possible role for allosteric interactions between ryanodine receptors},\
  }\href@noop {} {\bibfield  {journal} {\bibinfo  {journal} {The Journal of
  general physiology}\ }\textbf {\bibinfo {volume} {113}},\ \bibinfo {pages}
  {469} (\bibinfo {year} {1999})}\BibitemShut {NoStop}%
\bibitem [{\citenamefont {Rush}\ and\ \citenamefont
  {Larsen}(1978)}]{rush1978practical}%
  \BibitemOpen
  \bibfield  {author} {\bibinfo {author} {\bibfnamefont {S.}~\bibnamefont
  {Rush}}\ and\ \bibinfo {author} {\bibfnamefont {H.}~\bibnamefont {Larsen}},\
  }\bibfield  {title} {\bibinfo {title} {A practical algorithm for solving
  dynamic membrane equations},\ }\href@noop {} {\bibfield  {journal} {\bibinfo
  {journal} {IEEE Transactions on Biomedical Engineering}\ ,\ \bibinfo {pages}
  {389}} (\bibinfo {year} {1978})}\BibitemShut {NoStop}%
\bibitem [{\citenamefont {Marsh}\ \emph {et~al.}(2012)\citenamefont {Marsh},
  \citenamefont {Ziaratgahi},\ and\ \citenamefont
  {Spiteri}}]{marsh2012secrets}%
  \BibitemOpen
  \bibfield  {author} {\bibinfo {author} {\bibfnamefont {M.~E.}\ \bibnamefont
  {Marsh}}, \bibinfo {author} {\bibfnamefont {S.~T.}\ \bibnamefont
  {Ziaratgahi}},\ and\ \bibinfo {author} {\bibfnamefont {R.~J.}\ \bibnamefont
  {Spiteri}},\ }\bibfield  {title} {\bibinfo {title} {The secrets to the
  success of the rush--larsen method and its generalizations},\ }\href@noop {}
  {\bibfield  {journal} {\bibinfo  {journal} {IEEE transactions on biomedical
  engineering}\ }\textbf {\bibinfo {volume} {59}},\ \bibinfo {pages} {2506}
  (\bibinfo {year} {2012})}\BibitemShut {NoStop}%
\bibitem [{\citenamefont {Winslow}\ \emph {et~al.}(2011)\citenamefont
  {Winslow}, \citenamefont {Saltz}, \citenamefont {Foster}, \citenamefont
  {Carr}, \citenamefont {Ge}, \citenamefont {Miller}, \citenamefont {Younes},
  \citenamefont {Geman}, \citenamefont {Graniote}, \citenamefont {Kurc} \emph
  {et~al.}}]{winslow2011cardiovascular}%
  \BibitemOpen
  \bibfield  {author} {\bibinfo {author} {\bibfnamefont {R.}~\bibnamefont
  {Winslow}}, \bibinfo {author} {\bibfnamefont {J.}~\bibnamefont {Saltz}},
  \bibinfo {author} {\bibfnamefont {I.}~\bibnamefont {Foster}}, \bibinfo
  {author} {\bibfnamefont {J.}~\bibnamefont {Carr}}, \bibinfo {author}
  {\bibfnamefont {Y.}~\bibnamefont {Ge}}, \bibinfo {author} {\bibfnamefont
  {M.}~\bibnamefont {Miller}}, \bibinfo {author} {\bibfnamefont
  {L.}~\bibnamefont {Younes}}, \bibinfo {author} {\bibfnamefont
  {D.}~\bibnamefont {Geman}}, \bibinfo {author} {\bibfnamefont
  {S.}~\bibnamefont {Graniote}}, \bibinfo {author} {\bibfnamefont
  {T.}~\bibnamefont {Kurc}}, \emph {et~al.},\ }\bibfield  {title} {\bibinfo
  {title} {The cardiovascular research grid (cvrg) project},\ }\href@noop {}
  {\bibfield  {journal} {\bibinfo  {journal} {Proceedings of the AMIA Summit on
  Translational Bioinformatics}\ }\textbf {\bibinfo {volume} {2011}},\ \bibinfo
  {pages} {77} (\bibinfo {year} {2011})}\BibitemShut {NoStop}%
\bibitem [{\citenamefont {Fenton}\ \emph {et~al.}(2005)\citenamefont {Fenton},
  \citenamefont {Cherry}, \citenamefont {Karma},\ and\ \citenamefont
  {Rappel}}]{fenton2005modeling}%
  \BibitemOpen
  \bibfield  {author} {\bibinfo {author} {\bibfnamefont {F.~H.}\ \bibnamefont
  {Fenton}}, \bibinfo {author} {\bibfnamefont {E.~M.}\ \bibnamefont {Cherry}},
  \bibinfo {author} {\bibfnamefont {A.}~\bibnamefont {Karma}},\ and\ \bibinfo
  {author} {\bibfnamefont {W.-J.}\ \bibnamefont {Rappel}},\ }\bibfield  {title}
  {\bibinfo {title} {Modeling wave propagation in realistic heart geometries
  using the phase-field method},\ }\href@noop {} {\bibfield  {journal}
  {\bibinfo  {journal} {Chaos: An Interdisciplinary Journal of Nonlinear
  Science}\ }\textbf {\bibinfo {volume} {15}},\ \bibinfo {pages} {013502}
  (\bibinfo {year} {2005})}\BibitemShut {NoStop}%
\bibitem [{\citenamefont {Rajany}\ \emph {et~al.}(2021)\citenamefont {Rajany},
  \citenamefont {Majumder}, \citenamefont {Nayak},\ and\ \citenamefont
  {Pandit}}]{rajany2021effects}%
  \BibitemOpen
  \bibfield  {author} {\bibinfo {author} {\bibfnamefont {K.}~\bibnamefont
  {Rajany}}, \bibinfo {author} {\bibfnamefont {R.}~\bibnamefont {Majumder}},
  \bibinfo {author} {\bibfnamefont {A.~R.}\ \bibnamefont {Nayak}},\ and\
  \bibinfo {author} {\bibfnamefont {R.}~\bibnamefont {Pandit}},\ }\bibfield
  {title} {\bibinfo {title} {The effects of inhomogeneities on scroll-wave
  dynamics in an anatomically realistic mathematical model for canine
  ventricular tissue},\ }\href@noop {} {\bibfield  {journal} {\bibinfo
  {journal} {Physics Open}\ }\textbf {\bibinfo {volume} {9}},\ \bibinfo {pages}
  {100090} (\bibinfo {year} {2021})}\BibitemShut {NoStop}%
\bibitem [{\citenamefont {Majumder}\ \emph {et~al.}(2016)\citenamefont
  {Majumder}, \citenamefont {Pandit},\ and\ \citenamefont
  {Panfilov}}]{majumder2016scroll}%
  \BibitemOpen
  \bibfield  {author} {\bibinfo {author} {\bibfnamefont {R.}~\bibnamefont
  {Majumder}}, \bibinfo {author} {\bibfnamefont {R.}~\bibnamefont {Pandit}},\
  and\ \bibinfo {author} {\bibfnamefont {A.~V.}\ \bibnamefont {Panfilov}},\
  }\bibfield  {title} {\bibinfo {title} {Scroll-wave dynamics in the presence
  of ionic and conduction inhomogeneities in an anatomically realistic
  mathematical model for the pig heart},\ }\href@noop {} {\bibfield  {journal}
  {\bibinfo  {journal} {JETP letters}\ }\textbf {\bibinfo {volume} {104}},\
  \bibinfo {pages} {796} (\bibinfo {year} {2016})}\BibitemShut {NoStop}%
\bibitem [{\citenamefont {Mulimani}\ \emph {et~al.}(2022)\citenamefont
  {Mulimani}, \citenamefont {Zimik},\ and\ \citenamefont
  {Pandit}}]{mulimani2022silico}%
  \BibitemOpen
  \bibfield  {author} {\bibinfo {author} {\bibfnamefont {M.~K.}\ \bibnamefont
  {Mulimani}}, \bibinfo {author} {\bibfnamefont {S.}~\bibnamefont {Zimik}},\
  and\ \bibinfo {author} {\bibfnamefont {R.}~\bibnamefont {Pandit}},\
  }\bibfield  {title} {\bibinfo {title} {An in silico study of
  electrophysiological parameters that affect the spiral-wave frequency in
  mathematical models for cardiac tissue},\ }\href@noop {} {\bibfield
  {journal} {\bibinfo  {journal} {Frontiers in Physics}\ ,\ \bibinfo {pages}
  {840}} (\bibinfo {year} {2022})}\BibitemShut {NoStop}%
\bibitem [{\citenamefont {Kuznetsov}(2019)}]{kuznetsov2019codim}%
  \BibitemOpen
  \bibfield  {author} {\bibinfo {author} {\bibfnamefont {Y.~A.}\ \bibnamefont
  {Kuznetsov}},\ }\href {https://webspace.science.uu.nl/~kouzn101/USS2.pdf}
  {\bibinfo {title} {Codim 1 bifurcations of n-dimensional odes}} (\bibinfo
  {year} {2019})\BibitemShut {NoStop}%
\bibitem [{\citenamefont {Dhooge}\ \emph {et~al.}(2003)\citenamefont {Dhooge},
  \citenamefont {Govaerts},\ and\ \citenamefont
  {Kuznetsov}}]{dhooge2003matcont}%
  \BibitemOpen
  \bibfield  {author} {\bibinfo {author} {\bibfnamefont {A.}~\bibnamefont
  {Dhooge}}, \bibinfo {author} {\bibfnamefont {W.}~\bibnamefont {Govaerts}},\
  and\ \bibinfo {author} {\bibfnamefont {Y.~A.}\ \bibnamefont {Kuznetsov}},\
  }\bibfield  {title} {\bibinfo {title} {Matcont: a matlab package for
  numerical bifurcation analysis of odes},\ }\href@noop {} {\bibfield
  {journal} {\bibinfo  {journal} {ACM Transactions on Mathematical Software
  (TOMS)}\ }\textbf {\bibinfo {volume} {29}},\ \bibinfo {pages} {141} (\bibinfo
  {year} {2003})}\BibitemShut {NoStop}%
\bibitem [{\citenamefont {Ermentrout}(2001)}]{ermentrout2001xppaut}%
  \BibitemOpen
  \bibfield  {author} {\bibinfo {author} {\bibfnamefont {B.}~\bibnamefont
  {Ermentrout}},\ }\bibfield  {title} {\bibinfo {title} {Xppaut 5.0-the
  differential equations tool},\ }\href@noop {} {\bibfield  {journal} {\bibinfo
   {journal} {University of Pittsburgh, Pittsburgh}\ } (\bibinfo {year}
  {2001})}\BibitemShut {NoStop}%
\bibitem [{\citenamefont {Sobie}(2009)}]{sobie2009parameter}%
  \BibitemOpen
  \bibfield  {author} {\bibinfo {author} {\bibfnamefont {E.~A.}\ \bibnamefont
  {Sobie}},\ }\bibfield  {title} {\bibinfo {title} {Parameter sensitivity
  analysis in electrophysiological models using multivariable regression},\
  }\href@noop {} {\bibfield  {journal} {\bibinfo  {journal} {Biophysical
  journal}\ }\textbf {\bibinfo {volume} {96}},\ \bibinfo {pages} {1264}
  (\bibinfo {year} {2009})}\BibitemShut {NoStop}%
\bibitem [{\citenamefont {Shah}\ \emph {et~al.}(2019)\citenamefont {Shah},
  \citenamefont {Jiwani}, \citenamefont {Limbu}, \citenamefont {Weinberg},\
  and\ \citenamefont {Deo}}]{shah2019delayed}%
  \BibitemOpen
  \bibfield  {author} {\bibinfo {author} {\bibfnamefont {C.}~\bibnamefont
  {Shah}}, \bibinfo {author} {\bibfnamefont {S.}~\bibnamefont {Jiwani}},
  \bibinfo {author} {\bibfnamefont {B.}~\bibnamefont {Limbu}}, \bibinfo
  {author} {\bibfnamefont {S.}~\bibnamefont {Weinberg}},\ and\ \bibinfo
  {author} {\bibfnamefont {M.}~\bibnamefont {Deo}},\ }\bibfield  {title}
  {\bibinfo {title} {Delayed afterdepolarization-induced triggered activity in
  cardiac purkinje cells mediated through cytosolic calcium diffusion waves},\
  }\href@noop {} {\bibfield  {journal} {\bibinfo  {journal} {Physiological
  reports}\ }\textbf {\bibinfo {volume} {7}},\ \bibinfo {pages} {e14296}
  (\bibinfo {year} {2019})}\BibitemShut {NoStop}%
\bibitem [{\citenamefont {Zygmunt}\ \emph {et~al.}(1998)\citenamefont
  {Zygmunt}, \citenamefont {Goodrow},\ and\ \citenamefont
  {Weigel}}]{zygmunt1998naca}%
  \BibitemOpen
  \bibfield  {author} {\bibinfo {author} {\bibfnamefont {A.~C.}\ \bibnamefont
  {Zygmunt}}, \bibinfo {author} {\bibfnamefont {R.~J.}\ \bibnamefont
  {Goodrow}},\ and\ \bibinfo {author} {\bibfnamefont {C.~M.}\ \bibnamefont
  {Weigel}},\ }\bibfield  {title} {\bibinfo {title} {I naca and i cl (ca)
  contribute to isoproterenol-induced delayed afterdepolarizations in
  midmyocardial cells},\ }\href@noop {} {\bibfield  {journal} {\bibinfo
  {journal} {American Journal of Physiology-Heart and Circulatory Physiology}\
  }\textbf {\bibinfo {volume} {275}},\ \bibinfo {pages} {H1979} (\bibinfo
  {year} {1998})}\BibitemShut {NoStop}%
\bibitem [{\citenamefont {Catanzaro}\ \emph {et~al.}(2006)\citenamefont
  {Catanzaro}, \citenamefont {Nett}, \citenamefont {Rota},\ and\ \citenamefont
  {Vassalle}}]{catanzaro2006mechanisms}%
  \BibitemOpen
  \bibfield  {author} {\bibinfo {author} {\bibfnamefont {J.~N.}\ \bibnamefont
  {Catanzaro}}, \bibinfo {author} {\bibfnamefont {M.~P.}\ \bibnamefont {Nett}},
  \bibinfo {author} {\bibfnamefont {M.}~\bibnamefont {Rota}},\ and\ \bibinfo
  {author} {\bibfnamefont {M.}~\bibnamefont {Vassalle}},\ }\bibfield  {title}
  {\bibinfo {title} {On the mechanisms underlying diastolic voltage
  oscillations in the sinoatrial node},\ }\href@noop {} {\bibfield  {journal}
  {\bibinfo  {journal} {Journal of electrocardiology}\ }\textbf {\bibinfo
  {volume} {39}},\ \bibinfo {pages} {342} (\bibinfo {year} {2006})}\BibitemShut
  {NoStop}%
\bibitem [{\citenamefont {Song}\ \emph {et~al.}(2015)\citenamefont {Song},
  \citenamefont {Ko}, \citenamefont {Nivala}, \citenamefont {Weiss},\ and\
  \citenamefont {Qu}}]{song2015calcium}%
  \BibitemOpen
  \bibfield  {author} {\bibinfo {author} {\bibfnamefont {Z.}~\bibnamefont
  {Song}}, \bibinfo {author} {\bibfnamefont {C.~Y.}\ \bibnamefont {Ko}},
  \bibinfo {author} {\bibfnamefont {M.}~\bibnamefont {Nivala}}, \bibinfo
  {author} {\bibfnamefont {J.~N.}\ \bibnamefont {Weiss}},\ and\ \bibinfo
  {author} {\bibfnamefont {Z.}~\bibnamefont {Qu}},\ }\bibfield  {title}
  {\bibinfo {title} {Calcium-voltage coupling in the genesis of early and
  delayed afterdepolarizations in cardiac myocytes},\ }\href@noop {} {\bibfield
   {journal} {\bibinfo  {journal} {Biophysical journal}\ }\textbf {\bibinfo
  {volume} {108}},\ \bibinfo {pages} {1908} (\bibinfo {year}
  {2015})}\BibitemShut {NoStop}%
\bibitem [{\citenamefont {Fink}\ and\ \citenamefont
  {Noble}(2010)}]{fink2010pharmacodynamic}%
  \BibitemOpen
  \bibfield  {author} {\bibinfo {author} {\bibfnamefont {M.}~\bibnamefont
  {Fink}}\ and\ \bibinfo {author} {\bibfnamefont {D.}~\bibnamefont {Noble}},\
  }\bibfield  {title} {\bibinfo {title} {Pharmacodynamic effects in the
  cardiovascular system: the modeller’s view},\ }\href@noop {} {\bibfield
  {journal} {\bibinfo  {journal} {Basic \& clinical pharmacology \&
  toxicology}\ }\textbf {\bibinfo {volume} {106}},\ \bibinfo {pages} {243}
  (\bibinfo {year} {2010})}\BibitemShut {NoStop}%
\bibitem [{\citenamefont {Fowler}\ \emph {et~al.}(2018)\citenamefont {Fowler},
  \citenamefont {Kong}, \citenamefont {Hancox},\ and\ \citenamefont
  {Cannell}}]{fowler2018late}%
  \BibitemOpen
  \bibfield  {author} {\bibinfo {author} {\bibfnamefont {E.~D.}\ \bibnamefont
  {Fowler}}, \bibinfo {author} {\bibfnamefont {C.~H.}\ \bibnamefont {Kong}},
  \bibinfo {author} {\bibfnamefont {J.~C.}\ \bibnamefont {Hancox}},\ and\
  \bibinfo {author} {\bibfnamefont {M.~B.}\ \bibnamefont {Cannell}},\
  }\bibfield  {title} {\bibinfo {title} {Late ca2+ sparks and ripples during
  the systolic ca2+ transient in heart muscle cells},\ }\href@noop {}
  {\bibfield  {journal} {\bibinfo  {journal} {Circulation research}\ }\textbf
  {\bibinfo {volume} {122}},\ \bibinfo {pages} {473} (\bibinfo {year}
  {2018})}\BibitemShut {NoStop}%
\bibitem [{\citenamefont {Shiferaw}\ \emph {et~al.}(2012)\citenamefont
  {Shiferaw}, \citenamefont {Aistrup},\ and\ \citenamefont
  {Wasserstrom}}]{shiferaw2012intracellular}%
  \BibitemOpen
  \bibfield  {author} {\bibinfo {author} {\bibfnamefont {Y.}~\bibnamefont
  {Shiferaw}}, \bibinfo {author} {\bibfnamefont {G.~L.}\ \bibnamefont
  {Aistrup}},\ and\ \bibinfo {author} {\bibfnamefont {J.~A.}\ \bibnamefont
  {Wasserstrom}},\ }\href@noop {} {\bibinfo {title} {Intracellular ca2+ waves,
  afterdepolarizations, and triggered arrhythmias}} (\bibinfo {year}
  {2012})\BibitemShut {NoStop}%
\bibitem [{\citenamefont {Fowler}\ \emph {et~al.}(2020)\citenamefont {Fowler},
  \citenamefont {Wang}, \citenamefont {Hezzell}, \citenamefont {Chanoit},
  \citenamefont {Hancox},\ and\ \citenamefont
  {Cannell}}]{fowler2020arrhythmogenic}%
  \BibitemOpen
  \bibfield  {author} {\bibinfo {author} {\bibfnamefont {E.~D.}\ \bibnamefont
  {Fowler}}, \bibinfo {author} {\bibfnamefont {N.}~\bibnamefont {Wang}},
  \bibinfo {author} {\bibfnamefont {M.}~\bibnamefont {Hezzell}}, \bibinfo
  {author} {\bibfnamefont {G.}~\bibnamefont {Chanoit}}, \bibinfo {author}
  {\bibfnamefont {J.~C.}\ \bibnamefont {Hancox}},\ and\ \bibinfo {author}
  {\bibfnamefont {M.~B.}\ \bibnamefont {Cannell}},\ }\bibfield  {title}
  {\bibinfo {title} {Arrhythmogenic late ca2+ sparks in failing heart cells and
  their control by action potential configuration},\ }\href@noop {} {\bibfield
  {journal} {\bibinfo  {journal} {Proceedings of the National Academy of
  Sciences}\ }\textbf {\bibinfo {volume} {117}},\ \bibinfo {pages} {2687}
  (\bibinfo {year} {2020})}\BibitemShut {NoStop}%
\bibitem [{\citenamefont {Jost}\ \emph {et~al.}(2013)\citenamefont {Jost},
  \citenamefont {Nagy}, \citenamefont {Corici}, \citenamefont {Kohajda},
  \citenamefont {Horv{\'a}th}, \citenamefont {Acsai}, \citenamefont {Biliczki},
  \citenamefont {Levijoki}, \citenamefont {Pollesello}, \citenamefont
  {Koskelainen} \emph {et~al.}}]{jost2013orm}%
  \BibitemOpen
  \bibfield  {author} {\bibinfo {author} {\bibfnamefont {N.}~\bibnamefont
  {Jost}}, \bibinfo {author} {\bibfnamefont {N.}~\bibnamefont {Nagy}}, \bibinfo
  {author} {\bibfnamefont {C.}~\bibnamefont {Corici}}, \bibinfo {author}
  {\bibfnamefont {Z.}~\bibnamefont {Kohajda}}, \bibinfo {author} {\bibfnamefont
  {A.}~\bibnamefont {Horv{\'a}th}}, \bibinfo {author} {\bibfnamefont
  {K.}~\bibnamefont {Acsai}}, \bibinfo {author} {\bibfnamefont
  {P.}~\bibnamefont {Biliczki}}, \bibinfo {author} {\bibfnamefont
  {J.}~\bibnamefont {Levijoki}}, \bibinfo {author} {\bibfnamefont
  {P.}~\bibnamefont {Pollesello}}, \bibinfo {author} {\bibfnamefont
  {T.}~\bibnamefont {Koskelainen}}, \emph {et~al.},\ }\bibfield  {title}
  {\bibinfo {title} {Orm-10103, a novel specific inhibitor of the na+/ca2+
  exchanger, decreases early and delayed afterdepolarizations in the canine
  heart},\ }\href@noop {} {\bibfield  {journal} {\bibinfo  {journal} {British
  journal of pharmacology}\ }\textbf {\bibinfo {volume} {170}},\ \bibinfo
  {pages} {768} (\bibinfo {year} {2013})}\BibitemShut {NoStop}%
\bibitem [{\citenamefont {Voigt}\ \emph {et~al.}(2012)\citenamefont {Voigt},
  \citenamefont {Li}, \citenamefont {Wang}, \citenamefont {Wang}, \citenamefont
  {Trafford}, \citenamefont {Abu-Taha}, \citenamefont {Sun}, \citenamefont
  {Wieland}, \citenamefont {Ravens}, \citenamefont {Nattel} \emph
  {et~al.}}]{voigt2012enhanced}%
  \BibitemOpen
  \bibfield  {author} {\bibinfo {author} {\bibfnamefont {N.}~\bibnamefont
  {Voigt}}, \bibinfo {author} {\bibfnamefont {N.}~\bibnamefont {Li}}, \bibinfo
  {author} {\bibfnamefont {Q.}~\bibnamefont {Wang}}, \bibinfo {author}
  {\bibfnamefont {W.}~\bibnamefont {Wang}}, \bibinfo {author} {\bibfnamefont
  {A.~W.}\ \bibnamefont {Trafford}}, \bibinfo {author} {\bibfnamefont
  {I.}~\bibnamefont {Abu-Taha}}, \bibinfo {author} {\bibfnamefont
  {Q.}~\bibnamefont {Sun}}, \bibinfo {author} {\bibfnamefont {T.}~\bibnamefont
  {Wieland}}, \bibinfo {author} {\bibfnamefont {U.}~\bibnamefont {Ravens}},
  \bibinfo {author} {\bibfnamefont {S.}~\bibnamefont {Nattel}}, \emph
  {et~al.},\ }\bibfield  {title} {\bibinfo {title} {Enhanced sarcoplasmic
  reticulum ca2+ leak and increased na+-ca2+ exchanger function underlie
  delayed afterdepolarizations in patients with chronic atrial fibrillation},\
  }\href@noop {} {\bibfield  {journal} {\bibinfo  {journal} {Circulation}\
  }\textbf {\bibinfo {volume} {125}},\ \bibinfo {pages} {2059} (\bibinfo {year}
  {2012})}\BibitemShut {NoStop}%
\bibitem [{\citenamefont {B{\"o}geholz}\ \emph {et~al.}(2015)\citenamefont
  {B{\"o}geholz}, \citenamefont {Pauls}, \citenamefont {Bauer}, \citenamefont
  {Schulte}, \citenamefont {Dechering}, \citenamefont {Frommeyer},
  \citenamefont {Kirchhefer}, \citenamefont {Goldhaber}, \citenamefont
  {M{\"u}ller}, \citenamefont {Eckardt} \emph
  {et~al.}}]{bogeholz2015suppression}%
  \BibitemOpen
  \bibfield  {author} {\bibinfo {author} {\bibfnamefont {N.}~\bibnamefont
  {B{\"o}geholz}}, \bibinfo {author} {\bibfnamefont {P.}~\bibnamefont {Pauls}},
  \bibinfo {author} {\bibfnamefont {B.~K.}\ \bibnamefont {Bauer}}, \bibinfo
  {author} {\bibfnamefont {J.~S.}\ \bibnamefont {Schulte}}, \bibinfo {author}
  {\bibfnamefont {D.~G.}\ \bibnamefont {Dechering}}, \bibinfo {author}
  {\bibfnamefont {G.}~\bibnamefont {Frommeyer}}, \bibinfo {author}
  {\bibfnamefont {U.}~\bibnamefont {Kirchhefer}}, \bibinfo {author}
  {\bibfnamefont {J.~I.}\ \bibnamefont {Goldhaber}}, \bibinfo {author}
  {\bibfnamefont {F.~U.}\ \bibnamefont {M{\"u}ller}}, \bibinfo {author}
  {\bibfnamefont {L.}~\bibnamefont {Eckardt}}, \emph {et~al.},\ }\bibfield
  {title} {\bibinfo {title} {Suppression of early and late afterdepolarizations
  by heterozygous knockout of the na+/ca2+ exchanger in a murine model},\
  }\href@noop {} {\bibfield  {journal} {\bibinfo  {journal} {Circulation:
  Arrhythmia and Electrophysiology}\ }\textbf {\bibinfo {volume} {8}},\
  \bibinfo {pages} {1210} (\bibinfo {year} {2015})}\BibitemShut {NoStop}%
\bibitem [{\citenamefont {Myles}\ \emph {et~al.}(2015)\citenamefont {Myles},
  \citenamefont {Wang}, \citenamefont {Bers},\ and\ \citenamefont
  {Ripplinger}}]{myles2015decreased}%
  \BibitemOpen
  \bibfield  {author} {\bibinfo {author} {\bibfnamefont {R.~C.}\ \bibnamefont
  {Myles}}, \bibinfo {author} {\bibfnamefont {L.}~\bibnamefont {Wang}},
  \bibinfo {author} {\bibfnamefont {D.~M.}\ \bibnamefont {Bers}},\ and\
  \bibinfo {author} {\bibfnamefont {C.~M.}\ \bibnamefont {Ripplinger}},\
  }\bibfield  {title} {\bibinfo {title} {Decreased inward rectifying k+ current
  and increased ryanodine receptor sensitivity synergistically contribute to
  sustained focal arrhythmia in the intact rabbit heart},\ }\href@noop {}
  {\bibfield  {journal} {\bibinfo  {journal} {The Journal of physiology}\
  }\textbf {\bibinfo {volume} {593}},\ \bibinfo {pages} {1479} (\bibinfo {year}
  {2015})}\BibitemShut {NoStop}%
\bibitem [{\citenamefont {Xu}\ \emph {et~al.}(2007)\citenamefont {Xu},
  \citenamefont {Zhang},\ and\ \citenamefont {Chiamvimonvat}}]{xu2007ik1}%
  \BibitemOpen
  \bibfield  {author} {\bibinfo {author} {\bibfnamefont {Y.}~\bibnamefont
  {Xu}}, \bibinfo {author} {\bibfnamefont {Q.}~\bibnamefont {Zhang}},\ and\
  \bibinfo {author} {\bibfnamefont {N.}~\bibnamefont {Chiamvimonvat}},\
  }\bibfield  {title} {\bibinfo {title} {Ik1 and cardiac hypoxia: after the
  long and short qt syndromes, what else can go wrong with the inward rectifier
  k+ currents?},\ }\href@noop {} {\bibfield  {journal} {\bibinfo  {journal}
  {Journal of molecular and cellular cardiology}\ }\textbf {\bibinfo {volume}
  {43}},\ \bibinfo {pages} {15} (\bibinfo {year} {2007})}\BibitemShut {NoStop}%
\bibitem [{\citenamefont {Sato}\ \emph {et~al.}(2021)\citenamefont {Sato},
  \citenamefont {Uchinoumi},\ and\ \citenamefont {Bers}}]{sato2021increasing}%
  \BibitemOpen
  \bibfield  {author} {\bibinfo {author} {\bibfnamefont {D.}~\bibnamefont
  {Sato}}, \bibinfo {author} {\bibfnamefont {H.}~\bibnamefont {Uchinoumi}},\
  and\ \bibinfo {author} {\bibfnamefont {D.~M.}\ \bibnamefont {Bers}},\
  }\bibfield  {title} {\bibinfo {title} {Increasing serca function promotes
  initiation of calcium sparks and breakup of calcium waves},\ }\href@noop {}
  {\bibfield  {journal} {\bibinfo  {journal} {The Journal of Physiology}\
  }\textbf {\bibinfo {volume} {599}},\ \bibinfo {pages} {3267} (\bibinfo {year}
  {2021})}\BibitemShut {NoStop}%
\bibitem [{\citenamefont {Salazar-Cant{\'u}}\ \emph {et~al.}(2016)\citenamefont
  {Salazar-Cant{\'u}}, \citenamefont {P{\'e}rez-Trevi{\~n}o}, \citenamefont
  {Montalvo-Parra}, \citenamefont {Balderas-Villalobos}, \citenamefont
  {G{\'o}mez-V{\'\i}quez}, \citenamefont {Garc{\'\i}a},\ and\ \citenamefont
  {Altamirano}}]{salazar2016role}%
  \BibitemOpen
  \bibfield  {author} {\bibinfo {author} {\bibfnamefont {A.}~\bibnamefont
  {Salazar-Cant{\'u}}}, \bibinfo {author} {\bibfnamefont {P.}~\bibnamefont
  {P{\'e}rez-Trevi{\~n}o}}, \bibinfo {author} {\bibfnamefont {D.}~\bibnamefont
  {Montalvo-Parra}}, \bibinfo {author} {\bibfnamefont {J.}~\bibnamefont
  {Balderas-Villalobos}}, \bibinfo {author} {\bibfnamefont {N.~L.}\
  \bibnamefont {G{\'o}mez-V{\'\i}quez}}, \bibinfo {author} {\bibfnamefont
  {N.}~\bibnamefont {Garc{\'\i}a}},\ and\ \bibinfo {author} {\bibfnamefont
  {J.}~\bibnamefont {Altamirano}},\ }\bibfield  {title} {\bibinfo {title} {Role
  of serca and the sarcoplasmic reticulum calcium content on calcium waves
  propagation in rat ventricular myocytes},\ }\href@noop {} {\bibfield
  {journal} {\bibinfo  {journal} {Archives of biochemistry and biophysics}\
  }\textbf {\bibinfo {volume} {604}},\ \bibinfo {pages} {11} (\bibinfo {year}
  {2016})}\BibitemShut {NoStop}%
\bibitem [{\citenamefont {Davia}\ \emph {et~al.}(2001)\citenamefont {Davia},
  \citenamefont {Bernobich}, \citenamefont {Ranu}, \citenamefont {del Monte},
  \citenamefont {Terracciano}, \citenamefont {MacLeod}, \citenamefont
  {Adamson}, \citenamefont {Chaudhri}, \citenamefont {Hajjar},\ and\
  \citenamefont {Harding}}]{davia2001serca2a}%
  \BibitemOpen
  \bibfield  {author} {\bibinfo {author} {\bibfnamefont {K.}~\bibnamefont
  {Davia}}, \bibinfo {author} {\bibfnamefont {E.}~\bibnamefont {Bernobich}},
  \bibinfo {author} {\bibfnamefont {H.~K.}\ \bibnamefont {Ranu}}, \bibinfo
  {author} {\bibfnamefont {F.}~\bibnamefont {del Monte}}, \bibinfo {author}
  {\bibfnamefont {C.~M.}\ \bibnamefont {Terracciano}}, \bibinfo {author}
  {\bibfnamefont {K.~T.}\ \bibnamefont {MacLeod}}, \bibinfo {author}
  {\bibfnamefont {D.~L.}\ \bibnamefont {Adamson}}, \bibinfo {author}
  {\bibfnamefont {B.}~\bibnamefont {Chaudhri}}, \bibinfo {author}
  {\bibfnamefont {R.~J.}\ \bibnamefont {Hajjar}},\ and\ \bibinfo {author}
  {\bibfnamefont {S.~E.}\ \bibnamefont {Harding}},\ }\bibfield  {title}
  {\bibinfo {title} {Serca2a overexpression decreases the incidence of
  aftercontractions in adult rabbit ventricular myocytes},\ }\href@noop {}
  {\bibfield  {journal} {\bibinfo  {journal} {Journal of molecular and cellular
  cardiology}\ }\textbf {\bibinfo {volume} {33}},\ \bibinfo {pages} {1005}
  (\bibinfo {year} {2001})}\BibitemShut {NoStop}%
\bibitem [{\citenamefont {Singer}\ \emph {et~al.}(1967)\citenamefont {Singer},
  \citenamefont {Lazzara},\ and\ \citenamefont
  {HOFFMAN}}]{singer1967interrelationships}%
  \BibitemOpen
  \bibfield  {author} {\bibinfo {author} {\bibfnamefont {D.~H.}\ \bibnamefont
  {Singer}}, \bibinfo {author} {\bibfnamefont {R.}~\bibnamefont {Lazzara}},\
  and\ \bibinfo {author} {\bibfnamefont {B.~F.}\ \bibnamefont {HOFFMAN}},\
  }\bibfield  {title} {\bibinfo {title} {Interrelationships between
  automaticity and conduction in purkinje fibers},\ }\href@noop {} {\bibfield
  {journal} {\bibinfo  {journal} {Circulation research}\ }\textbf {\bibinfo
  {volume} {21}},\ \bibinfo {pages} {537} (\bibinfo {year} {1967})}\BibitemShut
  {NoStop}%
\bibitem [{\citenamefont {Liu}\ \emph {et~al.}(2015)\citenamefont {Liu},
  \citenamefont {de~Lange}, \citenamefont {Garfinkel}, \citenamefont {Weiss},\
  and\ \citenamefont {Qu}}]{liu2015delayed}%
  \BibitemOpen
  \bibfield  {author} {\bibinfo {author} {\bibfnamefont {M.~B.}\ \bibnamefont
  {Liu}}, \bibinfo {author} {\bibfnamefont {E.}~\bibnamefont {de~Lange}},
  \bibinfo {author} {\bibfnamefont {A.}~\bibnamefont {Garfinkel}}, \bibinfo
  {author} {\bibfnamefont {J.~N.}\ \bibnamefont {Weiss}},\ and\ \bibinfo
  {author} {\bibfnamefont {Z.}~\bibnamefont {Qu}},\ }\bibfield  {title}
  {\bibinfo {title} {Delayed afterdepolarizations generate both triggers and a
  vulnerable substrate promoting reentry in cardiac tissue},\ }\href@noop {}
  {\bibfield  {journal} {\bibinfo  {journal} {Heart rhythm}\ }\textbf {\bibinfo
  {volume} {12}},\ \bibinfo {pages} {2115} (\bibinfo {year}
  {2015})}\BibitemShut {NoStop}%
\bibitem [{\citenamefont {Langer}(1994)}]{langer1994myocardial}%
  \BibitemOpen
  \bibfield  {author} {\bibinfo {author} {\bibfnamefont {G.~A.}\ \bibnamefont
  {Langer}},\ }\bibfield  {title} {\bibinfo {title} {Myocardial calcium
  compartmentation},\ }\href@noop {} {\bibfield  {journal} {\bibinfo  {journal}
  {Trends in Cardiovascular Medicine}\ }\textbf {\bibinfo {volume} {4}},\
  \bibinfo {pages} {103} (\bibinfo {year} {1994})}\BibitemShut {NoStop}%
\bibitem [{\citenamefont {Chu}\ \emph {et~al.}(2016)\citenamefont {Chu},
  \citenamefont {Greenstein},\ and\ \citenamefont {Winslow}}]{chu2016modeling}%
  \BibitemOpen
  \bibfield  {author} {\bibinfo {author} {\bibfnamefont {L.}~\bibnamefont
  {Chu}}, \bibinfo {author} {\bibfnamefont {J.~L.}\ \bibnamefont
  {Greenstein}},\ and\ \bibinfo {author} {\bibfnamefont {R.~L.}\ \bibnamefont
  {Winslow}},\ }\bibfield  {title} {\bibinfo {title} {Modeling na+-ca2+
  exchange in the heart: Allosteric activation, spatial localization, sparks
  and excitation-contraction coupling},\ }\href@noop {} {\bibfield  {journal}
  {\bibinfo  {journal} {Journal of molecular and cellular cardiology}\ }\textbf
  {\bibinfo {volume} {99}},\ \bibinfo {pages} {174} (\bibinfo {year}
  {2016})}\BibitemShut {NoStop}%
\bibitem [{\citenamefont {Maruyama}\ \emph {et~al.}(2010)\citenamefont
  {Maruyama}, \citenamefont {Joung}, \citenamefont {Tang}, \citenamefont
  {Shinohara}, \citenamefont {On}, \citenamefont {Han}, \citenamefont {Choi},
  \citenamefont {Kim}, \citenamefont {Shen}, \citenamefont {Weiss} \emph
  {et~al.}}]{maruyama2010diastolic}%
  \BibitemOpen
  \bibfield  {author} {\bibinfo {author} {\bibfnamefont {M.}~\bibnamefont
  {Maruyama}}, \bibinfo {author} {\bibfnamefont {B.}~\bibnamefont {Joung}},
  \bibinfo {author} {\bibfnamefont {L.}~\bibnamefont {Tang}}, \bibinfo {author}
  {\bibfnamefont {T.}~\bibnamefont {Shinohara}}, \bibinfo {author}
  {\bibfnamefont {Y.-K.}\ \bibnamefont {On}}, \bibinfo {author} {\bibfnamefont
  {S.}~\bibnamefont {Han}}, \bibinfo {author} {\bibfnamefont {E.-K.}\
  \bibnamefont {Choi}}, \bibinfo {author} {\bibfnamefont {D.-H.}\ \bibnamefont
  {Kim}}, \bibinfo {author} {\bibfnamefont {M.~J.}\ \bibnamefont {Shen}},
  \bibinfo {author} {\bibfnamefont {J.~N.}\ \bibnamefont {Weiss}}, \emph
  {et~al.},\ }\bibfield  {title} {\bibinfo {title} {Diastolic intracellular
  calcium-membrane voltage coupling gain and postshock arrhythmias: role of
  purkinje fibers and triggered activity},\ }\href@noop {} {\bibfield
  {journal} {\bibinfo  {journal} {Circulation research}\ }\textbf {\bibinfo
  {volume} {106}},\ \bibinfo {pages} {399} (\bibinfo {year}
  {2010})}\BibitemShut {NoStop}%
\end{thebibliography}%

\end{document}